\documentclass[twocolumn]{aastex631}

\graphicspath{{./}{figures/}}
\usepackage[utf8]{inputenc}
\usepackage{float}
\hypersetup{
    colorlinks=true,
    linkcolor=blue,
    filecolor=magenta,      
    urlcolor=cyan,
    pdftitle={Overleaf Example},
    pdfpagemode=FullScreen,
    }
\usepackage{textcomp, gensymb}
\DeclareUnicodeCharacter{2212}{-}

\shorttitle{Spectroscopy of Aquarius II and \text{Bo\"{o}tes II}}
\shortauthors{Bruce et al.}

\begin{document}

\title{Spectroscopic analysis of Milky Way outer halo satellites: Aquarius II and \text{Bo\"{o}tes II}}

\author{Jordan Bruce}
\affiliation{David A. Dunlap Department of Astronomy \& Astrophysics, University of Toronto, 50 St. George Street, Toronto, ON, M5S3H4 Canada}

\author{Ting S. Li}\email{ting.li@astro.utoronto.ca}
\affiliation{David A. Dunlap Department of Astronomy \& Astrophysics, University of Toronto, 50 St. George Street, Toronto, ON, M5S3H4 Canada}
\affiliation{Dunlap Institute for Astronomy \& Astrophysics, University of Toronto, 50 St George Street, Toronto, ON M5S 3H4, Canada}

\author{Andrew B. Pace}
\affiliation{McWillliams Center for Cosmology, Carnegie Mellon University, 5000 Forbes Ave, Pittsburgh, PA 15213, USA}

\author{Mairead Heiger}
\affiliation{David A. Dunlap Department of Astronomy \& Astrophysics, University of Toronto, 50 St. George Street, Toronto, ON, M5S3H4 Canada}
\affiliation{Dunlap Institute for Astronomy \& Astrophysics, University of Toronto, 50 St George Street, Toronto, ON M5S 3H4, Canada}

\author{Ying-Yi Song}
\affiliation{David A. Dunlap Department of Astronomy \& Astrophysics, University of Toronto, 50 St. George Street, Toronto, ON, M5S3H4 Canada}
\affiliation{Dunlap Institute for Astronomy \& Astrophysics, University of Toronto, 50 St George Street, Toronto, ON M5S 3H4, Canada}

\author{Joshua D. Simon}
\affiliation{Observatories of the Carnegie Institution for Science, 813 Santa Barbara St., Pasadena, CA 91101, USA}




\begin{abstract}
\noindent In this paper we present a chemical and kinematic analysis of two ultra-faint dwarf galaxies (UFDs), Aquarius II (Aqu~II) and \text{Bo\"{o}tes II} (Boo~II), using Magellan/IMACS spectroscopy. We present the largest sample of member stars for Boo~II (12), and the largest sample of red-giant-branch members with metallicity measurements for Aqu~II (8).
In both UFDs, over 80\% of targets selected based on $Gaia$ proper motions turned out to be spectroscopic members. In order to maximize the accuracy of stellar kinematic measurements, we remove the identified binary stars and RR Lyrae variables. For Aqu~II we measure a systemic velocity of $-65.3 \pm 1.8$ km s$^{-1}$ and a metallicity of [Fe/H] = $-2.57^{+0.17}_{-0.17}$. When compared with previous measurements, these values display a $\sim 6$ km s$^{-1}$ difference in radial velocity and a decrease of 0.27 dex in metallicity. Similarly for Boo~II, we measure a systemic velocity of $-130.4^{+1.4}_{-1.1}$ km s$^{-1}$, more than 10 km s$^{-1}$ different from the literature, a metallicity almost 1 dex smaller at [Fe/H] = $-2.71^{+0.11}_{-0.10}$, and a velocity dispersion 3 times smaller at $\sigma_{v_{\rm hel}} = 2.9^{+1.6}_{-1.2}$ km s$^{-1}$. Additionally, we derive systemic proper motion parameters and model the orbits of both UFDs. 
Finally, we highlight the extremely dark matter dominated nature of Aqu~II and compute the J-factor for both galaxies to aid searches of dark matter annihilation.
Despite the small size and close proximity of Boo~II, it is an intermediate target for the  indirect detection of dark matter annihilation due to its low velocity dispersion and corresponding low dark matter density.
\end{abstract}



\section{Introduction}\label{sec:1}
Ultra-faint dwarf galaxies (UFDs) are among the oldest, most metal-poor, least luminous and most dark matter dominated systems known \citep[]{doi:10.1146/annurev-astro-091918-104453}. They provide excellent conditions for a diverse array of research including generating or assessing models of galaxy evolution, galaxy formation, dark matter, dark matter interactions, and early universe star formation. The extremely low metallicity of these UFDs provides ideal conditions to study early universe chemical evolution \citep[]{2014ApJ...786...74F}. With stellar ages nearly as old as the universe \citep[]{oay+08, 2012ApJ...753L..21B, weisz_2014, brown_2014, Gallart_2021}, UFDs provide compact and nearby examples of ancient star formation in some of the oldest galaxies, acting as relics of galaxy formation before reionization \citep[]{2009ApJ...693.1859B, 2012ApJ...753L..21B, 2021ApJ...914L..10T}. UFDs also present many opportunities for studies related to dark matter and dark matter interactions, such as detecting dark-matter annihilation and modeling dark-matter distributions using resolved kinematics \citep[]{2018RPPh...81e6901S, Pace2019MNRAS.482.3480P, 2020ApJ...893...21S, 2020ApJ...904...45H}. Additionally, analyzing the tidal disruption of UFDs gives insight on the structure and shape of the Milky Way's dark matter halo \citep[]{2016ApJ...833...31B, 2018ApJ...866...22L}.

Understanding the kinematics and chemical composition of UFDs can be approached using photometry and astrometry such as in \citet[]{2022arXiv220505699P}, which prioritizes large samples of data but is limited by reduced precision. Alternatively, spectroscopy generally offers a much deeper analysis with significantly increased depth and accuracy, with the limitation of a reduced sample size of observed member stars \citep[]{Simon_2017, Li_2017, 2022arXiv220604580C, 2022arXiv220311788C}. Furthermore, spectroscopy can be used to gain insight on the nature of unclassified satellites as either a UFD or globular cluster (GC) through kinematic and chemical analysis. Previous spectroscopic studies of UFDs like Aquarius II (Aqu~II) and \text{Bo\"{o}tes II} (Boo~II) have been hindered by this limited sample size preventing the determination of reliable systemic properties. These two galaxies are ideal candidates for further spectroscopic study because of (1) their uncertain characteristics in the previous studies, and (2) the large number of bright unobserved high probability member candidates \citep[]{2022arXiv220505699P}. These high probability member candidates were identified using astrometric data from the third data release (DR3) of $Gaia$ \citep[]{2022arXiv220800211G}, and could significantly improve the uncertainties in characteristics.

Aqu~II was discovered by \citet[]{Torrealba_2016} using the VST ATLAS \citep[]{2015MNRAS.451.4238S} and SDSS \citep[]{2012ApJS..203...21A} surveys and is of scientific interest due to being one of the most difficult UFDs to find. With an absolute magnitude of $M_V = -4.36 \pm 0.14$ and a distance of $\mathrm{D}_{\odot} = 107.9 \pm 3.3$ kpc, Aqu~II lies close to the current detection limit and supports the claim that there are still further and fainter UFDs yet unidentified (e.g., \citet[]{drlica_2021, msc+21}). \citet[]{Torrealba_2016} combined deep imaging with the IMACS camera on the 6.5m Baade telescope with spectroscopy obtained with DEIMOS on Keck to measure initial structural parameters such as the relatively large half-light radius ($r_{1/2} = 159 \pm 24$ pc). This large half-light radius generated the initial classification of Aqu~II as a dwarf galaxy since GCs never have such large sizes (GC $r_{1/2} < 20$ pc at $M_V \sim -5$) \citep[]{harris_2010}. \citet[]{Torrealba_2016} also reports global properties using 9 spectroscopically confirmed member stars (4 red giant branch (RGB) and 5 blue horizontal branch (BHB) members). They measure a systemic velocity of $v_{\rm hel} = -71.1 \pm 2.5 \;\mathrm{km\;s}^{-1}$ with a dispersion of $\sigma_{v_{\rm hel}} = 5.4^{+3.4}_{-0.9} \;\mathrm{km\;s}^{-1}$, a metallicity of [Fe/H] = $-2.3 \pm 0.5$ (only RGB members were used for systemic metallicity) and no reported metallicity dispersion. Based on proper motion and photometric data, \citet[]{2022arXiv220505699P} identified 15 likely Aqu~II candidate member stars (with membership probability P $>$ 0.5) from {\it Gaia} DR3, 12 of which have not been spectroscopically observed. Further spectroscopic investigation is therefore warranted to increase the sample size of RGB member stars, improve systemic values, and resolve a metallicity dispersion that provides insight on whether Aqu~II has undergone multiple epochs of star formation.

Boo~II was discovered in SDSS imaging by \citet[]{2007ApJ...662L..83W} and initially unclassified as either a UFD or GC. However, a further spectroscopic study by \citet[]{Koch_2008} argued that Boo~II was likely a UFD. Additional high-resolution spectroscopy was conducted by \citet[]{koch_2014} and \citet[]{2016ApJ...817...41J} which provided chemical evidence that supported this classification. An additional photometric study by \citet[]{2018ApJ...860...66M} further confirmed Boo~II as a dwarf galaxy candidate due to its relatively large half-light radius ($r_{1/2} = 38.7 \pm 5.1$ pc). A photometric and spectroscopic study conducted by \citet[]{Koch_2008} identified five member stars with a systemic velocity of $v_{\rm hel} = -117 \pm 5.2 \;\mathrm{km\;s}^{-1}$, an unexpectedly inflated dispersion of $\sigma_{v_{\rm hel}} = 10.5 \pm 7.4 \;\mathrm{km\;s}^{-1}$, and a systemic metallicity of [Fe/H] $= -1.79$. This dispersion is extremely large compared to other UFDs with similar properties \citep[]{doi:10.1146/annurev-astro-091918-104453}. Using a photometric halo mass estimation technique, \citet[]{2022arXiv221202948Z} estimate the velocity dispersion to be much smaller at $\sigma_{v_{\rm hel}} = 3.4$ km s$^{-1}$. \citet[]{2016ApJ...817...41J} conducted high-resolution spectroscopy of the four brightest member stars and determined that one of the member stars is a spectroscopic binary which can cause an artificially inflated velocity dispersion when considered in kinematic analysis \citep[]{Pianta2022ApJ...939....3P}. Using the results from \citet[]{2016ApJ...817...41J}, \citet[]{doi:10.1146/annurev-astro-091918-104453} reports a systemic metallicity of [Fe/H] = $-2.79 \pm 0.08$ with a dispersion of $\sigma_{\mathrm{[Fe/H]}} < 0.35$. Despite identifying the binary star which most likely contributed to the inflated the velocity dispersion, further follow-up observations are required as the limited sample size prevented an improved velocity dispersion from being measured. According to \citet[]{2022arXiv220505699P}, there are 21 likely Boo~II member candidates located in {\it Gaia} DR3, and 17 have not been spectroscopically observed.

\begin{deluxetable*}{c c c c c c c c c}
\tablecolumns{9}
\centering
\tablewidth{0pt}
\tablecaption{\hypertarget{Table 1}{Observation Details}.}\label{tab:1}
\tablehead{
\colhead{Mask} & \colhead{Date} & \colhead{MJD\tablenotemark{a}} & \colhead{R.A. (h:m:s)\tablenotemark{b}} & \colhead{Decl. (d:m:s)\tablenotemark{b}} & \colhead{Exposure (s)} & \colhead{Seeing} & \colhead{\# Observed Spectra} & \colhead{\# Useful Spectra\tablenotemark{c}}}
\startdata
Aqu II & June 18, 2021 & 59384.35 & 22:33:53.079 & -09:18:08.63 & 9600 & 1.0"-1.4" & 55 & 12 \\
Aqu II & July 13, 2021 & 59409.26 & 22:33:53.079 & -09:18:08.63 & 4800 & 0.8" & 55 & 12 \\
Boo II & June 18, 2021 & 59383.80 & 13:57:56.000 & 12:50:50.00 & 9600 & 0.8" & 43 & 17 \\
\enddata
\tablenotetext{a}{We list the midpoint MJD of observations for each night.}
\tablenotetext{b}{We list the central mask location coordinates.}
\tablenotetext{c}{Useful spectra are those generating reliable velocity measurements. Most of the faintest targets do not have reliable velocity measurements due to the low signal-to-noise ratio of the spectra.}
\end{deluxetable*} 
\begin{deluxetable*}{l c c c c}
\tablecolumns{5}
\tablewidth{0pt}
\tabletypesize{\small}\label{tab:2}
\tablecaption{Properties of Aquarius II and Bo\"{o}tes II}
\tablehead{\colhead{Property} & \colhead{Description} &  \colhead{Aquarius II} & \colhead{Bo\"{o}tes II} & \colhead{Units}}
\startdata
$\alpha_{2000}$ & Right Ascension &  $338.4813 \pm 0.005$\tablenotemark{a} & $209.5141 \pm 0.005$\tablenotemark{b} & deg \\
$\delta_{2000}$ & Declination &  $-9.3274 \pm 0.005$\tablenotemark{a} & $12.8553 \pm 0.006$\tablenotemark{b} & deg \\
$r_h$ & Angular Half-Light Radius & $5.1 \pm 0.8$\tablenotemark{a} & $3.17 \pm 0.42$\tablenotemark{b} & arcmin \\
$r_{1/2}$ & Physical Half-Light Radius & $159 \pm 24$\tablenotemark{a}  & $ 38.7 \pm 5.1$\tablenotemark{b} & pc \\
$\epsilon$ & Ellipticity & $0.39 \pm 0.09$\tablenotemark{a} & $0.25 \pm 0.11$\tablenotemark{b} & --\\
P.A. & Position Angle & $ 121 \pm 9$\tablenotemark{a} & $−68 \pm 27$\tablenotemark{b} & deg \\
$m-M$  & Distance Modulus & $20.16 \pm 0.07$\tablenotemark{a} & 18.12 $\pm$ 0.06\tablenotemark{b} & mag\\
$D_{\odot}$  & Distance (Heliocentric) & $107.9 \pm 3.3$\tablenotemark{a} & 42.0 $\pm$ 2.0\tablenotemark{b} & kpc \\
$M_v$  & Absolute Magnitude (V-band) & $-4.36 \pm 0.14$\tablenotemark{a} & $-2.9 \pm 0.7$\tablenotemark{b} & mag \\
\hline
$N_{reported}$ &  Number of Observed Members &  8 & 12 & -- \\ 
$E(B-V)$ & Average Reddening & 0.07\tablenotemark{c} & 0.03\tablenotemark{c} & mag \\
$v_{\rm hel}$ &  Systemic Velocity (Heliocentric) &  $-65.3 \pm 1.8$ & $-130.4^{+1.4}_{-1.1}$ & $\mathrm{km\;s}^{-1}$ \\
$\sigma_{v_{\rm hel}}$ & Velocity Dispersion &  $4.7^{+1.8}_{-1.2}$ & $2.9^{+1.6}_{-1.2}$ & $\mathrm{km\;s}^{-1}$ \\
$[\rm Fe/H]$ & Systemic Metallicity & $-2.57^{+0.17}_{-0.17}$ & $-2.71^{+0.11}_{-0.10}$ & dex \\
$\sigma_{[\rm Fe/H]}$ &  Metallicity Dispersion & $0.36^{+0.20}_{-0.14}$ & $< 0.37$ & dex \\
$\mathrm{M}_{1/2}$ & Dynamical Mass Inside $r_{1/2}$ & $3.1^{+3.0}_{-1.4} \times 10^6$ & $3.1^{+4.4}_{-2.1} \times 10^5$ & $M_{\odot}$ \\
$M_{1/2}/L_{v, 1/2}$ & Mass-to-Light Ratio & $1300^{+1300}_{-600}$ & $460^{+1000}_{-440}$ & $M_{\odot}/L_{\odot}$\\ 
\hline 
$\mu_{\alpha}$ & Systemic Proper Motion In R.A. & $-0.27\pm0.12$ & $-2.50 \pm 0.07$ & $\mathrm{mas\;yr}^{-1}$ \\
$\mu_{\delta}$ & Systemic Proper Motion In Decl. &  $-0.44 \pm 0.10$ & $-0.46 \pm 0.06$& $\mathrm{mas\;yr}^{-1}$ \\
$r_{\rm apo}$ & Orbit Apocenter & $148^{+119}_{-35}$ & $203^{+61}_{-56}$ & kpc \\
$r_{\rm peri}$ & Orbit Pericenter & $96^{+7}_{-35}$ & $38^{+1}_{-2}$ & kpc \\
$ e $ & Orbit Eccentricity & $0.30^{+0.20}_{-0.11}$ & $0.69^{+0.06}_{-0.08}$ & -- \\
$r^{\rm LMC}_{\rm apo}$ & Orbit Apocenter (With LMC) & $159^{+114}_{-31}$ & $194^{+64}_{-53}$ & kpc \\
$r^{\rm LMC}_{\rm peri}$ & Orbit Pericenter (With LMC) & $88^{+14}_{-39}$ & $36 \pm 2$ & kpc \\
$ e^{\rm LMC} $ & Orbit Eccentricity (With LMC) & $0.37^{+0.18}_{-0.08}$ & $0.68^{+0.06}_{-0.08}$ & -- \\
\hline
$\log_{10}{J(0.2\degree)}$ &  Integrated J-factor within a solid angle of $0.2\degree$ & $17.7^{+0.6}_{-0.5}$ & $18.1^{+0.8}_{-1.1}$ & ${\rm GeV^{2}~cm^{-5}}$\\
$\log_{10}{J(0.5\degree)}$ & Integrated J-factor within a solid angle of $0.5\degree$ & $17.8^{+0.6}_{-0.5}$ & $18.3^{+0.8}_{-1.1}$ & ${\rm GeV^{2}~cm^{-5}}$ \\
\enddata
\tablenotetext{a}{Most recent measurements reported in the Aqu II discovery paper \citep[]{Torrealba_2016}.}
\tablenotetext{b}{Most recent measurements for Boo II reported in the 2018 photometric study by \citet[]{2018ApJ...860...66M}.}
\tablenotetext{c}{The reddening was computed using the average of individual star reddening based on values from \citet[]{schlafly_2011}.}
\end{deluxetable*}

In this paper, we use spectroscopic data from Magellan/IMACS to identify member stars and resolve systemic properties in both Aqu~II and Boo~II. We spectroscopically confirm the identity of 8 members in Aqu~II and 12 members in Boo~II. These member stars are used to support the classification of these satellites as UFDs, improve measurements of systemic characteristics, and investigate orbital properties. 

The paper is organized in the following way: Section \ref{sec:2}
describes the observation process, instrumental setup, and data reduction methods. Section \ref{sec:3} describes the processes used to identify members and determine radial velocities and metallicities. Section \ref{sec:4} discusses the scientific results and conducts an orbit analysis of the two satellites. Section \ref{sec:5} summarizes the findings and concludes the paper. Throughout the paper, unless specified otherwise, we use the astrometric data from $Gaia$ DR3 \citep[]{2022arXiv220800211G}, and photometric data from Dark Energy Camera Legacy Survey (DECaLS) DR9 \citep[]{2019AJ....157..168D}. The photometry is dereddened using the color excess $E(B-V)$ from \citet{schlafly_2011} multiplied by the extinction coefficients $R_g = 3.186$ and $R_r = 2.140$ \citep{desdr1}.

\section{OBSERVATIONS AND DATA}\label{sec:2}
\subsection{Observation Details}\label{subsec:2.1}

\hypertarget{OBSERVATIONS AND DATA}{Observations} were taken using the IMACS spectrograph \citep[]{10.1117/12.670573} on the Magellan-Baade telescope in Chile on June 18 and July 13, 2021. Following the instrument setup outlined in \citet[]{Simon_2017} and \citet[]{Li_2017}, we used the f/4 camera to attain a 15.4\arcmin\ by 7\arcmin\ field-of-view, coupled with the 1200 $\ell$ mm$^{-1}$ grating tilted at an angle of 32.4$^\circ$ and a WB5600-9200 filter. This setup provided a resolution of R $\sim$ 11,000 for the 0.7" x 5" slit size used. The wavelength range 7550-8750 \text{\AA} was chosen to adequately cover the calcium triplet (CaT) absorption lines ($\sim$ 8500 \text{\AA}) used to measure radial velocities and metallicities of target stars, and the telluric A-band absorption lines ($\sim$ 7600 \text{\AA}) used to correct for spectral shifts resulting from a mis-centering of stars within their slits.

The observations included four 2400 second science exposures each for Aqu~II and Boo~II on June 18 with an average seeing of $\sim$ 1.1", and two 2400 second science exposures for Aqu~II on July 13 with an average seeing of 0.8" (refer to \hyperlink{Table 1}{Table 1}). Following the methods detailed in \citet[]{Li_2017}, a wavelength calibration frame and a flat-field frame were captured immediately following each pair of science exposures. For wavelength calibration, we used He, Ne, and Ar comparison lamps as well as a Kr lamp, which has strong lines in the region neighbouring the telluric absorption lines \citep[]{Li_2017}.

\begin{figure*}
    \centering
    \includegraphics[width = 0.7\textwidth]{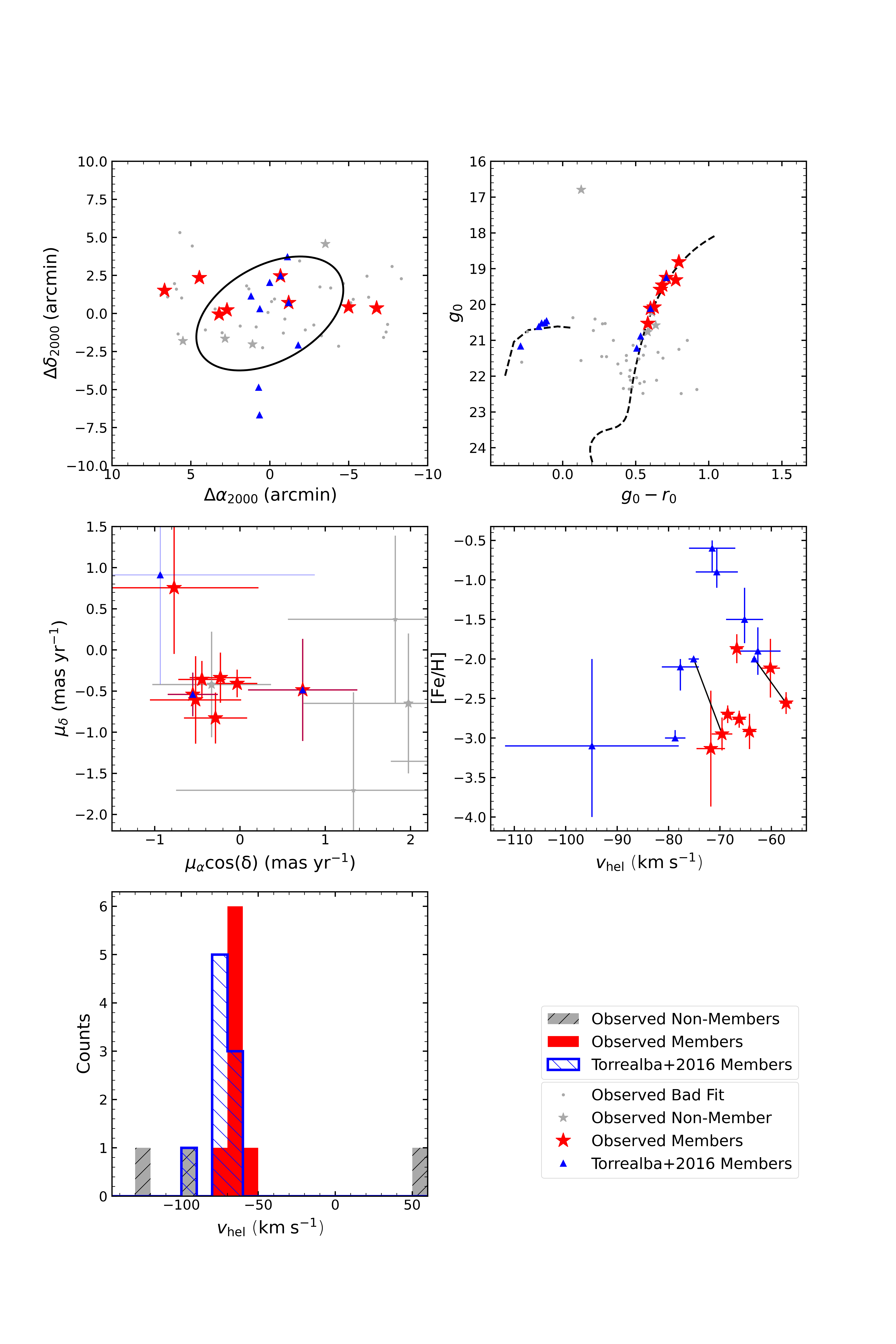}
    \caption{\hypertarget{Figure 1}{Top} left: Spatial position of 55 observed stars and 9 member stars from \citet[]{Torrealba_2016} for Aqu~II. The dashed ellipse displays the half-light radius plotted using the ellipticity, P.A., and $r_h$ from \citet[]{Torrealba_2016}. Top right: Color-magnitude diagram displaying a Dartmouth isochrone \citep[]{2008ApJS..178...89D} with an age of 12.5 Gyr and a metallicity of [Fe/H] = $-2.49$ plotted using the distance modulus of 20.16 from \citet[]{Torrealba_2016}. Middle left: Proper motion distribution using {\it Gaia} DR3 data \citep[]{2016A&A...595A...1G, 2021A&A...649A...1G, 2021A&A...649A...2L} for stars when available. Middle right: Velocity ($v_{\rm hel}$) and metallicity ([Fe/H]) distribution, with black lines connecting identical stars observed in this work and \citet[]{Torrealba_2016}. Bottom left: Histogram of observed velocities ($v_{\rm hel}$) and \citet[]{Torrealba_2016} velocities.}
    \label{fig:1}
\end{figure*}

We selected the targets from three categories. First, we selected all member candidates with membership probability $P>0.2$ from \citet[]{2022arXiv220505699P} as the highest priority targets. We then picked other possible member candidates whose proper motions are consistent with the UFDs (but are not included in \citet[]{2022arXiv220505699P}) as the next priority. Finally, we included fainter candidates in DECaLS guided by an old metal-poor Dartmouth isochrone \citep[]{2008ApJS..178...89D} as the lowest priority, whose astrometric information is not available in $Gaia$. We designed one slitmask for each UFD, placing the center and slit position angle of the mask that maximized the number of targets in the highest priority category. In total, we selected 55 stars for Aqu~II and 43 stars for Boo~II. However, we note that many of the targets are in the lowest priority with no $Gaia$ data available. Due to the low signal-to-noise ratio (S/N) on these faint targets, we were not able to obtain good velocity measurements for stars in the lowest category, returning a low number in the useful spectra in \hyperlink{Table 1}{Table~1}.
    
\subsection{Data Reduction}\label{subsec:2.2}
The IMACS spectra were reduced following the methods outlined in \citet[]{Li_2017}. This process involves subtracting bias, removing read-out pattern noise, and using the COSMOS reduction pipeline \citep[]{Dressler_2011, oemler_2017} to map the slits and generate a precursory wavelength solution. We then generated an enhanced wavelength calibration and spectral extraction using an IMACS pipeline \citep[]{Simon_2017} derived from the DEEP2 data reduction pipeline \citep[]{2012ascl.soft03003C}. Each set of science exposures was uniquely reduced, resulting in extracted 1-D spectra that were combined using a weighted average to attain the final coadded spectra for each night. These spectra were subsequently normalized to unity by fitting a quadratic polynomial. For the 55 Aqu~II targets and 43 Boo~II targets, 52 and 39 spectra were extracted respectively, with several stars falling into a chip gap. The complete observation and slit mask details are outlined in \hyperlink{Table 1}{Table~1}. 

\begin{figure*}
    \centering
    \includegraphics[width = 0.7\textwidth]{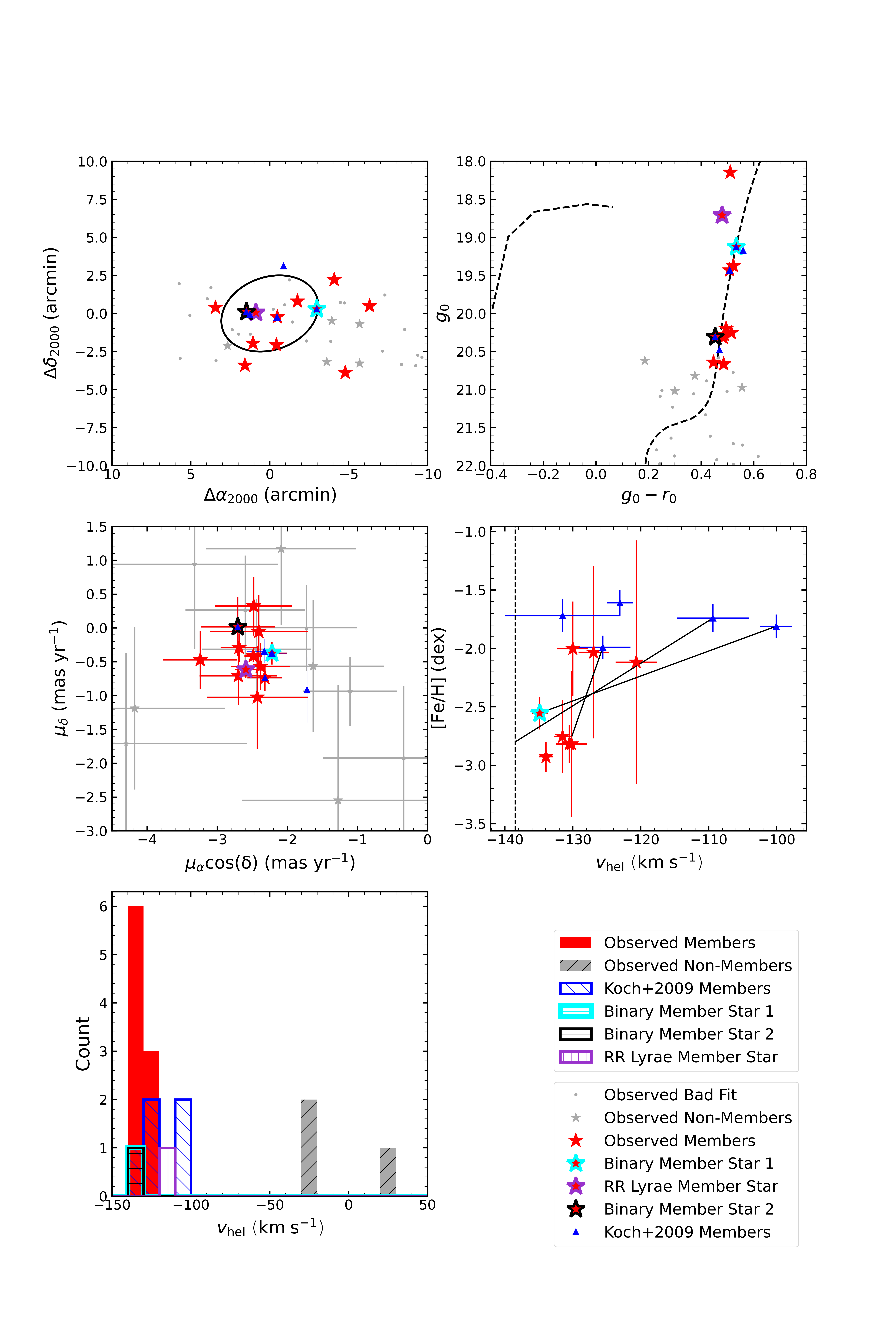}
    \caption{\hypertarget{Figure 2}{Top} left: Spatial position of 43 observed stars and 5 member stars from \citet[]{Koch_2008} for Boo~II. The dashed ellipse displays the half-light radius plotted using the ellipticity, P.A., and $r_h$ from \citet[]{2018ApJ...860...66M}. Top right: Color-magnitude diagram displaying a Dartmouth isochrone \citep[]{2008ApJS..178...89D} with an age of 12.5 Gyr and a metallicity of [Fe/H] = $-2.49$ plotted using the distance modulus of 18.12 from \citet[]{Walsh_2008}. Middle left: Proper motion distribution using {\it Gaia} DR3 data \citep[]{2016A&A...595A...1G, 2021A&A...649A...1G, 2021A&A...649A...2L} for stars when available. Middle right: Velocity ($v_{\rm hel}$) and metallicity ([Fe/H]) distribution of members from this work and \citet[]{Koch_2008} with well fit metallicity measurements, with black lines connecting stars observed in both studies. The black vertical line represents binary member star 2 which had a low S/N ratio that prevented accurate metallicity measurements. Bottom left: Histogram of observed velocity ($v_{\rm hel}$) and \citet[]{Koch_2008} velocities. Binary star 1 refers to the binary identified by \citet[]{2016ApJ...817...41J}, whereas binary star 2 refers to the suspected binary identified in this work.}
    \label{fig:2}
\end{figure*}

\section{Analysis}\label{sec:3}
\subsection{Radial Velocities}\label{subsec:3.1}
\hypertarget{Analysis}{We} determined stellar radial velocities following the methods detailed in \citet[]{Li_2017}, fitting a set of templates to the reduced spectra using \texttt{$emcee$}, a Markov Chain Monte Carlo (MCMC) sampler \citep[]{2013PASP..125..306F}. Three spectral templates were fit for each star; the metal-poor RGB star HD 122563, the BHB star HD 161817, and the moderately metal-poor RGB star HD 26297. Each template was shifted through a range of velocities to find the radial velocity $v_{\rm rad}$ value that maximized the log likelihood function 
    \begin{equation} \label{eq:1}
        \rm{log}L = -\frac{1}{2}\Sigma^{\lambda_2}_{\lambda = \lambda_1}\frac{[S(\lambda) - T(\lambda(1 + \frac{v_{\rm rad}}{c}))]^2}{\sigma^2_{\rm spec}(\lambda)}.
    \end{equation}
    
\noindent Above, S($\lambda$) and $\sigma^2_{\rm spec}(\lambda)$ reflect the normalized spectrum and its variance and T($\lambda(1-\frac{v_{\rm rad}}{c})$) represents the normalized template. The wavelength bounds of the fit were set as $\lambda_1= 8450$ \text{\AA} to $\lambda_2 = 8685$ \text{\AA} since the CaT spectral feature was predominantly used to measure radial velocity.

For each star, we ran the MCMC sampler with 20 walkers each making 1000 steps, and a burn in stage of 50 steps which was omitted. We then removed 5$\sigma$ outliers, and report the median and standard deviation of the posterior distributions as the radial velocity ($v_{\rm rad}$) and its error ($\sigma_{v_{\rm rad}}$) respectively. 

Next we measured and applied a telluric correction ($v_{\rm tell}$) to account for radial velocity shifts arising from a mis-centering of stars within their slits. The MCMC sampler was run again using the methods previously described, using a telluric template of the hot, rapidly rotating star HR 4781 \citep[]{Simon_2017} over the range $\lambda_1 = 7550$ \text{\AA} to $\lambda_2 = 7700$ \text{\AA}. Similarly, the median and standard deviation of the posterior distribution are reported as the telluric correction ($v_{\rm tell}$) and its uncertainty ($\sigma_{v_{\rm tell}}$) respectively.

The observed radial velocity ($v$) was measured by subtracting the A-band correction, $v = v_{\rm rad} - v_{\rm tell}$. We note that $\sigma_{v_{\rm rad}}$ and $\sigma_{v_{\rm tell}}$ are solely statistical and uniquely associated with the measurements and S/N of the 1-D spectra. We also include a systematic uncertainty ($\sigma_{\rm sys}$) of 1.0 km s$^{-1}$ previously identified in \citet[]{Li_2017} and \citet[]{Simon_2017} by comparing multiple successive IMACS observations using similar instrumental setups. The total velocity uncertainty is defined as the square root of the sum of the squares, $\sigma_v = \sqrt{\sigma_{v_{\rm rad}}^2 + \sigma_{v_{\rm tell}}^2 + \sigma_{\rm sys}^2}$.

Aqu~II was observed on two epochs, June 18, 2021 and July 13, 2021, respectively. The two epochs are far enough apart that the heliocentric corrections from the two epochs are significantly different. Considering the low S/N property of the data, we performed a joint fit with data from both epochs.
The radial velocity templates were simultaneously fit to spectra from each night, and the MCMC sampler was run as previously described. We measured the observed velocity as the velocity that generated the largest overall likelihood from Equation \ref{eq:1} on both nights combined. Individual telluric corrections were applied for each spectrum prior to the simultaneous fitting, and the median and standard deviation of the combined posterior distribution are reported as the observed radial velocity ($v_{\rm rad}$) and its error ($\sigma_{v_{\rm rad}}$) respectively. The statistical uncertainties were similarly summed in quadrature, $\sigma_{v_{\rm stat}} = \sqrt{\sigma_{v_{\rm tell,\;June\;18}}^2 + \sigma_{v_{\rm tell,\;July\;13}}^2 + \sigma_{v_{\rm rad}}^2}$, along with the systemic uncertainty from \citet[]{Li_2017, Simon_2017}, $\sigma_v = \sqrt{\sigma_{v_{\rm stat}}^2 + \sigma_{v_{\rm sys}}^2}$. 

We note that a S/N cut of 2.0 was applied to our set of spectra as the large amount of noise prevents an accurate template fitting. This reduction, as well as visual inspection to remove imprecise measurements, resulted in 12 and 17 useful spectra to be reported for Aqu~II and Boo~II respectively. 

The observed radial velocities were transformed to account for a heliocentric correction based on the location of the dwarf galaxy, the time of observation, and the location of the observatory. The resulting radial velocities in the heliocentric frame ($v_{\rm hel}$) are reported in Tables \ref{tab:3} and \ref{tab:4}, and hereafter referred to as the stellar velocity. Table \ref{tab:3} presents the velocity measurements of the observed stars for Aqu~II, while Table \ref{tab:4} displays the velocity measurements of the observed stars for Boo~II. Our observations of Aqu~II resulted in velocities for 12 stars to be computed, 8 of which were identified as member stars. 17 stars were observed with good velocity measurements during the Boo~II exposures, 12 of which were identified to be members.
    
\subsection{Metallicity Measurements}\label{subsec:3.2}
We determined metallicities of RGB stars using the equivalent widths (EWs) of the CaT absorption lines. Following the method described in \citet[]{Li_2017} and \citet[]{Simon_2017}, we uniquely measured the EW of each CaT line by fitting a Gaussian plus Lorentzian function. We then converted the summed EWs and absolute V-band magnitude of each star into metallicity ([Fe/H]) using a well defined conversion relation from \citet[]{10.1093/mnras/stt1126}. Using photometry from DECaLS, the dereddened g-band ($g_0$) and r-band ($r_0$) magnitudes for each star were converted into apparent V-band magnitude using Equation (5) from \citet[]{Bechtol_2015}, using the distance modulus from previous studies (Table \ref{tab:2}). We determined the metallicity generated from the conversion relation in \citet[]{10.1093/mnras/stt1126}, with the uncertainties measured by summing the statistical and systemic uncertainties in quadrature as outlined in \citet[]{Li_2017}. The statistical uncertainty stems primarily from the Gaussian plus Lorentzian fitting, along with uncertainty in the DECaLS DR9 photometry and distance modulus, which are all propagated through the conversion relation. Uncertainties in the relation parameters as outlined in \citet[]{10.1093/mnras/stt1126} are also included. A systematic uncertainty of $0.2$ \text{\AA} in EWs was also incorporated as determined from repeat measurements in \citet[]{Li_2017}.

Target selection prioritized potential RGB members, resulting in the majority of member stars providing metallicity measurements. A S/N cut of 2.0 was made for spectra once again, and the EW fit of each CaT line was visually inspected to remove stars with bad fitting. For Aqu~II, when both epochs produced verified metallicities a weighted average is reported. In the case of two stars ({\it Gaia} DR3 source IDs: 2609061687357323776 and 2609107798126218880) only the first epoch provided accurate measurements, preventing a weighted average from being determined. Visual inspection yielded 8 RGB stars with good fit EW measurements in Aqu~II observations, and 8 in Boo~II observations. Since the metallicity relation relies on the distance modulus and solely applies to RGB stars, we only report the metallicity of RGB member stars in each UFD. The metallicity values for Aqu~II members are listed in Table \ref{tab:3}, while the metallicity values for Boo~II members are listed in Table \ref{tab:4}.

\begin{deluxetable*}{l c c c c c c c c c c}
\tabletypesize{\scriptsize}
\tablecolumns{9}
\centering
\tablewidth{0pt}
\label{tab:3}
\tablecaption{\hypertarget{Table 3}{Aquarius II} Observed Spectra.}
\tablehead{
\colhead{$Gaia$ DR3 Source ID} & \colhead{R.A. (deg)} & \colhead{Decl. (deg)} & \colhead{$g_0$ (mag)} & \colhead{$r_0$ (mag)} & \colhead{S/N} & \colhead{$v_{\rm hel}$ (km s$^{-1}$)} & \colhead{EW 1\tablenotemark{b}} & \colhead{EW 2\tablenotemark{b}} & \colhead{[Fe/H] (dex)} & \colhead{Member\tablenotemark{a}}}
\startdata
2609060660860398208 & 338.57394 & $-9.35700$ & 20.58 & 19.94 & 5.1 & 50.6 $\pm$ 2.7 & -- & -- & -- & NM \\
2609061068882022784 & 338.52885 & $-9.35453$ & 20.22 & 19.61 & 8.1 & $-92.5$ $\pm$ 1.8 & -- & -- & -- & NM \\
2609061141896425088 & 338.49945 & $-9.36063$ & 20.77 & 20.18 & 4.7 & $-126.0$ $\pm$ 3.5 & -- & -- & -- & NM \\
2609061687357323776 & 338.53523 & $-9.32780$ & 19.32 & 18.54 & 20.5 & $-66.7$ $\pm$ 1.2 & 4.74 $\pm$ 0.39 & -- & $-1.87$ $\pm$ 0.18 & M \\
2609061760371960704 & 338.52684 & $-9.32317$ & 19.58 & 18.91 & 15.0 & $-64.2$ $\pm$ 1.3 & 2.17 $\pm$ 0.41 & 2.13 $\pm$ 1.27 & $-2.92$ $\pm$ 0.22 & M \\
2609107798126218880 & 338.39674 & $-9.31998$ & 20.53 & 19.95 & 5.3 & $-71.8$ $\pm$ 2.8 & 1.47 $\pm$ 1.00 & -- & $-3.13$ $\pm$ 0.73 & M \\
2609107935565172224 & 338.36675 & $-9.32105$ & 20.08 & 19.45 & 8.8 & $-60.2$ $\pm$ 1.8 & 3.55 $\pm$ 1.34 & 3.75 $\pm$ 1.08 & $-2.12$ $\pm$ 0.37 & M \\
2609108588400203264 & 338.46083 & $-9.31525$ & 20.11 & 19.51 & 8.2 & $-69.6$ $\pm$ 2.0 & 1.85 $\pm$ 0.57 & 1.92 $\pm$ 0.40 & $-2.95$ $\pm$ 0.21 & M \\
2609109756631321472 & 338.46956 & $-9.28585$ & 19.25 & 18.54 & 21.0 & $-57.1$ $\pm$ 1.2 & 3.29 $\pm$ 0.42 & 2.94 $\pm$ 0.36 & $-2.56$ $\pm$ 0.14 & M \\
2609113055166221056 & 338.42154 & $-9.25084$ & 16.79 & 16.67 & 51.0 & $-57.4$ $\pm$ 1.1 & -- & -- & -- & NM \\
2609155489443129088 & 338.59370 & $-9.30185$ & 18.81 & 18.02 & 29.9 & $-68.5$ $\pm$ 1.1 & 3.08 $\pm$ 0.25 & 2.92 $\pm$ 0.33 & $-2.70$ $\pm$ 0.11 & M \\
2609202665363907328 & 338.55648 & $-9.28793$ & 19.46 & 18.78 & 16.0 & $-66.2$ $\pm$ 1.3 & 2.59 $\pm$ 0.40 & 2.52 $\pm$ 0.14 & $-2.76$ $\pm$ 0.11 & M \\
\enddata
\tablecomments{Astrometry and source IDs are taken from $Gaia$ DR3, and the photometry is taken from LS DR9 (DECaLS). The S/N ratio refers to the spectra observed with the IMACS data.}
\tablenotetext{a}{M refers to confirmed member stars, NM are confirmed non-member stars.}
\tablenotetext{b}{EW 1 refers to the equivalent width measurements of the CaT lines from the first epoch of observations (June 18, 2021). EW 2 refers to the equivalent width measurements of the CaT lines from the second epoch of observations (July 13, 2021).}
\end{deluxetable*}

\subsection{Determining Membership}\label{subsec:3.3}
For each dwarf, the membership status of observed stars was determined by examining their location on a color-magnitude diagram (CMD), spatial position, proper motion, and velocity. We use DECaLS DR9 photometry, with proper motion and astrometry retrieved through a positional cross-match with {\it Gaia} DR3. By jointly analyzing these parameters, the observed members were distinctly identified.

In Figure \ref{fig:1} we compare chemical and kinematic properties of all stars observed during the Aqu~II observations. Of the 12 useful Aqu~II spectra, 9 stars comprise a velocity grouping between $-$55 km s$^{-1}$ and $-$70 km s$^{-1}$ consistent with the expected velocity from \citet[]{Torrealba_2016}. Of these potential members, 8 are tightly grouped in proper motion space, located within 2 half-light radii ($r_h$) of the observed center, and lie along the projected Dartmouth isochrone \citep[]{2008ApJS..178...89D} plotted using the reported distance modulus from \citet[]{Torrealba_2016}, a metallicity of [Fe/H] $= -2.49$, and an age of 12.5 Gyr. We note that one member star slightly deviates from the isochrone ({\it Gaia} Dr3 source ID: 2609061687357323776). This star's redder color may be due to a high carbon abundance causing excess absorption in blue bands. We also note the velocity consistent non-member ({\it Gaia} DR3 source ID: 2609113055166221056) that is located $>3\sigma$ away from the clustering in proper motion space, and therefore is not a member. This non-member is also not included in the high probability sample from \citet[]{2022arXiv220505699P}, since the parallax was non-zero at 3$\sigma$ (parallax $-$ 3$\times$parallax$\_$error $>$ 0).

In Figure \ref{fig:2} we compare chemical and kinematic properties of all stars observed during the Boo~II observations. Of the 17 useful Boo~II spectra, 12 stars form a clear velocity grouping between $-$115 km s$^{-1}$ and $-$140 km s$^{-1}$, resulting in the 5 remaining stars being immediately classified as non-members. These 12 candidates were similarly grouped near the center of Boo~II in position space, along the projected isochrone, and within 1-2$\sigma$ of each other in proper motion space. We therefore classify these 12 stars as members of Boo~II. We note that in addition to the RR Lyrae member, there is another star that deviates from the isochrone ({\it Gaia} DR3 source ID: 3727837759579718272) and there is a possibility that this star is a non-member based on this offset in photometry. As discussed later in Section \ref{subsec:4.1}, when this star is removed the velocity dispersion decreases from $\sigma_{v_{\rm hel}} = 2.9^{+1.6}_{-1.2}$ km s$^{-1}$ to $\sigma_{v_{\rm hel}} = 1.3^{+2.3}_{-1.3}$ km s$^{-1}$. However, we still consider the star a member due to its extremely metal-poor nature ([Fe/H] $\sim $ -3) and close proximity in proper motion, position and velocity plots (Figure \ref{fig:2}).

One Boo~II member star ({\it Gaia} DR3 source ID: 3727826519650056576), is in the {\it Gaia} DR3 RR Lyrae star (RRL) catalog \citep[]{2022arXiv220606278C}. This star is outlined in green in Figure \ref{fig:2} and its characteristics are detailed in Table \ref{tab:5}. This star is reported as a member but it is excluded from calculations determining systemic radial velocity and velocity dispersion in Section \ref{subsec:4.1}, as well as metallicity and metallicity dispersion in Section \ref{subsec:4.1}. 

We also highlight the presence of two binary stars in Boo~II ({\it Gaia} DR3 source IDs: 3727827382938924800 and 3727826519650056832) which are similarly detailed in Table \ref{tab:5}. The first binary star (binary star 1 in this work and BooII-15 in \citet[]{Koch_2008}) was first identified by \citet[]{2016ApJ...817...41J} using high resolution spectroscopy and is outlined in cyan in Figure \ref{fig:2}. The second binary star (binary star 2 in this work and BooII-3 in \citet[]{Koch_2008}) is outlined in black in Figure \ref{fig:2}. This second binary star was identified when we noticed the velocity from the \citet[]{Koch_2008} sample differed from this work by 29.1 $\pm$ 8.0 km s$^{-1}$. This large variation in velocity indicates its likely identity as a binary star. Both of these stars are excluded from kinematic analysis to prevent an artificial inflation of the velocity dispersion or a shift in systemic velocity measurements.

From the above analysis, we confirm 8 Aqu~II members and 12 Boo~II members (three of which are excluded from kinematic analysis). These member stars are unambiguous in their classification due to tightly grouped locations in position, velocity, metallicity, proper motion and color-magnitude diagrams.

Among the 10 (13) high probability member candidates ($P > 0.5$ in \citet[]{2022arXiv220505699P}, using the $\texttt{mem\_fixed}$ column) we observed in Aqu~II (Boo~II), 8 (11) are confirmed to be members. Only 2 (2) are non-members in these galaxies. This shows that the proper motions and astrometric information from {\it Gaia} largely improved the target selection efficiency. We note that there remain 5 high probability member candidates in Aqu~II\footnote{{\it Gaia} DR3 source IDs for additional high probability member candidates in Aqu~II: 2609057770347087104, 2609059179096375808, 2609059213456115584, 2609100170264245376, 2609108760198902144} and 8 in Boo~II\footnote{{\it Gaia} DR3 source IDs for additional high probability member candidates in Boo~II: 3727823118035927936, 3727826107333175936, 3727826416570828800, 3727826519650055168, 3727827309924042624, 3727827997118825856, 3727835075224948352, 3727837239888417792} that either we did not observe due to slit conflicts in the mask design, or we observed but could not obtain $v_{\rm hel}$ due to the low S/N of the spectra. These stars should be prioritized in future observations to complete the sample of {\it Gaia} members in these two dwarf galaxies. We also note that for the RRL star in Boo~II ({\it Gaia} DR3 source ID: 3727826519650056576) we confirmed its membership despite the fact that it was not included in the high probability member candidates list from \citet[]{2022arXiv220505699P}. This star was not considered in the input data for the modelling in \citet[]{2022arXiv220505699P} due to it failing an uncommon {\it Gaia} flag ($\texttt{ipd\_gof\_harmonic\_amplitude} > 0$).

\begin{deluxetable*}{l c c c c c c c c c}
\tablecolumns{9}
\centering
\tablewidth{0pt}
\label{tab:4}
\tablecaption{\hypertarget{Table 4}{Bo\"{o}tes II} Observed Spectra.}
\tablehead{
\colhead{$Gaia$ DR3 Source ID} & \colhead{R.A. (deg)} & \colhead{Decl. (deg)} & \colhead{$g_0$ (mag)} & \colhead{$r_0$ (mag)} & \colhead{S/N} & \colhead{$v_{hel}$ (km s$^{-1}$)} & \colhead{EW\_CaT} & \colhead{[Fe/H] (dex)} & \colhead{Member\tablenotemark{a}}}
\startdata
3727825076541007232 & 209.43238 & 12.79053 & 20.20 & 19.70 & 7.6 & $-126.9$ $\pm$ 2.2 & 2.79 $\pm$ 1.44 & $-2.03$ $\pm$ 0.74 & M \\
3727825145260749952 & 209.41684 & 12.80064 & 22.32 & 20.94 & 6.0 & $-23.5$ $\pm$ 2.6 & -- & -- & NM \\
3727825488857874816 & 209.45241 & 12.80229 & 20.82 & 20.44 & 3.3 & 110.6 $\pm$ 6.6 & -- & -- & NM \\
3727826283427161728 & 209.50698 & 12.82085 & 20.64 & 20.19 & 4.4 & $-128.4$ $\pm$ 2.9 & -- & -- & M \\
3727826004253947776 & 209.54107 & 12.79852 & 20.28 & 19.79 & 5.8 & $-130.2$ $\pm$ 2.3 & 1.43 $\pm$ 0.78 & $-2.82$ $\pm$ 0.63 & M \\
3727826111628452224 & 209.53242 & 12.82263 & 20.66 & 20.18 & 4.1 & $-120.7$ $\pm$ 3.1 & 2.44 $\pm$ 1.92 & $-2.12$ $\pm$ 1.04 & M \\
3727826141692913792 & 209.55987 & 12.82006 & 21.02 & 20.72 & 2.4 & 29.8 $\pm$ 5.3 & -- & -- & NM \\
3727837587781240320 & 209.41683 & 12.84375 & 20.62 & 20.43 & 2.6 & 98.8 $\pm$ 4.5 & -- & -- & NM \\
3727837759579718272 & 209.40610 & 12.86346 & 18.14 & 17.63 & 35.0 & $-134.0$ $\pm$ 1.0 & 1.89 $\pm$ 0.04 & $-2.93$ $\pm$ 0.13 & M \\
3727825901174760832 & 209.44692 & 12.84726 & 20.97 & 20.42 & 3.5 & $-26.4$ $\pm$ 5.5 & -- & -- & NM \\
3727827382938924800\tablenotemark{b} & 209.46324 & 12.86025 & 19.12 & 18.59 & 18.6 & $-134.9$ $\pm$ 1.2 & 2.22 $\pm$ 0.14 & $-2.55$ $\pm$ 0.14 & M \\
3727827516082474880 & 209.48409 & 12.86877 & 19.37 & 18.85 & 14.8 & $-131.5$ $\pm$ 1.3 & 1.78 $\pm$ 0.46 & $-2.75$ $\pm$ 0.32 & M \\
3727827241204558720 & 209.50591 & 12.85140 & 19.43 & 18.92 & 14.7 & $-130.5$ $\pm$ 1.3 & 1.66 $\pm$ 0.10 & $-2.82$ $\pm$ 0.16 & M \\
3727839301472744192 & 209.44428 & 12.89235 & 20.25 & 19.74 & 6.5 & $-130.0$ $\pm$ 2.3 & 2.83 $\pm$ 0.77 & $-2.00$ $\pm$ 0.40 & M \\
3727826519650056576\tablenotemark{c} & 209.52933 & 12.85634 & 18.71 & 18.23 & 18.3 & $-118.4$ $\pm$ 1.4 & -- & -- & M \\
3727826519650056832\tablenotemark{d} & 209.53947 & 12.85719 & 20.31 & 19.86 & 6.0 & $-138.5$ $\pm$ 2.6 & -- & -- & M \\
3727826657089011840 & 209.57290 & 12.86183 & 20.32 & 19.83 & 5.5 & $-135.8$ $\pm$ 2.4 & -- & -- & M \\
\enddata
\tablecomments{Astrometry and source IDs are taken from $Gaia$ DR3, and the photometry is taken from LS DR9 (DECaLS). The S/N ratio refers to the spectra observed with the IMACS data.}
\tablenotetext{a}{M refers to confirmed member stars, NM are confirmed non-member stars.}
\tablenotetext{b}{The binary star identified in \citet[]{2016ApJ...817...41J}.}
\tablenotetext{c}{The RRL star that was also not included in the high probability list from \citet[]{2022arXiv220505699P}.}
\tablenotetext{d}{The potential binary star identified due to variation in observed velocity measurements between \citet[]{Koch_2008} and this work.}
\end{deluxetable*}
\begin{deluxetable*}{c c c c c c}
\tabletypesize{\scriptsize}
\tablecolumns{9}
\centering
\tablewidth{0pt}
\tablecaption{Binary and RR Lyrae Stars in Boo II}\label{tab:5}
\tablehead{
\colhead{ID} & \colhead{$v_{\rm hel}$ (km s$^{-1}$)\tablenotemark{a}} & \colhead{$v_{\rm hel}$ (km s$^{-1}$)\tablenotemark{b}} & \colhead{$v_{\rm hel}$ (km s$^{-1}$)\tablenotemark{c}} & \colhead{$v_{\rm hel}$ (km s$^{-1}$)\tablenotemark{d}} & \colhead{Classification}}
\startdata
3727826519650056576 & -- & $-$118.4 $\pm$ 1.4 & -- & -- & RR Lyrae Star \\
3727827382938924800 & $-$100.07 $\pm$ 2.33 & $-$130.5 $\pm$ 1.3 & $-$104.8 & $\sim -120$ & Binary Star (Ji+2016) \\
3727826519650056832 & $-$109.38 $\pm$ 5.28 & $-$138.5 $\pm$ 2.7 & -- & -- & Suspected Binary Star \\
\enddata
\tablenotetext{a}{Values taken from \citet[]{Koch_2008}. Observations occurred on April 17, 2007.}
\tablenotetext{b}{Values from this work. Observations occurred on June 18, 2021.}
\tablenotetext{c}{Values taken from this \citet[]{koch_2014}. Observations occurred on May 23, 2014.}
\tablenotetext{d}{Values taken from this \citet[]{2016ApJ...817...41J}. Observations occurred on March 21, 2014 and June 17-18, 2015.}
\end{deluxetable*}
    
\section{Discussion}\label{sec:4}
\hypertarget{Discussion}{In this section} we evaluate systemic properties of Aqu~II and Boo~II, and discuss their nature as UFDs. We consider the global velocity dispersions, metallicity dispersions, and orbital characteristics as well as investigate the dark matter annihilation.

\begin{figure*}
    \centering
    \includegraphics[width = .45\textwidth]{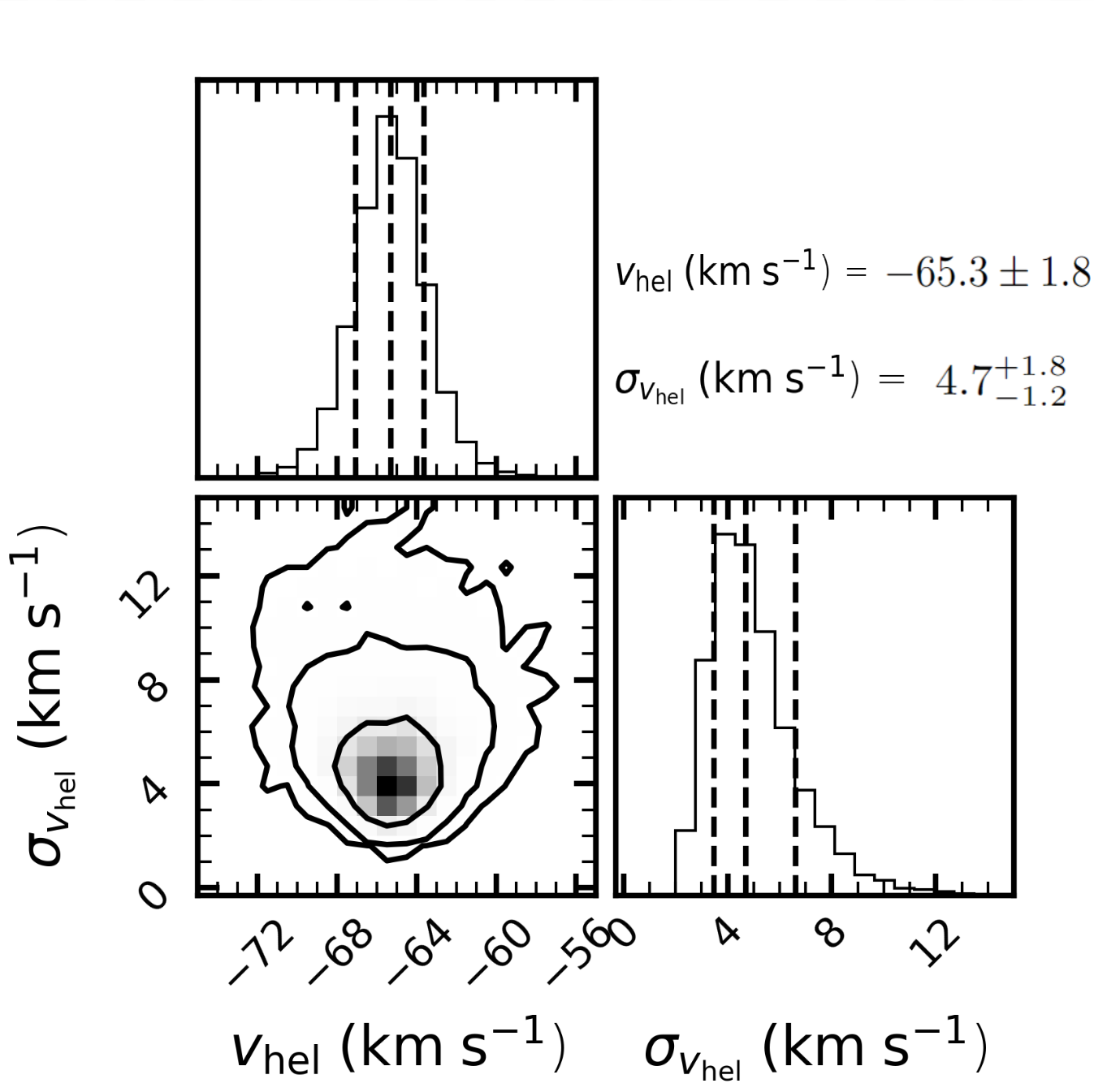}
    \includegraphics[width = .45\textwidth]{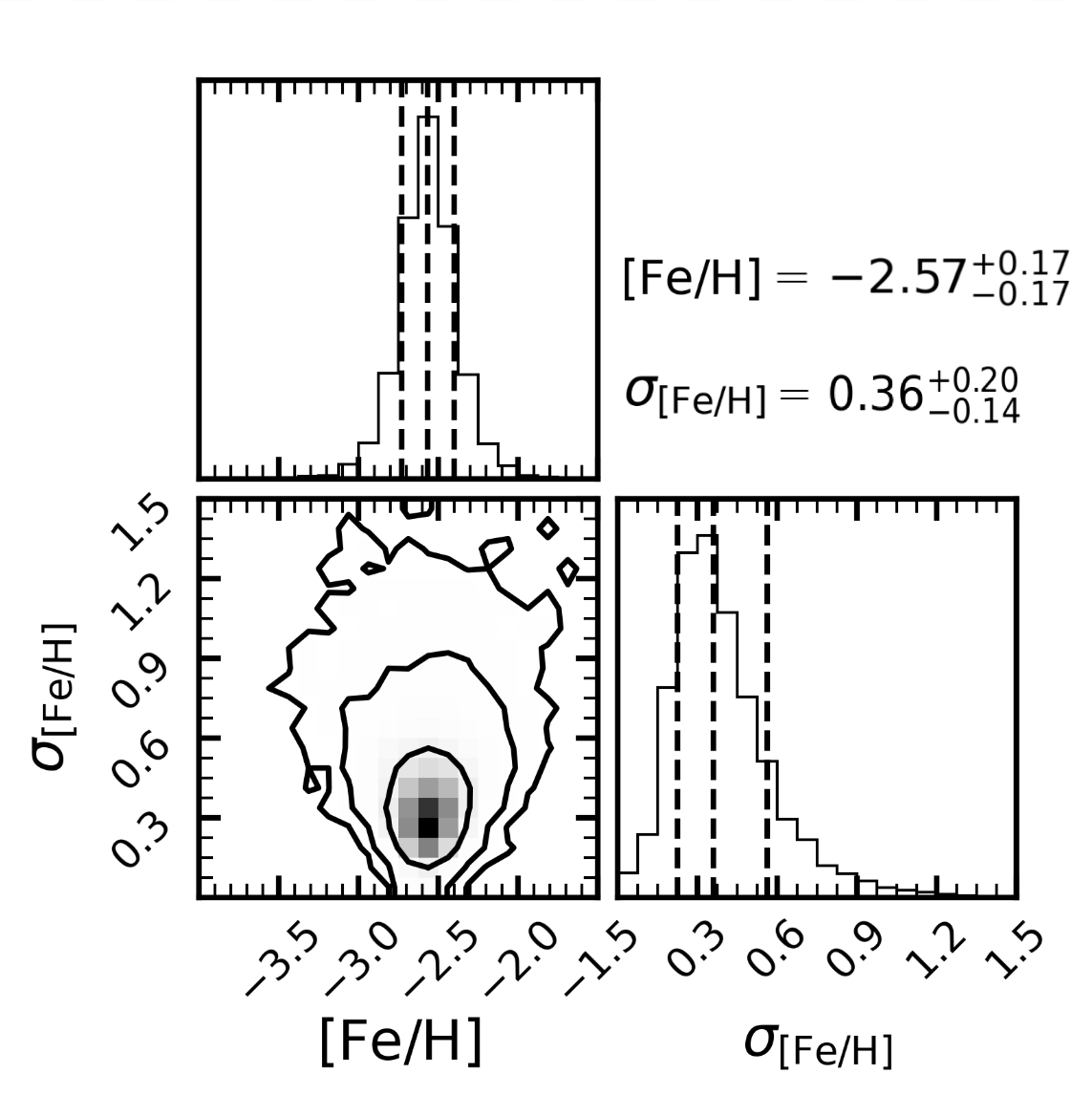}
    \caption{\hypertarget{Figure 3}{The} resulting 2-D posterior probability distributions from running an MCMC sampler and likelihood function as outlined in section \ref{subsec:4.1} for Aqu~II. In the 1-D histogram, the dashed lines display the 16th, 50th and 84th percentiles, while the 2-D histogram contours display the 68\%, 95.5\%, and 99.7\% confidence intervals from the peak density. (Left) The resulting distribution for the systemic velocity ($v_{\rm hel}$) and velocity dispersion ($\sigma_{v_{\rm hel}}$). (Right) The resulting distribution for the systemic metallicity ([Fe/H]) and metallicity dispersion ($\sigma_{\rm [Fe/H]}$).}
    \label{fig:3}
\end{figure*}

\begin{figure*}
    \centering
    \includegraphics[width = .45\textwidth]{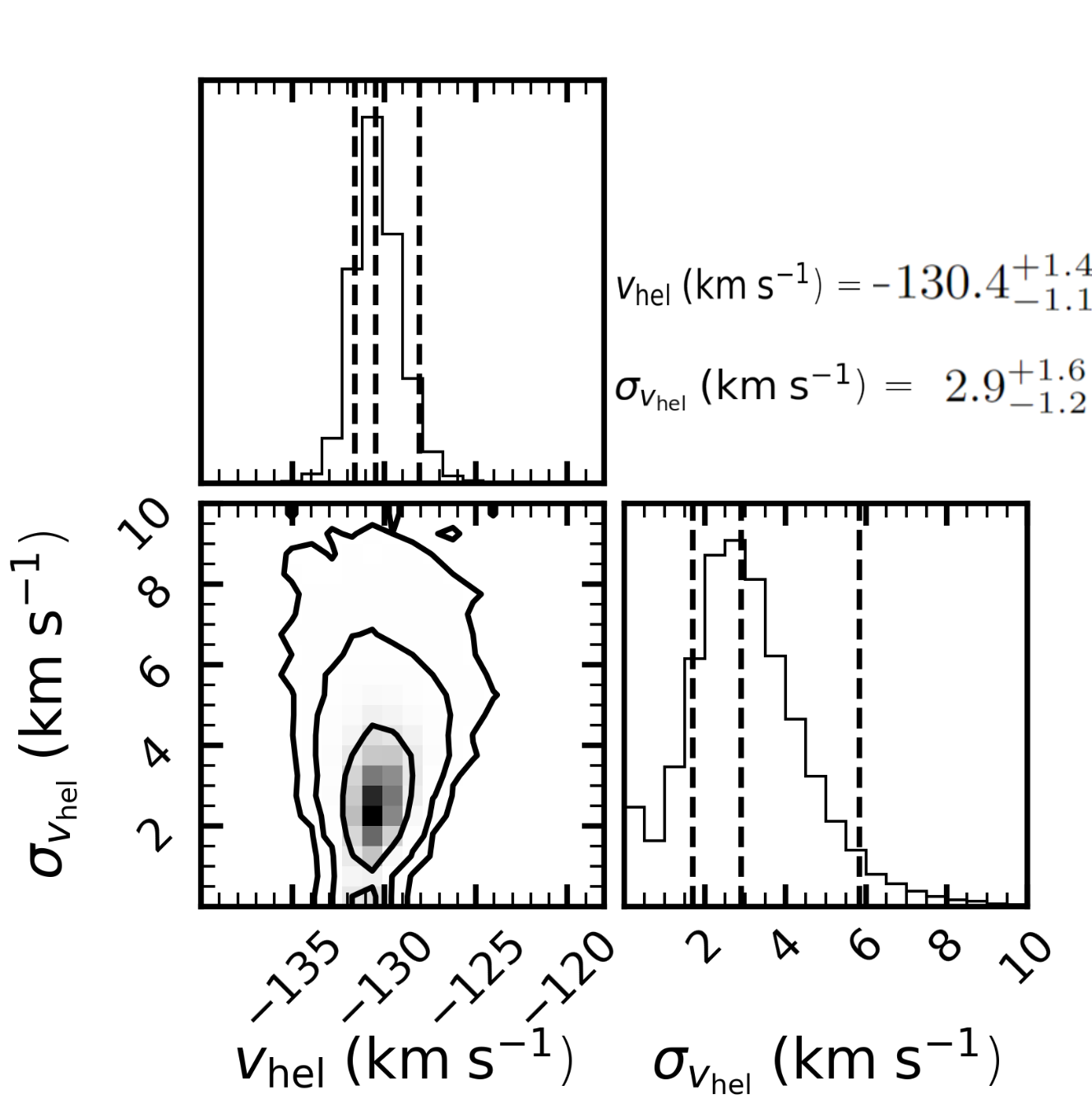}
    \includegraphics[width = .45\textwidth]{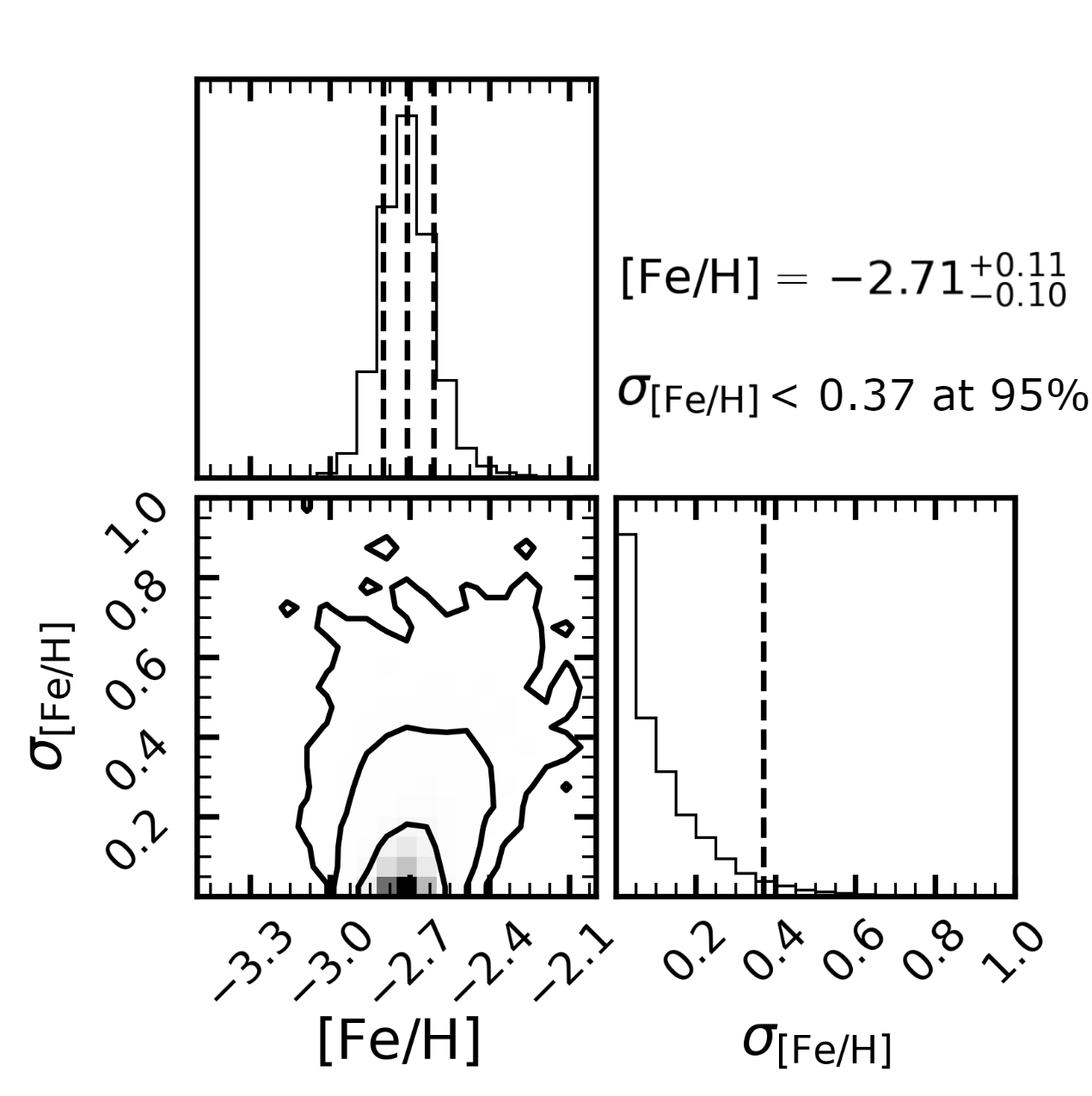}
    \caption{\hypertarget{Figure 4}{The} resulting 2-D posterior probability distributions from running an MCMC sampler and likelihood function as outlined in section \ref{subsec:4.1} for Boo~II. (Left) The resulting distribution for the systemic velocity ($v_{\rm hel}$) and velocity dispersion ($\sigma_{v_{\rm hel}}$). In the 1-D histogram. the dashed lines display the 16th, 50th and 84th percentiles, while the 2-D histogram contours display the 68\%, 95.5\%, and 99.7\% confidence intervals from the peak density. (Right) The resulting distribution for the systemic metallicity ([Fe/H]) and metallicity dispersion ($\sigma_{\rm [Fe/H]}$). In the upper left 1-D histogram the dashed lines display the 16th, 50th and 84th percentiles, while the dashed line in the lower right 1-D histogram displays the 95th percentile location. The contour lines in the 2-D histogram display the 68\%, 95.5\%, and 99.7\% confidence intervals from the peak density.}
    \label{fig:4}
\end{figure*}

\subsection{Velocity and Metallicity Dispersion}\label{subsec:4.1}
With 8 and 9 (excluding RRL and binary stars) spectroscopically confirmed members of Aqu~II and Boo~II, we evaluate the systemic velocity and its dispersion following the methods in \citet[]{Li_2017}. Similarly we use the 8 and 11 (excluding the RRL star) spectroscopically confirmed members of Aqu~II and Boo~II to measure the systemic metallicity and its dispersion. We use a 2-parameter Gaussian likelihood function (equation (2) from \citet[]{Li_2017}) derived from \citet[]{Walker_2006}. Using \texttt{$emcee$} \citep[]{2013PASP..125..306F}, an MCMC sampler was run to sample this Gaussian using 50 walkers making 1000 steps with an additional burn-in period of 100 steps. The systemic velocity or metallicity is determined as the median of the respective posterior distribution with uncertainties derived from the 16th and 84th percentiles.

In Figure \ref{fig:3} we determine the systemic velocity, metallicity, and the respective dispersions for Aqu~II. Using the 8 observed Aqu~II members, we measure a systemic velocity of $v_{\rm hel} = -65.3 \pm 1.8$ km s$^{-1}$ and a dispersion of $\sigma_{v_{\rm hel}} = 4.7^{+1.8}_{-1.2}$ km s$^{-1}$. The systemic metallicity is measured as [Fe/H] $= -2.57^{+0.17}_{-0.17}$ with a dispersion of $\sigma_{\rm [Fe/H]} = 0.36^{+0.20}_{-0.14}$.

In Figure \ref{fig:4} we similarly determine the systemic velocity, velocity dispersion, metallicity, and metallicity dispersion for Boo~II. The 9 confirmed members of Boo~II in this work (excluding RRL and binary stars) generate a systemic velocity of $v_{\rm hel} = -130.4^{+1.4}_{-1.1}$ km s$^{-1}$ with a dispersion of $\sigma_{v_{\rm hel}} = 2.9^{+1.6}_{-1.2}$ km s$^{-1}$. The systemic metallicity from the 8 RGB member stars in this work with well-fit metallicity measurements (excluding the RRL star, but including the binary stars) is [Fe/H] $= -2.71^{+0.11}_{-0.10}$. The metallicity dispersion of Boo~II is not resolved with these 8 RGB members, we instead report a 95\% upper limit of $\sigma_{\rm [Fe/H]} < 0.37$.

As discussed in \citet[]{2010ApJ...722L.209M}, \citet[]{Li_2017}, and \citet[]{Simon_2017}, it is possible for binary member stars to artificially inflate the velocity dispersion, especially with the small time span of observations for Boo~II. To test the robustness we performed a jackknife test for each dwarf, repeatedly removing a single star at a time to measure the effect it has on the global properties. 

For Aqu~II the majority of stars had minimal effects on any property of the galaxy. However, the most metal-rich star ({\it Gaia} DR3 source ID: 2609061687357323776), which was also noted to deviate from the isochrone in Section \ref{subsec:3.3}, dramatically affects the metallicity dispersion. When this star is not considered in calculations, the metallicity dispersion decreases to $\sigma_{\rm [Fe/H]} = 0.04^{+0.08}_{-0.02}$ (from the initial $\sigma_{\rm [Fe/H]} = 0.36^{+0.20}_{-0.14}$ dex) and therefore becomes unresolved. Additionally, a separate star ({\it Gaia} DR3 source ID: 2609109756631321472) significantly lowers the velocity dispersion from $\sigma_{v_{\rm hel}} = 4.7^{+1.8}_{-1.2}$ km s$^{-1}$ to $\sigma_{v_{\rm hel}} = 2.7^{+1.6}_{-1.2}$ km s$^{-1}$ when removed. These stars are still considered members but are noted here due to the large impacts on metallicity, metallicity dispersion and velocity dispersion.

Similarly for Boo~II we performed a jackknife test, and we note that no star dominantly affected the systemic metallicity, metallicity dispersion or systemic velocity (the known RRL and binary stars were not included for velocity jackknife tests). We do notice that two stars ({\it Gaia} DR3 source IDs: 3727826111628452224 and 3727837759579718272, which was the star discussed in Section \ref{subsec:3.3} due to it's deviation from the isochrone) lowered the velocity dispersion from $\sigma_{v_{\rm hel}} = 2.9^{+1.6}_{-1.2}$ km s$^{-1}$ to $\sigma_{v_{\rm hel}} = 0.8^{+1.5}_{-0.8}$ km s$^{-1}$ and $\sigma_{v_{\rm hel}} = 1.3^{+2.3}_{-1.3}$ km s$^{-1}$ respectively when removed. These stars are still considered members but are noted here due to the large impact on the velocity dispersion. 

The variation in the velocity dispersion noticed during the jackknife tests shows how one star can have a significant impact. With a limited sample size there is a possibility these stars are binaries or non-members, causing inflation in the velocity dispersion. However the derived values could also purely be the true dispersions of the galaxies. For example, when we include the two binary stars in our calculations for Boo~II, the velocity dispersion increases to $\sigma_{v_{\rm hel}} = 3.6^{+1.4}_{-1.1}$ km s$^{-1}$, which is consistent within the uncertainty but still $\sim 25 \%$ larger than the dispersion derived without binaries. This emphasizes the need for larger sample sizes and/or multi-epoch observations to resolve this uncertainty, which could be achieved by observing the high probability {\it Gaia} member candidates from Section \ref{subsec:3.3} that were not included in this study.

Both dwarf galaxies are extremely metal-poor (Aqu~II [Fe/H] = $-2.57^{+0.17}_{-0.17}$, Boo~II [Fe/H] = $-2.71^{+0.11}_{-0.10}$), placing them amongst the lowest known UFD metallicities alongside Horologium I, Reticulum II, Segue 1, Draco II, and Tucana II \citep[]{doi:10.1146/annurev-astro-091918-104453}. Specifically, Boo~II can be classified as the third most metal-poor dwarf galaxy after Tucana II and Horologium I \citep[]{doi:10.1146/annurev-astro-091918-104453}.

\subsection{Dynamical Mass and Mass-to-Light Ratio}\label{subsec:4.2}
As described in \citet[]{2010MNRAS.406.1220W}, assuming both Aqu~II and Boo~II are in dynamical equilibrium we can use their resolved velocity dispersions to estimate the system's enclosed mass within 1 $r_{1/2}$ using the relation:
\begin{equation}\label{eq:2}
    \rm M_{1/2} \approx 930(\frac{\langle \sigma^2_{v_{\rm hel}} \rangle }{km^2\;s^{-2}})(\frac{r_{1/2}}{pc})M_{\odot}.
\end{equation}
We use the reported half-light radii from \citet[]{Torrealba_2016} for Aqu~II and \citet[]{2018ApJ...860...66M} for Boo~II respectively. Additionally, we treat these half-light radii and the previously determined velocity dispersions from section \ref{subsec:4.1} as Gaussians with a mean of their systemic value and a standard deviation of their uncertainty. We then sampled the estimation relation (Equation \ref{eq:2}) 1000 times using these Gaussians, and report the median as the systemic mass and the 16th and 84th percentiles as the uncertainties. We then estimate the mass-to-light ratio using the absolute V-band magnitude ($M_V$) from \citet[]{Torrealba_2016} for Aqu~II and \citet[]{2018ApJ...860...66M} for Boo~II.

For Aqu~II, we estimated a dynamical half-light mass of $3.1^{+3.0}_{-1.4} \times 10^6 M_{\odot}$, and a mass-to-light ratio of $1300^{+1300}_{-600} M_{\odot}/L_{\odot}$. This large mass-to-light ratio distinctly indicates the dark matter dominated nature of Aqu~II, as expected due to its identity as a UFD. For Boo~II, we estimate a dynamical half-mass of $3.1^{+4.4}_{-2.1} \times 10^5 M_{\odot}$, and a mass-to-light ratio of $460^{+1000}_{-440} M_{\odot}/L_{\odot}$. Boo~II is therefore consistent with a large dark matter content as well, but we cannot confirm at high confidence that it is dark matter dominated because of the large uncertainties.

\begin{figure*}
    \centering
    \includegraphics[width = .45\textwidth]{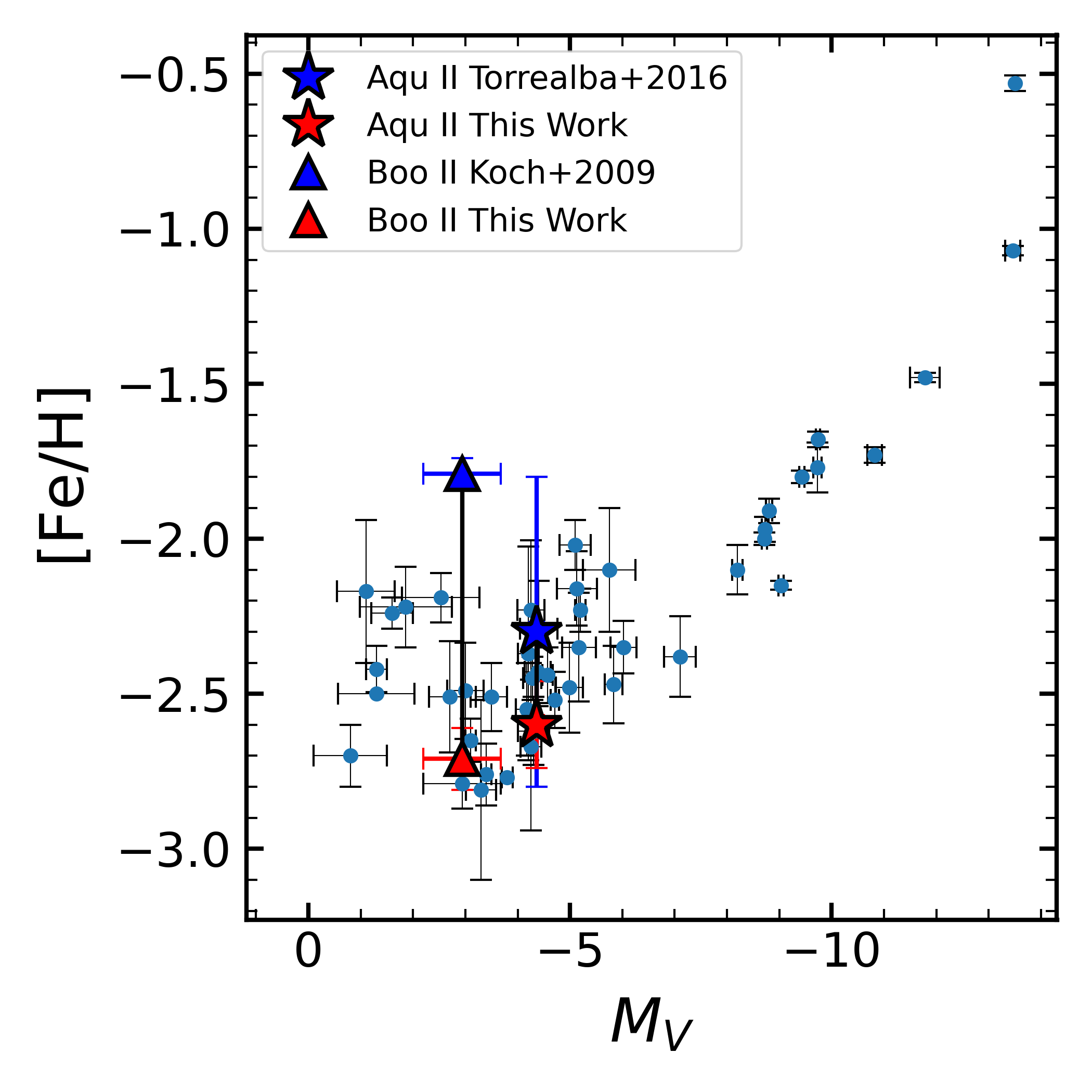}
    \includegraphics[width = .46\textwidth]{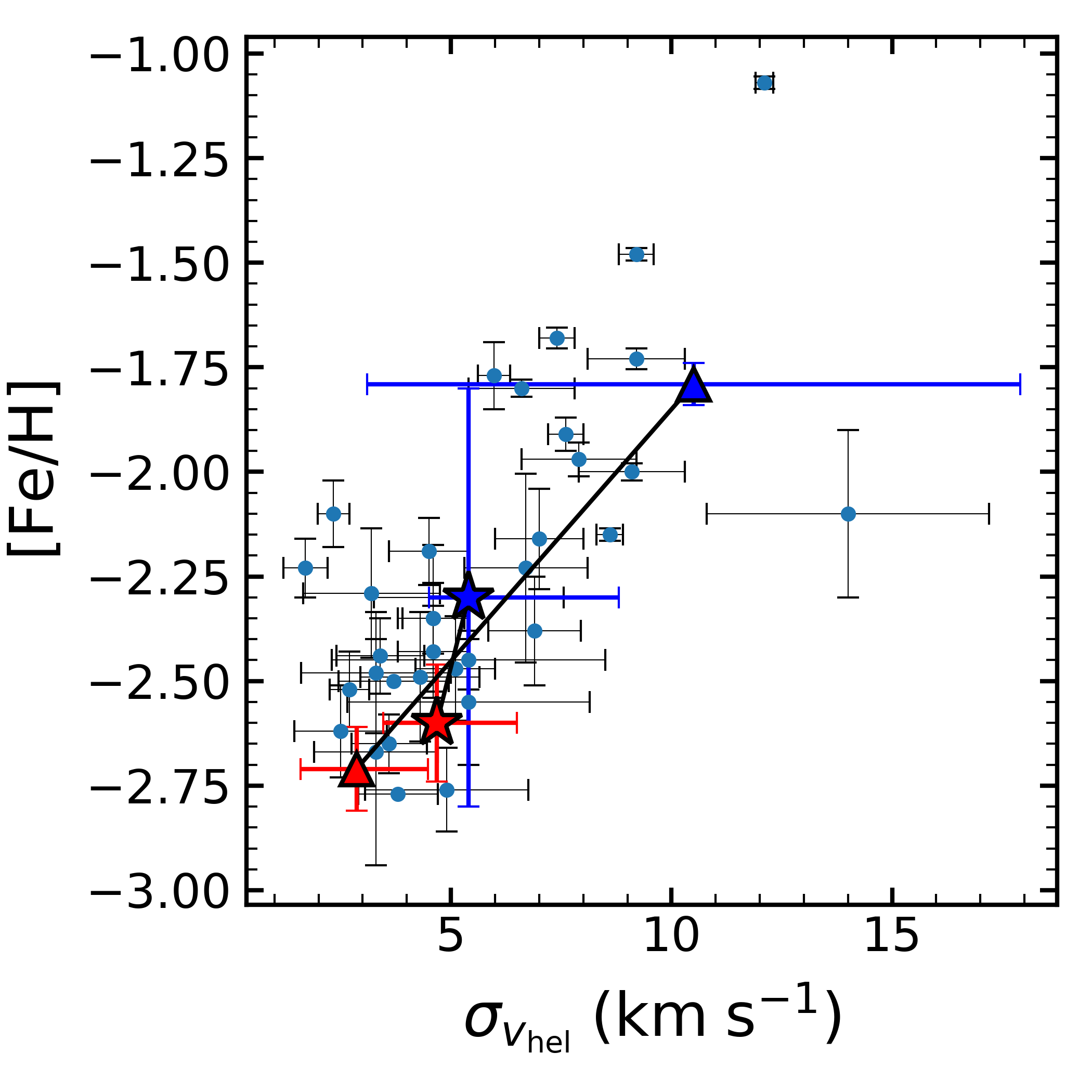}
    \caption{\hypertarget{Figure 6}{Left:} Comparison of systemic metallicity ([Fe/H]) as a function of absolute V-band magnitude ($M_V$) from this work and previous literature measurements for both UFDs. Right: A comparison of the systemic metallicity ([Fe/H]) as a function of velocity dispersion ($\sigma_{v_{\rm hel}}$). The properties of all known UFDs are shown as blue dots using the luminosities and velocity dispersions from \citet[]{2022arXiv220505699P} and the metallicities from \citet[]{doi:10.1146/annurev-astro-091918-104453} (A full list of references can be found in Appendix \ref{app:refs}). The values reported in this work agree with the trend of other UFDs, resolving the inconsistency presented by the previous Boo~II literature velocity dispersion.}
    \label{fig:5}
\end{figure*}

\begin{figure*}
    \centering
    \includegraphics[width = .45\textwidth]{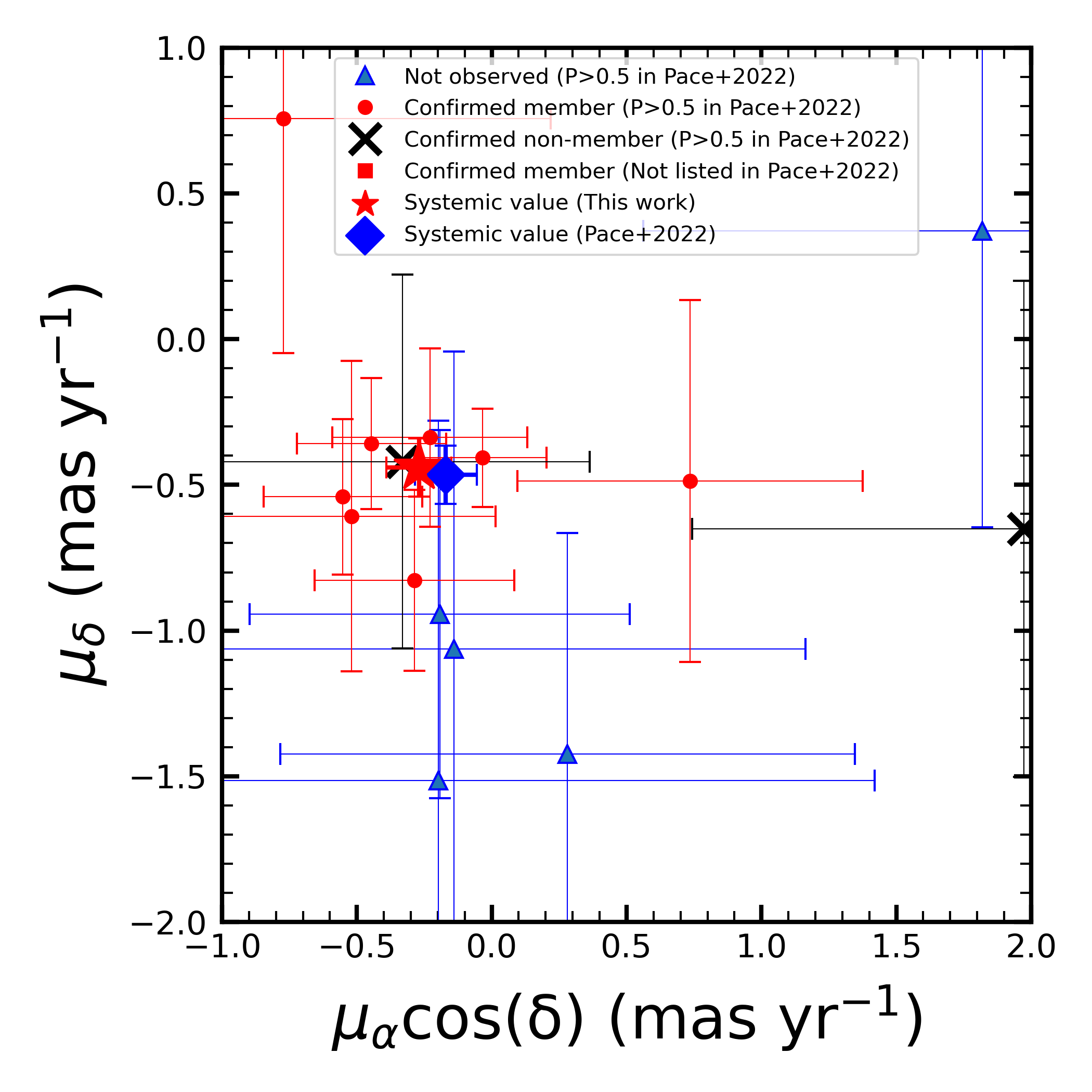}
    \includegraphics[width = .45\textwidth]{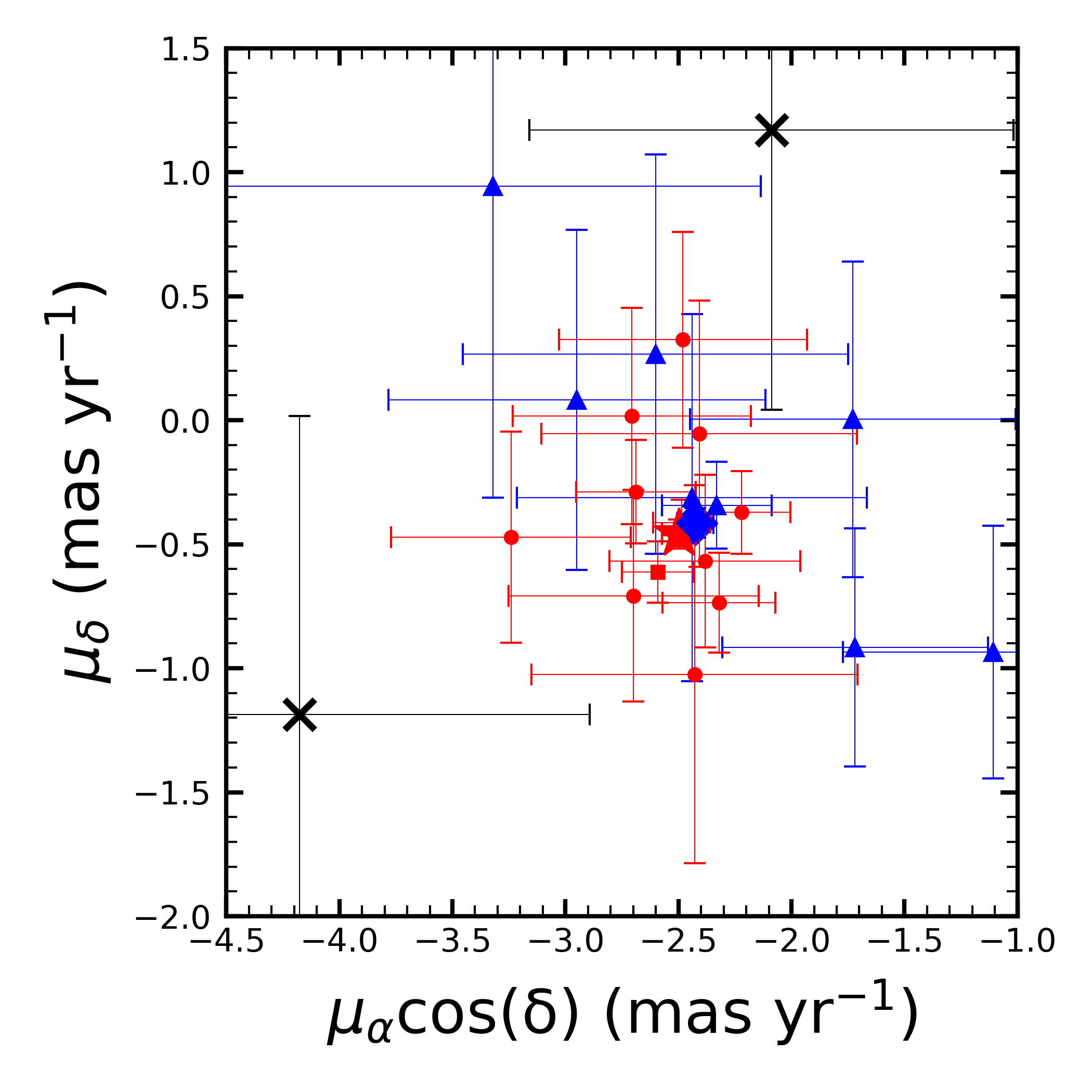}
    \caption{\hypertarget{Figure 6}{Comparison} of member star and systemic proper motion for Aqu~II (left) and Boo~II (right) between this work and \citet[]{2022arXiv220505699P} using astrometry from {\it Gaia} DR3 \citep[]{2016A&A...595A...1G, 2021A&A...649A...1G, 2021A&A...649A...2L}. The systemic value for this work is the  weighted average of spectroscopically confirmed member stars, while the systemic value from \citet[]{2022arXiv220505699P} is from a mixture model based on photometric and astrometric data only.}
    \label{fig:6}
\end{figure*}

\subsection{Literature Comparison in Spectroscopic Members}\label{subsec:4.3}
The increased sample size of confirmed bright RGB members provides smaller uncertainties and significant improvements to the systemic velocities, velocity dispersions, metallicities and dynamical mass estimates of Aqu~II and Boo~II. In Figure \ref{fig:5} we show the comparison of the metallicity and velocity dispersions from this work with previous measurements from the literature on these two UFDs. 

\citet[]{Torrealba_2016} provides the most recent measurements for Aqu~II, using 9 member stars (4 RGB and 5 BHB) to report a systemic velocity of $v_{\rm hel} = -71.1 \pm 2.5$ km s$^{-1}$ with a dispersion of $\sigma_{v_{\rm hel}} = 5.4^{+3.4}_{-0.9}$ km s$^{-1}$ generating a mass estimate of $2.7^{+6.6}_{-0.5} \times 10^6 M_{\odot}$, a metallicity of [Fe/H] $= -2.3 \pm 0.5$ and no resolved metallicity dispersion. We measured a systemic velocity difference of $\sim 6$ km s$^{-1}$ at $v_{\rm hel} = -65.3 \pm 1.8$ km s$^{-1}$, with a $28 \%$ smaller uncertainty. Despite the slightly larger sample size in \citet[]{Torrealba_2016} of 9 stars, our increased precision most likely originates from our sample of RGB stars rather than BHB stars. These BHB stars have a fainter magnitude and broader absorption line features and therefore a larger velocity uncertainty. Similarly, we resolve a smaller velocity dispersion at $\sigma_{v_{\rm hel}} = 4.7^{+1.8}_{-1.2}$ with the upper error decreased by almost a factor of two. Most noticeably however, we measure an average metallicity $0.27$ dex lower at [Fe/H] $= -2.57 ^{+0.17}_{-0.17}$ with an uncertainty smaller by a factor of 3. This decreased average metallicity further identifies the extremely metal-poor nature of Aqu~II. Furthermore, while \citet[]{Torrealba_2016} does not report a metallicity dispersion, we are able to resolve it due to our RGB sample size. This dispersion at [Fe/H] = $0.36^{+0.20}_{-0.14}$ further displays the spread of metallicity amongst members and highlights the nature of Aqu~II as a UFD.

Using the GMOS-N spectrograph on the Gemini North Telescope whose spectral resolution is $R \sim 4,000$, \citet[]{Koch_2008} provided the first spectroscopic study of Boo~II. Using 5 RGB members, \citet[]{Koch_2008} reports a mean systemic velocity of $v_{\rm hel} = -117.0 \pm 5.2$ km s$^{-1}$ with an extremely large dispersion of $\sigma_{v_{\rm hel}} = 10.5\pm 7.4$ km s$^{-1}$, and a mean metallicity of [Fe/H] $= -1.79\pm 0.05$ with a dispersion of $\sigma_{\rm [Fe/H]} = 0.14$. Contrastingly, the IMACS/Magellan spectrograph used in this study provided a resolution of more than double (R $\sim 11,000$). This increased resolution combined with our sample size more than doubling the number of confirmed RGB members significantly affected systemic properties and drastically reduced uncertainties.

The systemic velocity of Boo~II measured in this work differed from \citet[]{Koch_2008} by more than 10 km s$^{-1}$ at $-130.4^{+1.4}_{-1.1}$ km s$^{-1}$ while decreasing the uncertainty by a factor of 4. Similarly, we measure a decrease in Boo~II metallicity by almost 1 dex to $-2.71^{+0.11}_{-0.10}$. A main difference in metallicity calculation is that \citet[]{Koch_2008} used a CaT to metallicity conversion that was not calibrated for metallicities below [Fe/H] = $-2$. \citet[]{10.1093/mnras/stt1126} later extended the conversion relation to [Fe/H] = $-4$ dex, and this extended relation is what was used in our study. The difference in the conversion relations used most likely accounts for the large difference in systemic metallicity between this work and \citet[]{Koch_2008}. Most notably, we measure a velocity dispersion of Boo~II that is smaller by a factor of 3 at $2.9^{+1.6}_{-1.2}$ km s$^{-1}$ with an uncertainty smaller by a factor of 5. This decrease in velocity dispersion would significantly affect the mass estimation and therefore the mass-to-light ratio, which were not measured by \citet[]{Koch_2008}. Our increased sample size and improved velocity measurements for each star were the main factor for this reduced velocity dispersion.

\citet[]{2010ApJ...722L.209M} hypothesized that the velocity dispersion in Boo~II could be explained by the presence of binary stars. Due to their low intrinsic velocity dispersion ($\sim 2-8 ~{\rm km~s^{-1}}$), ultra-faint dwarf galaxies are susceptible to unidentified  binary stars \citep{2010ApJ...722L.209M, 2010ApJ...721.1142M, Pianta2022ApJ...939....3P}. In particular, small samples sizes ($N \lesssim 10$) are quite susceptible to the inflation of the velocity dispersion if only single-epoch velocity data is available \citep{2010ApJ...722L.209M, 2010ApJ...721.1142M}, as the initial spectroscopy of Boo~II has demonstrated. As previously mentioned, \citet[]{2016ApJ...817...41J} identified that one star of the five members in the \citet[]{Koch_2008} sample was a spectroscopic binary and had inflated the velocity dispersion. Similarly, we identify a second plausible binary member star based on measurements from \citet[]{Koch_2008} and this work (see section \ref{subsec:3.3}). The larger sample size of members in this work is less vulnerable to binary stars (although with 9 stars it remains a small size). We do however have multi-epoch data by comparing observations in this work to older GMOS data that can be used to detect the presence of further binaries using methods from \citet[]{2010ApJ...721.1142M}. We note many ultra-faint dwarfs have had binary stars identified with multi-epoch data \citep[e.g., Carina~II, Grus~I, Grus~II][]{Li2018ApJ...857..145L, Simon2020ApJ...892..137S, 2022arXiv220604580C} which has led to an inflated velocity dispersion in some other dwarf galaxies \citep[e.g., Triangulum~II; ][]{Kirby2015ApJ...814L...7K, Kirby2017ApJ...838...83K, Buttry2022MNRAS.514.1706B}.

A separate study of Boo~II by \citet[]{2016ApJ...817...41J} used high-resolution spectroscopy to observe four bright RGB stars in order to measure chemical abundances of the system. \citet[]{doi:10.1146/annurev-astro-091918-104453} used these results to report a metallicity of [Fe/H] $= -2.79^{+0.06}_{-0.10}$ consistent with the extremely metal-poor nature derived in this work. Additionally, \citet[]{doi:10.1146/annurev-astro-091918-104453} report a metallicity dispersion of $\sigma_{\rm [Fe/H]} < 0.35$ which is very similar to the upper limit of $\sigma_{\rm [Fe/H]} < 0.37$ measured in this work. However, \citet[]{2016ApJ...817...41J} stated that their observations spanned a limited range and had a small sample size, half of what was considered in this work. 

\subsection{Proper Motion and Orbital Analysis}\label{subsec:4.4}
We measured the systemic proper motion of both UFDs through a weighted average of the 8 Aqu~II and 12 Boo~II confirmed member stars using {\it Gaia} DR3 astrometry. We note that in the case of Boo~II, both the RRL star and the spectroscopic binaries are included in proper motion calculations. For Aqu~II we measure $\mu_{\alpha} = -0.27 \pm 0.12$ mas yr$^{-1}$ and $\mu_{\delta} = -0.44 \pm 0.10$ mas yr$^{-1}$, and for Boo~II we measure $\mu_{\alpha} = -2.50 \pm 0.07$ mas yr$^{-1}$ and $\mu_{\delta} = -0.46 \pm 0.06$ mas yr$^{-1}$. 

In Figure \ref{fig:6} we compare the proper motion of Aqu~II and Boo~II member stars with the results from \citet[]{2022arXiv220505699P}. While \citet[]{2022arXiv220505699P} contained a larger sample size of high probability member candidates (7 more for Aqu~II, 9 more for Boo~II), our proper motion measurements were able to exclude non-members based on kinematic information from spectroscopy which were included in the measurements from \citet[]{2022arXiv220505699P}. In fact, we confirm that 4 high probability member candidates (2 in each UFD) in \citet[]{2022arXiv220505699P} are non-members, which indeed slightly shifted the systemic proper motion.

These contaminant stars resemble members in both proper motion and photometry space, but are clearly non-members when observed in velocity space. By removing these contaminants, we determine a more robust measurement of systemic proper motion which in turn generates a more robust orbital analysis. We specifically note that the Boo~II RRL member star ({\it Gaia} DR3 source ID: 3727826519650056576) is not included in \citet[]{2022arXiv220505699P} but was spectroscopically confirmed as a member. This star is the second brightest member star (in $g_0$ magnitude), and has one of the most precise proper motions resulting in it having a large influence on the systemic proper motion. 

Our proper motion measurements also generally agree within 1$\sigma$ of additional studies using {\it Gaia} DR3 measurements \citep{McConnachie2020RNAAS...4..229M, Battaglia2022A&A...657A..54B,2022arXiv220505699P}. The exceptions are that \citet[]{McConnachie2020RNAAS...4..229M} reports $\mu_{\alpha}$ for Boo~II consistent only within 2$\sigma$ of the value in this work, and \citet[]{Battaglia2022A&A...657A..54B} report a $\mu_{\alpha}$ for Aqu~II only consistent within 2$\sigma$ of this work.

We model the orbits using the 6-D phase space (R.A., Decl., distance, proper motions, and systemic velocity) parameters listed in Table \ref{tab:2} and \texttt{galpy} \citep[]{2015ApJS..216...29B}. We treat each trait as the mean of a Gaussian distribution with a standard deviation equal to the uncertainty and sample from them 1000 times. We integrate backwards in time 5 Gyr, recording the eccentricity and nearest apocenter and pericenter. The nearest apocenter or pericenter is described as being the first local minimum or maximum recorded in our orbit integration. The choice to make this constraint is motivated by the results of \citet[]{2022arXiv220505699P} and \citet[]{2022MNRAS.512..739D} which highlight the unreliability of considering preceeding values. We also excluded instances where the orbit was unbound and resulted in no local minima or maxima being available. In this work we initialized the 1000 orbits using the \texttt{McMillan17} gravitational potential \citep[]{2017MNRAS.465...76M}, considering scenarios with and without the inclusion of the Large Magellanic Cloud (LMC) potential. When included, the LMC potential has a Hernquist profile with a mass of $1.38 \times 10^{11} M_{\odot}$ \citep[]{2019MNRAS.487.2685E} and a scale radius of 10.2 kpc, following \citet[]{2022arXiv220505699P}.

\begin{figure*}
    \centering
    \includegraphics[width = .45\textwidth]{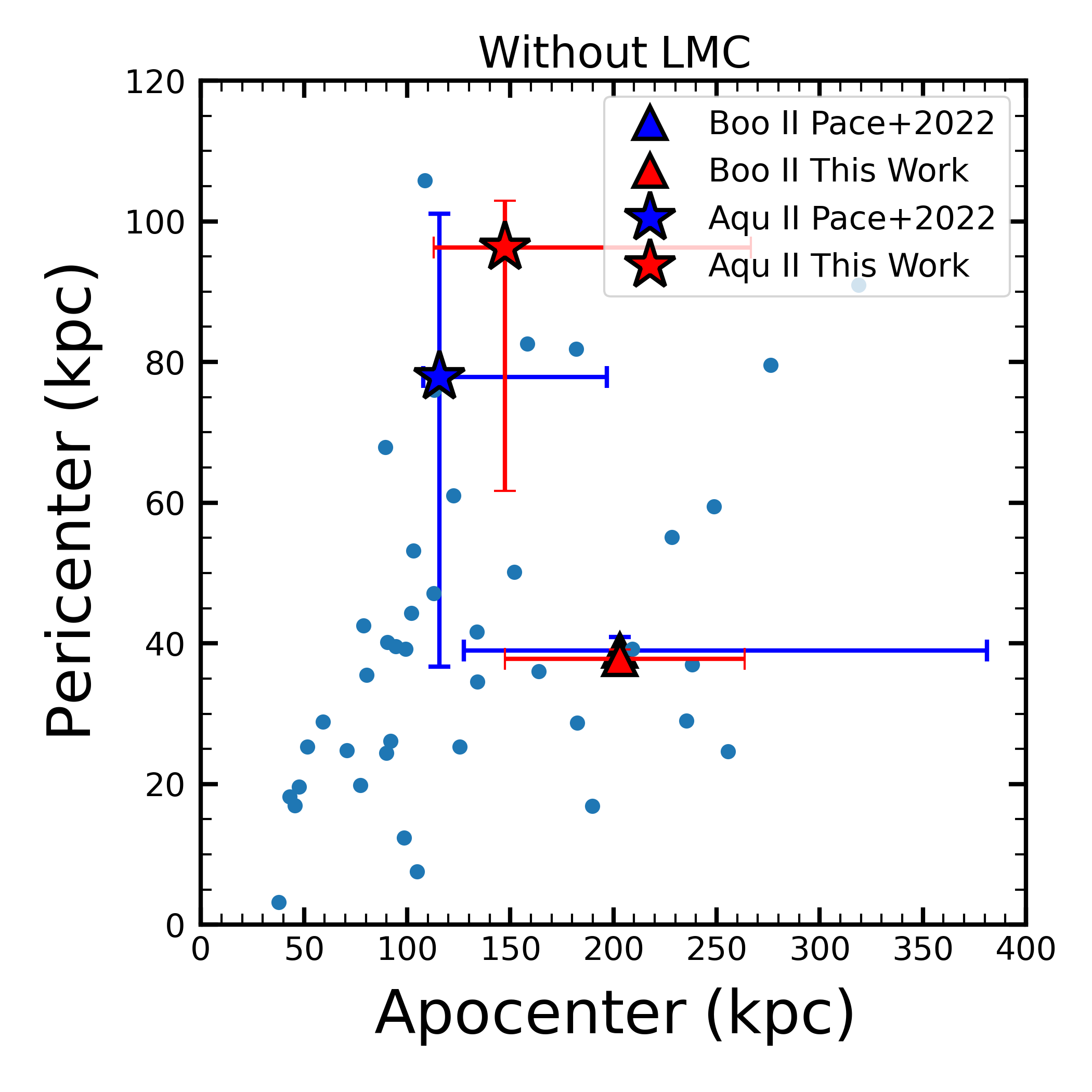}
    \includegraphics[width = .45\textwidth]{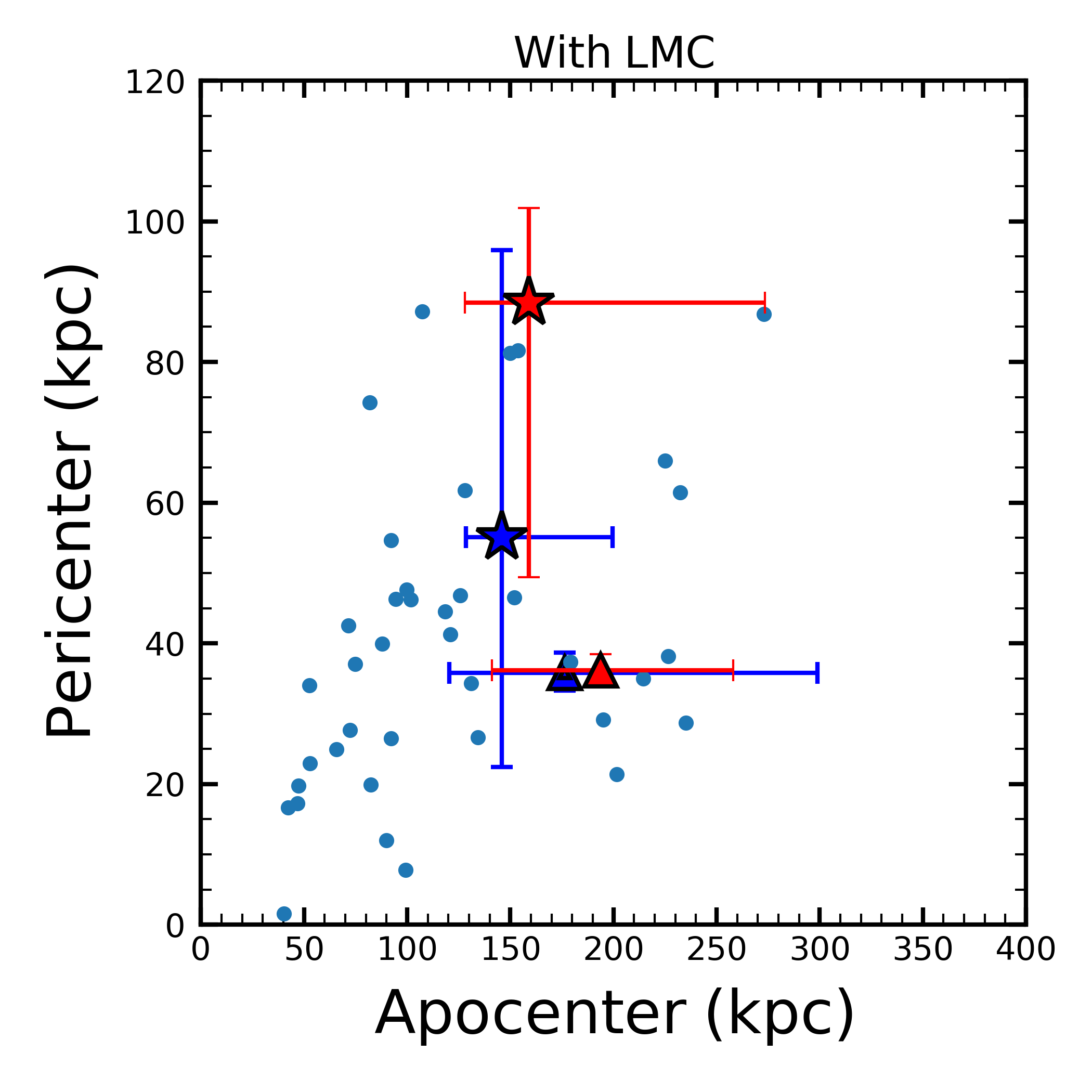}
    \caption{\hypertarget{Figure 7}{Comparison} of orbital characteristics for both UFDs between this work and \citet[]{2022arXiv220505699P}, with all known UFDs displayed as blue dots using values from \citet[]{2022arXiv220505699P}. Left: Comparison of the nearest apocenter and pericenter distances for both UFDs without the inclusion of the LMC. Right: Comparison of the apocenter and pericenter distances for both UFDs with the inclusion of the LMC.}
    \label{fig:7}
\end{figure*}

In Figure \ref{fig:7}, we show the orbital characteristics of these two UFDs along with all other known UFDs using values from \citet[]{2022arXiv220505699P}. 
For Aqu~II we measure a pericenter of $96^{+7}_{-35}$ kpc, an apocenter of $148^{+119}_{-35}$ kpc,  and an eccentricity of $0.30^{+0.20}_{-0.11}$ without the inclusion of the LMC. When including the LMC,  the pericenter decreases to $88^{+14}_{-39}$ kpc, the apocenter increases to $156^{+114}_{-31}$ kpc, and the eccentricity increases to $0.37^{+0.18}_{-0.08}$. These variances are insignificant ($<9 \%$ change) designating that Aqu~II is not significantly affected by the LMC.

For Boo~II we measure a pericenter of $38^{+1}_{-2}$ kpc, an apocenter of $203^{+61}_{-56}$ kpc, and an eccentricity of $0.69^{+0.06}_{-0.08}$ without the LMC potential. When including the LMC, the pericenter decreases to $36 \pm 2$ kpc, the apocenter decreases to $194^{+64}_{-53}$ kpc, and the eccentricity decreases to $0.68^{+0.06}_{-0.08}$. Once again, the insignificant change caused by considering the LMC ($<5\%$) dictates that Boo~II is not significantly affected by the LMC.

For Aqu~II, \citet[]{2022arXiv220505699P} reports a pericenter, apocenter and eccentricity of $77.9^{+23.2}_{-41.2}$ kpc, $115.6^{+81.3}_{-7.8}$ kpc and $0.31^{+0.24}_{-0.13}$ respectively without inclusion of LMC. Similarly for Boo~II, they report a pericenter, apocenter, and eccentricty of $39.0 \pm 1.9$ kpc, $203.0^{+178.1}_{-75.4}$ kpc, and $0.68^{+0.10}_{-0.11}$ without the LMC respectively. These values are consistent within 1$\sigma$ of our spectroscopic member sample which utilized improved systemic velocity measurements and excluded contaminant stars affecting proper motions. 

A study by \citet[]{Battaglia2022A&A...657A..54B} used {\it Gaia} EDR3 values to model the orbits of 74 dwarf galaxies or dwarf galaxy candidates in the Local Group using multiple gravitational potentials. The values attained using the heavy model (a larger total virial mass than in \texttt{McMillan17}) for Boo~II agree within 1$\sigma$ of the measurements of this work, however the values for Aqu~II only agree within 2$\sigma$ of the values in this work. \citet[]{Battaglia2022A&A...657A..54B} also used a light model (a smaller virial mass than in \texttt{McMillan17}) to analyze Aqu~II (but not Boo~II) which generated a pericenter and apocenter that agreed within 1$\sigma$ of the systemic values in this work, however the eccentricity still only agreed within 2$\sigma$. The main reason for this inconsistency is most likely due to the our use of the \texttt{McMillan17} potential as opposed to the potentials used in \citet[]{Battaglia2022A&A...657A..54B}.

Another study by \citet[]{2021ApJ...916....8L} similarly uses data from {\it Gaia} EDR3 to estimate the proper motion and model the orbits of 46 dwarf galaxies in the Milky Way. Comparing the values from their model using the most massive Milky Way mass available ($15 \times 10^{11} M_{\odot}$), the values for apocenter and pericenter of both Boo~II and Aqu~II are consistent within 1$\sigma$ of the values measured in this work. The eccentricity for Boo~II also agrees within 1$\sigma$ of our measurements, however the eccentricity of Aqu II only agrees within 2$\sigma$. 

\subsection{Astrophysical J-Factor}\label{subsec:4.5}

The UFDs are an excellent target for searches of dark matter annihilation or decay due to their close distance, large dark matter densities, and low astrophysical backgrounds \citep[e.g.,][]{Baltz2008JCAP...07..013B, 2015PhRvL.115w1301A}. 
The astrophysical component of the dark matter annihilation (decay) rate is referred to as the J-factor (D-factor) and depends on the squared (linear) dark matter density along the line-of-sight. 
To compute the dark matter density of Aqu~II and Boo~II we follow the methodology in  \citet{Pace2019MNRAS.482.3480P}. Briefly, we compute the line-of-sight velocity dispersion from the spherical Jeans equation and compare the solutions to the observed velocity dispersions to determine the dark matter distribution \citep[e.g.,][]{GeringerSameth2015ApJ...801...74G, Bonnivard2015MNRAS.453..849B}. 
For the Jeans modeling, we assume an NFW profile for the dark matter, a Plummer profile for the stellar distribution, and that the stellar anisotopy is constant with radius.  We assume Gaussian errors for the distance and structural parameters ($r_h$, $\epsilon$). For additional details see \citet{Pace2019MNRAS.482.3480P}. 

We apply this framework to  the 8 star sample of Aqu~II and the 9 star sample of Boo~II (removing the RRL and binary stars). 
For Aqu~II, we find $\log_{10}{J(\theta)}=17.5^{+0.6}_{-0.5}$, $17.7^{+0.6}_{-0.5}$,  $17.8^{+0.6}_{-0.5}$, and $17.8^{+0.6}_{-0.6}$   for solid angles of $\theta=0.1\degree, 0.2\degree, 0.5\degree, 1\degree$ in logarithmic units of ${\rm GeV^{2}~cm^{-5}}$. 
With the same methodology,  \citet{Pace2019MNRAS.482.3480P} found $\log_{10}{J(0.5^{\circ})}\sim18.3$ with the \citet{Torrealba_2016} data set. The difference is due to the  larger velocity dispersion in the \citet{Torrealba_2016} data set. 
For Boo~II, we find $\log_{10}{J(\theta)}=18.0^{+0.8}_{-1.1}$, $18.1^{+0.8}_{-1.1}$,  $18.3^{+0.8}_{-1.1}$, and $18.4^{+0.8}_{-1.1}$   for solid angles of $\theta=0.1\degree, 0.2\degree, 0.5\degree, 1\degree$. There is a small tail in the velocity dispersion towards low values which leads to a similar low value tail in the  J-factor. We note if we reject values below $\log_{10}{J}=16.5$ then the median J-factor increases by $\sim0.1$. In the future with a larger data set that fully resolves the dispersion we expect the J-factor to be larger than the median value quoted here. Boo~II has an intermediate J-factor and will be an important target for searches for dark matter annihilation.  Aqu~II will be useful for stacked analyses searching for dark matter annihilation products.

\section{Conclusion}\label{sec:5}
\hypertarget{Conclusion}{In this work} we present a refined  study of two  ultra-faint dwarf galaxies, Aquarius II and \text{Bo\"{o}tes II}, using medium resolution spectroscopy obtained from Magellan/IMACS. 
With a high efficiency target selection utilizing {\it Gaia} astrometry \citep{2022arXiv220505699P}, we have roughly doubled the number of spectroscopic members in both UFDs.

With our larger samples, we provide precise systemic velocities, velocity dispersions, metallicities, metallicity dispersions, and orbital characteristics such as apocenters and pericenters. Most notably we derive a velocity dispersion of \text{Bo\"{o}tes II} that is not affected by the presence of a known binary star and a further suspected binary. Combining the removal of these binary stars with our improved stellar velocity measurements resulted in the velocity dispersion decreasing by a factor of 3. We also confirm the extreme metal-poor nature of both satellites ([Fe/H] $<-2.50$), making them two of the least metal-enriched dwarfs known. Additionally, based on the large mass-to-light ratio ($M_{\odot}/L_{\odot} \sim  1300$), we conclude that Aquarius II is highly dark matter dominated. We do note that our observations are mostly focused on stars not previously observed, preventing us from testing for binary stars located in our sample (aside from the two binary stars described Table \ref{tab:5}). Further observations of high probability member candidates could improve upon this and provide opportunities for binarity to be tested and further increase the accuracy of our
results.

These observations also show the high efficiency in target selection based on the high probability member candidates constructed with the photometric and astrometric data from \citet{2022arXiv220505699P}. For both Aquarius II and \text{Bo\"{o}tes II} over $80\%$ of these well-observed candidates turned out to be members when observed with sufficient S/N. There still remain 5 (8) high-probability member candidates for Aqu~II (Boo~II) that can be prioritized in future spectroscopic follow ups due to this high membership success rate. 

\section{Acknowledgements}\label{sec:6}

JB \& TSL acknowledges financial support from Natural Sciences and Engineering Research Council of Canada (NSERC) through grant RGPIN-2022-04794.
ABP is supported by NSF grant AST-1813881. 

This paper includes data gathered with the 6.5 meter Magellan Telescopes located at Las Campanas Observatory, Chile.

This work has made use of data from the European Space Agency (ESA) mission
{\it Gaia} (\url{https://www.cosmos.esa.int/gaia}), processed by the {\it Gaia}
Data Processing and Analysis Consortium (DPAC,
\url{https://www.cosmos.esa.int/web/gaia/dpac/consortium}). Funding for the DPAC
has been provided by national institutions, in particular the institutions
participating in the {\it Gaia} Multilateral Agreement.

The Legacy Surveys consist of three individual and complementary projects: the Dark Energy Camera Legacy Survey (DECaLS; NOAO Proposal ID \# 2014B-0404; PIs: David Schlegel and Arjun Dey), the Beijing-Arizona Sky Survey (BASS; NOAO Proposal ID \# 2015A-0801; PIs: Zhou Xu and Xiaohui Fan), and the Mayall $z$-band Legacy Survey (MzLS; NOAO Proposal ID \# 2016A-0453; PI: Arjun Dey). DECaLS, BASS, and MzLS together include data obtained, respectively, at the Blanco telescope, Cerro Tololo Inter-American Observatory, National Optical Astronomy Observatory (NOAO); the Bok telescope, Steward Observatory, University of Arizona; and the Mayall telescope, Kitt Peak National Observatory, NOAO. The Legacy Surveys project is honored to be permitted to conduct astronomical research on Iolkam Du'ag (Kitt Peak), a mountain with particular significance to the Tohono O'odham Nation.

NOAO is operated by the Association of Universities for Research in Astronomy (AURA) under a cooperative agreement with the National Science Foundation.

This project used data obtained with the Dark Energy Camera (DECam), which was constructed by the DES collaboration. Funding for the DES Projects has been provided by the US Department of Energy, the US National Science Foundation, the Ministry of Science and Education of Spain, the Science and Technology Facilities Council of the United Kingdom, the Higher Education Funding Council for England, the National Center for Supercomputing Applications at the University of Illinois at Urbana-Champaign, the Kavli Institute of Cosmological Physics at the University of Chicago, Center for Cosmology and Astro-Particle Physics at the Ohio State University, the Mitchell Institute for Fundamental Physics and Astronomy at Texas A\&M University, Financiadora de Estudos e Projetos, \text{Fundação} Carlos Chagas Filho de Amparo, Financiadora de Estudos e Projetos, \text{Fundação} Carlos Chagas Filho de Amparo \text{à} Pesquisa do Estado do Rio de Janeiro, Conselho Nacional de Desenvolvimento Cientifico e Tecnologico and the Ministerio da Ciencia, Tecnologia e Inovacao, the Deutsche Forschungsgemeinschaft and the Collaborating Institutions in the Dark Energy Survey. The Collaborating Institutions are Argonne National Laboratory, the University of California at Santa Cruz, the University of Cambridge, Centro de Investigaciones Energeticas, Medioambientales y Tecnologicas-Madrid, the University of Chicago, University College London, the DES-Brazil Consortium, the University of Edinburgh, the Eidgenossische Technische Hochschule (ETH) Zurich, Fermi National Accelerator Laboratory, the University of Illinois at Urbana-Champaign, the Institut de Ciencies de l'Espai (IEEC/CSIC), the Institut de Fisica d'Altes Energies, Lawrence Berkeley National Laboratory, the Ludwig-Maximilians Universitat Munchen, and the associated Excellence Cluster Universe, the University of Michigan, the National Optical Astronomy Observatory, the University of Nottingham, the Ohio State University, the University of Pennsylvania, the University of Portsmouth, SLAC National Accelerator Laboratory, Stanford University, the University of Sussex, and Texas A\&M University.

BASS is a key project of the Telescope Access Program (TAP), which has been funded by the National Astronomical Observatories of China, the Chinese Academy of Sciences (the Strategic Priority Research Program ``The Emergence of Cosmological Structures'' grant \# XDB09000000), and the Special Fund for Astronomy from the Ministry of Finance. The BASS is also supported by the External Cooperation Program of Chinese Academy of Sciences (grant \# 114A11KYSB20160057), and Chinese National Natural Science Foundation (grant \# 11433005).

The Legacy Survey team makes use of data products from the Near-Earth Object Wide-field Infrared Survey Explorer (NEOWISE), which is a project of the Jet Propulsion Laboratory/California Institute of Technology. NEOWISE is funded by the National Aeronautics and Space Administration.

The Legacy Surveys imaging of the DESI footprint is supported by the Director, Office of Science, Office of High Energy Physics of the US Department of Energy under contract No. DE-AC02-05CH1123, by the National Energy Research Scientific Computing Center, a DOE Office of Science User Facility under the same contract; and by the US National Science Foundation, Division of Astronomical Sciences under contract No. AST-0950945 to NOAO.



\appendix

\section{References for dwarf galaxy data in Figures~5~and~7}
\label{app:refs}

Here we provide a list the references for the velocity dispersions, metallicities, and luminosities of the dwarf galaxies plotted in Figure~\ref{fig:5}: 
  \citet{msw+03}; \citet{bth+06}; \citet{sg+07}; \citet{bic+08}; \citet{dhc+08}; \citet{mow+08}; \citet{oay+08}; \citet{Koch_2008}; \citet{wmo+09b}; \citet{wmo+09}; \citet{sgm+11}; \citet{wgs+11}; \citet{fmt+12}; \citet{kbc+13}; \citet{kcg+13}; \citet{2014ApJ...786...74F}; \citet{Bechtol_2015}; \citet{dbr+15}; \citet{kcb+15}; \citet{kjm+15}; \citet{ksc+15}; \citet{sdl+15}; \citet{csz+16}; \citet{jfs+16_ret}; \citet{kjg+16}; \citet{tkb+16b}; \citet{Torrealba_2016}; \citet{cts+17};  \citet{cwm+17}; \citet{Kirby2017ApJ...838...83K}; \citet{Li_2017}; \citet{mbi+17}; \citet{Simon_2017}; \citet{smw+17}; \citet{cfj+18}; \citet{kwb+18}; \citet{lms+18}; \citet{Li2018ApJ...857..145L}; \citet{2018ApJ...860...66M};  \citet{msc+18}; \citet{tbk+18}; \citet{doi:10.1146/annurev-astro-091918-104453}; \citet{Simon2020ApJ...892..137S}; \citet{jlp+21}; \citet{lmi+21}; \citet{cfj+22}; \citet{2022arXiv220311788C};.

\bibliography{main}

\newcommand{\noop}[1]{}
\begin{thebibliography}{}
\expandafter\ifx\csname natexlab\endcsname\relax\def\natexlab#1{#1}\fi
\providecommand{\url}[1]{\href{#1}{#1}}
\providecommand{\dodoi}[1]{doi:~\href{http://doi.org/#1}{\nolinkurl{#1}}}
\providecommand{\doeprint}[1]{\href{http://ascl.net/#1}{\nolinkurl{http://ascl.net/#1}}}
\providecommand{\doarXiv}[1]{\href{https://arxiv.org/abs/#1}{\nolinkurl{https://arxiv.org/abs/#1}}}

\bibitem[{{Ackermann} {et~al.}(2015){Ackermann}, {Albert}, {Anderson},
  {Atwood}, {Baldini}, {Barbiellini}, {Bastieri}, {Bechtol}, {Bellazzini},
  {Bissaldi}, {Blandford}, {Bloom}, {Bonino}, {Bottacini}, {Brandt}, {Bregeon},
  {Bruel}, {Buehler}, {Caliandro}, {Cameron}, {Caputo}, {Caragiulo}, {Caraveo},
  {Cecchi}, {Charles}, {Chekhtman}, {Chiang}, {Chiaro}, {Ciprini}, {Claus},
  {Cohen-Tanugi}, {Conrad}, {Cuoco}, {Cutini}, {D'Ammando}, {de Angelis}, {de
  Palma}, {Desiante}, {Digel}, {Di Venere}, {Drell}, {Drlica-Wagner}, {Essig},
  {Favuzzi}, {Fegan}, {Ferrara}, {Focke}, {Franckowiak}, {Fukazawa}, {Funk},
  {Fusco}, {Gargano}, {Gasparrini}, {Giglietto}, {Giordano}, {Giroletti},
  {Glanzman}, {Godfrey}, {Gomez-Vargas}, {Grenier}, {Guiriec}, {Gustafsson},
  {Hays}, {Hewitt}, {Horan}, {Jogler}, {J{\'o}hannesson}, {Kuss}, {Larsson},
  {Latronico}, {Li}, {Li}, {Llena Garde}, {Longo}, {Loparco}, {Lubrano},
  {Malyshev}, {Mayer}, {Mazziotta}, {McEnery}, {Meyer}, {Michelson}, {Mizuno},
  {Moiseev}, {Monzani}, {Morselli}, {Murgia}, {Nuss}, {Ohsugi}, {Orienti},
  {Orlando}, {Ormes}, {Paneque}, {Perkins}, {Pesce-Rollins}, {Piron}, {Pivato},
  {Porter}, {Rain{\`o}}, {Rando}, {Razzano}, {Reimer}, {Reimer}, {Ritz},
  {S{\'a}nchez-Conde}, {Schulz}, {Sehgal}, {Sgr{\`o}}, {Siskind}, {Spada},
  {Spandre}, {Spinelli}, {Strigari}, {Tajima}, {Takahashi}, {Thayer},
  {Tibaldo}, {Torres}, {Troja}, {Vianello}, {Werner}, {Winer}, {Wood}, {Wood},
  {Zaharijas}, {Zimmer}, \& {Fermi-LAT Collaboration}}]{2015PhRvL.115w1301A}
{Ackermann}, M., {Albert}, A., {Anderson}, B., {et~al.} 2015, \prl, 115,
  231301, \dodoi{10.1103/PhysRevLett.115.231301}

\bibitem[{{Ahn} {et~al.}(2012){Ahn}, {Alexandroff}, {Allende Prieto},
  {Anderson}, {Anderton}, {Andrews}, {Aubourg}, {Bailey}, {Balbinot}, {Barnes},
  {Bautista}, {Beers}, {Beifiori}, {Berlind}, {Bhardwaj}, {Bizyaev}, {Blake},
  {Blanton}, {Blomqvist}, {Bochanski}, {Bolton}, {Borde}, {Bovy}, {Brandt},
  {Brinkmann}, {Brown}, {Brownstein}, {Bundy}, {Busca}, {Carithers}, {Carnero},
  {Carr}, {Casetti-Dinescu}, {Chen}, {Chiappini}, {Comparat}, {Connolly},
  {Crepp}, {Cristiani}, {Croft}, {Cuesta}, {da Costa}, {Davenport}, {Dawson},
  {de Putter}, {De Lee}, {Delubac}, {Dhital}, {Ealet}, {Ebelke}, {Edmondson},
  {Eisenstein}, {Escoffier}, {Esposito}, {Evans}, {Fan}, {Femen{\'\i}a
  Castell{\'a}}, {Fern{\'a}ndez Alvar}, {Ferreira}, {Filiz Ak}, {Finley},
  {Fleming}, {Font-Ribera}, {Frinchaboy}, {Garc{\'\i}a-Hern{\'a}ndez},
  {Garc{\'\i}a P{\'e}rez}, {Ge}, {G{\'e}nova-Santos}, {Gillespie}, {Girardi},
  {Gonz{\'a}lez Hern{\'a}ndez}, {Grebel}, {Gunn}, {Guo}, {Haggard}, {Hamilton},
  {Harris}, {Hawley}, {Hearty}, {Ho}, {Hogg}, {Holtzman}, {Honscheid},
  {Huehnerhoff}, {Ivans}, {Ivezi{\'c}}, {Jacobson}, {Jiang}, {Johansson},
  {Johnson}, {Kauffmann}, {Kirkby}, {Kirkpatrick}, {Klaene}, {Knapp}, {Kneib},
  {Le Goff}, {Leauthaud}, {Lee}, {Lee}, {Long}, {Loomis}, {Lucatello},
  {Lundgren}, {Lupton}, {Ma}, {Ma}, {MacDonald}, {Mack}, {Mahadevan}, {Maia},
  {Majewski}, {Makler}, {Malanushenko}, {Malanushenko}, {Manchado},
  {Mandelbaum}, {Manera}, {Maraston}, {Margala}, {Martell}, {McBride},
  {McGreer}, {McMahon}, {M{\'e}nard}, {Meszaros}, {Miralda-Escud{\'e}},
  {Montero-Dorta}, {Montesano}, {Morrison}, {Muna}, {Munn}, {Murayama},
  {Myers}, {Neto}, {Nguyen}, {Nichol}, {Nidever}, {Noterdaeme}, {Nuza},
  {Ogando}, {Olmstead}, {Oravetz}, {Owen}, {Padmanabhan},
  {Palanque-Delabrouille}, {Pan}, {Parejko}, {Parihar}, {P{\^a}ris},
  {Pattarakijwanich}, {Pepper}, {Percival}, {P{\'e}rez-Fournon},
  {P{\'e}rez-R{\`a}fols}, {Petitjean}, {Pforr}, {Pieri}, {Pinsonneault}, {Porto
  de Mello}, {Prada}, {Price-Whelan}, {Raddick}, {Rebolo}, {Rich}, {Richards},
  {Robin}, {Rocha-Pinto}, {Rockosi}, {Roe}, {Ross}, {Ross}, {Rossi},
  {Rubi{\~n}o-Martin}, {Samushia}, {Sanchez Almeida}, {S{\'a}nchez},
  {Santiago}, {Sayres}, {Schlegel}, {Schlesinger}, {Schmidt}, {Schneider},
  {Schultheis}, {Schwope}, {Sc{\'o}ccola}, {Seljak}, {Sheldon}, {Shen}, {Shu},
  {Simmerer}, {Simmons}, {Skibba}, {Skrutskie}, {Slosar}, {Sobreira}, {Sobeck},
  {Stassun}, {Steele}, {Steinmetz}, {Strauss}, {Streblyanska}, {Suzuki},
  {Swanson}, {Tal}, {Thakar}, {Thomas}, {Thompson}, {Tinker}, {Tojeiro},
  {Tremonti}, {Vargas Maga{\~n}a}, {Verde}, {Viel}, {Vikas}, {Vogt}, {Wake},
  {Wang}, {Weaver}, {Weinberg}, {Weiner}, {West}, {White}, {Wilson},
  {Wisniewski}, {Wood-Vasey}, {Yanny}, {Y{\`e}che}, {York}, {Zamora},
  {Zasowski}, {Zehavi}, {Zhao}, {Zheng}, {Zhu}, \&
  {Zinn}}]{2012ApJS..203...21A}
{Ahn}, C.~P., {Alexandroff}, R., {Allende Prieto}, C., {et~al.} 2012, \apjs,
  203, 21, \dodoi{10.1088/0067-0049/203/2/21}

\bibitem[{{Baltz} {et~al.}(2008){Baltz}, {Berenji}, {Bertone}, {Bergstr{\"o}m},
  {Bloom}, {Bringmann}, {Chiang}, {Cohen-Tanugi}, {Conrad}, {Edmonds},
  {Edsj{\"o}}, {Godfrey}, {Hughes}, {Johnson}, {Lionetto}, {Moiseev},
  {Morselli}, {Moskalenko}, {Nuss}, {Ormes}, {Rando}, {Sander}, {Sellerholm},
  {Smith}, {Strong}, {Wai}, {Wang}, \& {Winer}}]{Baltz2008JCAP...07..013B}
{Baltz}, E.~A., {Berenji}, B., {Bertone}, G., {et~al.} 2008, \jcap, 2008, 013,
  \dodoi{10.1088/1475-7516/2008/07/013}

\bibitem[{{Battaglia} {et~al.}(2022){Battaglia}, {Taibi}, {Thomas}, \&
  {Fritz}}]{Battaglia2022A&A...657A..54B}
{Battaglia}, G., {Taibi}, S., {Thomas}, G.~F., \& {Fritz}, T.~K. 2022, \aap,
  657, A54, \dodoi{10.1051/0004-6361/202141528}

\bibitem[{{Battaglia} {et~al.}(2006){Battaglia}, {Tolstoy}, {Helmi}, {Irwin},
  {Letarte}, {Jablonka}, {Hill}, {Venn}, {Shetrone}, {Arimoto}, {Primas},
  {Kaufer}, {Francois}, {Szeifert}, {Abel}, \& {Sadakane}}]{bth+06}
{Battaglia}, G., {Tolstoy}, E., {Helmi}, A., {et~al.} 2006, \aap, 459, 423,
  \dodoi{10.1051/0004-6361:20065720}

\bibitem[{Bechtol {et~al.}(2015)Bechtol, Drlica-Wagner, Balbinot, Pieres,
  Simon, Yanny, Santiago, Wechsler, Frieman, Walker, Williams, Rozo, Rykoff,
  Queiroz, Luque, Benoit-L{\'{e}}vy, Tucker, Sevilla, Gruendl, da~Costa, Neto,
  Maia, Abbott, Allam, Armstrong, Bauer, Bernstein, Bernstein, Bertin, Brooks,
  Buckley-Geer, Burke, Rosell, Castander, Covarrubias, D'Andrea, DePoy, Desai,
  Diehl, Eifler, Estrada, Evrard, Fernandez, Finley, Flaugher, Gaztanaga,
  Gerdes, Girardi, Gladders, Gruen, Gutierrez, Hao, Honscheid, Jain, James,
  Kent, Kron, Kuehn, Kuropatkin, Lahav, Li, Lin, Makler, March, Marshall,
  Martini, Merritt, Miller, Miquel, Mohr, Neilsen, Nichol, Nord, Ogando,
  Peoples, Petravick, Plazas, Romer, Roodman, Sako, Sanchez, Scarpine,
  Schubnell, Smith, Soares-Santos, Sobreira, Suchyta, Swanson, Tarle, Thaler,
  Thomas, Wester, \& and}]{Bechtol_2015}
Bechtol, K., Drlica-Wagner, A., Balbinot, E., {et~al.} 2015, The Astrophysical
  Journal, 807, 50, \dodoi{10.1088/0004-637x/807/1/50}

\bibitem[{{Bellazzini} {et~al.}(2008){Bellazzini}, {Ibata}, {Chapman},
  {Mackey}, {Monaco}, {Irwin}, {Martin}, {Lewis}, \& {Dalessandro}}]{bic+08}
{Bellazzini}, M., {Ibata}, R.~A., {Chapman}, S.~C., {et~al.} 2008, \aj, 136,
  1147, \dodoi{10.1088/0004-6256/136/3/1147}

\bibitem[{{Bonnivard} {et~al.}(2015){Bonnivard}, {Combet}, {Daniel}, {Funk},
  {Geringer-Sameth}, {Hinton}, {Maurin}, {Read}, {Sarkar}, {Walker}, \&
  {Wilkinson}}]{Bonnivard2015MNRAS.453..849B}
{Bonnivard}, V., {Combet}, C., {Daniel}, M., {et~al.} 2015, \mnras, 453, 849,
  \dodoi{10.1093/mnras/stv1601}

\bibitem[{{Bovill} \& {Ricotti}(2009)}]{2009ApJ...693.1859B}
{Bovill}, M.~S., \& {Ricotti}, M. 2009, \apj, 693, 1859,
  \dodoi{10.1088/0004-637X/693/2/1859}

\bibitem[{{Bovy}(2015)}]{2015ApJS..216...29B}
{Bovy}, J. 2015, \apjs, 216, 29, \dodoi{10.1088/0067-0049/216/2/29}

\bibitem[{{Bovy} {et~al.}(2016){Bovy}, {Bahmanyar}, {Fritz}, \&
  {Kallivayalil}}]{2016ApJ...833...31B}
{Bovy}, J., {Bahmanyar}, A., {Fritz}, T.~K., \& {Kallivayalil}, N. 2016, \apj,
  833, 31, \dodoi{10.3847/1538-4357/833/1/31}

\bibitem[{{Brown} {et~al.}(2012){Brown}, {Tumlinson}, {Geha}, {Kirby},
  {VandenBerg}, {Mu{\~n}oz}, {Kalirai}, {Simon}, {Avila}, {Guhathakurta},
  {Renzini}, \& {Ferguson}}]{2012ApJ...753L..21B}
{Brown}, T.~M., {Tumlinson}, J., {Geha}, M., {et~al.} 2012, \apjl, 753, L21,
  \dodoi{10.1088/2041-8205/753/1/L21}

\bibitem[{{Brown} {et~al.}(2014){Brown}, {Tumlinson}, {Geha}, {Simon},
  {Vargas}, {VandenBerg}, {Kirby}, {Kalirai}, {Avila}, {Gennaro}, {Ferguson},
  {Mu{\~n}oz}, {Guhathakurta}, \& {Renzini}}]{brown_2014}
---. 2014, \apj, 796, 91, \dodoi{10.1088/0004-637X/796/2/91}

\bibitem[{{Buttry} {et~al.}(2022){Buttry}, {Pace}, {Koposov}, {Walker},
  {Caldwell}, {Kirby}, {Martin}, {Mateo}, {Olszewski}, {Starkenburg},
  {Badenes}, \& {Daher}}]{Buttry2022MNRAS.514.1706B}
{Buttry}, R., {Pace}, A.~B., {Koposov}, S.~E., {et~al.} 2022, \mnras, 514,
  1706, \dodoi{10.1093/mnras/stac1441}

\bibitem[{{Caldwell} {et~al.}(2017){Caldwell}, {Walker}, {Mateo}, {Olszewski},
  {Koposov}, {Belokurov}, {Torrealba}, {Geringer-Sameth}, \&
  {Johnson}}]{cwm+17}
{Caldwell}, N., {Walker}, M.~G., {Mateo}, M., {et~al.} 2017, \apj, 839, 20,
  \dodoi{10.3847/1538-4357/aa688e}

\bibitem[{Carrera {et~al.}(2013)Carrera, Pancino, Gallart, \& del
  Pino}]{10.1093/mnras/stt1126}
Carrera, R., Pancino, E., Gallart, C., \& del Pino, A. 2013, Monthly Notices of
  the Royal Astronomical Society, 434, 1681, \dodoi{10.1093/mnras/stt1126}

\bibitem[{{Cerny} {et~al.}(2022){Cerny}, {Simon}, {Li}, {Drlica-Wagner},
  {Pace}, {Mart{\i}nez-Vazquez}, {Riley}, {Mutlu-Pakdil}, {Mau}, {Ferguson},
  {Erkal}, {Munoz}, {Bom}, {Carlin}, {Carollo}, {Choi}, {Ji},
  {Mart{\i}nez-Delgado}, {Manwadkar}, {Miller}, {Noel}, {Sakowska}, {Sand},
  {Stringfellow}, {Tollerud}, {Vivas}, {Carballo-Bello}, {Hernandez-Lang},
  {James}, {Nilo Castellon}, {Olsen}, \& {Zenteno}}]{2022arXiv220311788C}
{Cerny}, W., {Simon}, J.~D., {Li}, T.~S., {et~al.} 2022, arXiv e-prints,
  arXiv:2203.11788.
\newblock \doarXiv{2203.11788}

\bibitem[{{Chiti} {et~al.}(2018){Chiti}, {Frebel}, {Ji}, {Jerjen}, {Kim}, \&
  {Norris}}]{cfj+18}
{Chiti}, A., {Frebel}, A., {Ji}, A.~P., {et~al.} 2018, \apj, 857, 74,
  \dodoi{10.3847/1538-4357/aab4fc}

\bibitem[{{Chiti} {et~al.}(2022{\natexlab{a}}){Chiti}, {Simon}, {Frebel},
  {Pace}, {Ji}, \& {Li}}]{2022arXiv220604580C}
{Chiti}, A., {Simon}, J.~D., {Frebel}, A., {et~al.} 2022{\natexlab{a}}, arXiv
  e-prints, arXiv:2206.04580.
\newblock \doarXiv{2206.04580}

\bibitem[{{Chiti} {et~al.}(2022{\natexlab{b}}){Chiti}, {Frebel}, {Ji},
  {Mardini}, {Ou}, {Simon}, {Rasmussen}, {Jerjen}, {Kim}, \& {Norris}}]{cfj+22}
{Chiti}, A., {Frebel}, A., {Ji}, A.~P., {et~al.} 2022{\natexlab{b}}, arXiv
  e-prints, arXiv:2205.01740.
\newblock \doarXiv{2205.01740}

\bibitem[{{Clementini} {et~al.}(2022){Clementini}, {Ripepi}, {Garofalo},
  {Molinaro}, {Muraveva}, {Leccia}, {Rimoldini}, {Holl}, {Jevardat de
  Fombelle}, {Sartoretti}, {Marchal}, {Audard}, {Nienartowicz}, {Andrae},
  {Marconi}, {Szabados}, {Evans}, {Lecoeur-Taibi}, {Mowlavi}, {Musella}, \&
  {Eyer}}]{2022arXiv220606278C}
{Clementini}, G., {Ripepi}, V., {Garofalo}, A., {et~al.} 2022, arXiv e-prints,
  arXiv:2206.06278.
\newblock \doarXiv{2206.06278}

\bibitem[{{Collins} {et~al.}(2017){Collins}, {Tollerud}, {Sand}, {Bonaca},
  {Willman}, \& {Strader}}]{cts+17}
{Collins}, M. L.~M., {Tollerud}, E.~J., {Sand}, D.~J., {et~al.} 2017, \mnras,
  467, 573, \dodoi{10.1093/mnras/stx067}

\bibitem[{{Cooper} {et~al.}(2012){Cooper}, {Newman}, {Davis}, {Finkbeiner}, \&
  {Gerke}}]{2012ascl.soft03003C}
{Cooper}, M.~C., {Newman}, J.~A., {Davis}, M., {Finkbeiner}, D.~P., \& {Gerke},
  B.~F. 2012, {spec2d: DEEP2 DEIMOS Spectral Pipeline}, Astrophysics Source
  Code Library, record ascl:1203.003.
\newblock \doeprint{1203.003}

\bibitem[{{Crnojevi{\'c}} {et~al.}(2016){Crnojevi{\'c}}, {Sand}, {Zaritsky},
  {Spekkens}, {Willman}, \& {Hargis}}]{csz+16}
{Crnojevi{\'c}}, D., {Sand}, D.~J., {Zaritsky}, D., {et~al.} 2016, \apjl, 824,
  L14, \dodoi{10.3847/2041-8205/824/1/L14}

\bibitem[{{de Jong} {et~al.}(2008){de Jong}, {Harris}, {Coleman}, {Martin},
  {Bell}, {Rix}, {Hill}, {Skillman}, {Sand}, {Olszewski}, {Zaritsky},
  {Thompson}, {Giallongo}, {Ragazzoni}, {DiPaola}, {Farinato}, {Testa}, \&
  {Bechtold}}]{dhc+08}
{de Jong}, J.~T.~A., {Harris}, J., {Coleman}, M.~G., {et~al.} 2008, \apj, 680,
  1112, \dodoi{10.1086/587835}

\bibitem[{{DES Collaboration} {et~al.}(2018){DES Collaboration}, {Abbott},
  {Abdalla}, {Allam}, {Amara}, {Annis}, {Asorey}, {Avila}, {Ballester},
  {Banerji}, {Barkhouse}, {Baruah}, {Baumer}, {Bechtol}, {Becker},
  {Benoit-L{\'e}vy}, {Bernstein}, {Bertin}, {Blazek}, {Bocquet}, {Brooks},
  {Brout}, {Buckley-Geer}, {Burke}, {Busti}, {Campisano}, {Cardiel-Sas},
  {Carnero Rosell}, {Carrasco Kind}, {Carretero}, {Castander}, {Cawthon},
  {Chang}, {Chen}, {Conselice}, {Costa}, {Crocce}, {Cunha}, {D'Andrea}, {da
  Costa}, {Das}, {Daues}, {Davis}, {Davis}, {De Vicente}, {DePoy}, {DeRose},
  {Desai}, {Diehl}, {Dietrich}, {Dodelson}, {Doel}, {Drlica-Wagner}, {Eifler},
  {Elliott}, {Evrard}, {Farahi}, {Fausti Neto}, {Fernandez}, {Finley},
  {Flaugher}, {Foley}, {Fosalba}, {Friedel}, {Frieman}, {Garc{\'\i}a-Bellido},
  {Gaztanaga}, {Gerdes}, {Giannantonio}, {Gill}, {Glazebrook}, {Goldstein},
  {Gower}, {Gruen}, {Gruendl}, {Gschwend}, {Gupta}, {Gutierrez}, {Hamilton},
  {Hartley}, {Hinton}, {Hislop}, {Hollowood}, {Honscheid}, {Hoyle}, {Huterer},
  {Jain}, {James}, {Jeltema}, {Johnson}, {Johnson}, {Kacprzak}, {Kent},
  {Khullar}, {Klein}, {Kovacs}, {Koziol}, {Krause}, {Kremin}, {Kron}, {Kuehn},
  {Kuhlmann}, {Kuropatkin}, {Lahav}, {Lasker}, {Li}, {Li}, {Liddle}, {Lima},
  {Lin}, {L{\'o}pez-Reyes}, {MacCrann}, {Maia}, {Maloney}, {Manera}, {March},
  {Marriner}, {Marshall}, {Martini}, {McClintock}, {McKay}, {McMahon},
  {Melchior}, {Menanteau}, {Miller}, {Miquel}, {Mohr}, {Morganson}, {Mould},
  {Neilsen}, {Nichol}, {Nogueira}, {Nord}, {Nugent}, {Nunes}, {Ogand o}, {Old},
  {Pace}, {Palmese}, {Paz-Chinch{\'o}n}, {Peiris}, {Percival}, {Petravick},
  {Plazas}, {Poh}, {Pond}, {Porredon}, {Pujol}, {Refregier}, {Reil}, {Ricker},
  {Rollins}, {Romer}, {Roodman}, {Rooney}, {Ross}, {Rykoff}, {Sako}, {Sanchez},
  {Sanchez}, {Santiago}, {Saro}, {Scarpine}, {Scolnic}, {Serrano},
  {Sevilla-Noarbe}, {Sheldon}, {Shipp}, {Silveira}, {Smith}, {Smith}, {Smith},
  {Soares-Santos}, {Sobreira}, {Song}, {Stebbins}, {Suchyta}, {Sullivan},
  {Swanson}, {Tarle}, {Thaler}, {Thomas}, {Thomas}, {Troxel}, {Tucker},
  {Vikram}, {Vivas}, {Walker}, {Wechsler}, {Weller}, {Wester}, {Wolf}, {Wu},
  {Yanny}, {Zenteno}, {Zhang}, {Zuntz}, {DES Collaboration}, {Juneau},
  {Fitzpatrick}, {Nikutta}, {Nidever}, {Olsen}, {Scott}, \& {NOAO Data
  Lab}}]{desdr1}
{DES Collaboration}, {Abbott}, T.~M.~C., {Abdalla}, F.~B., {et~al.} 2018,
  \apjs, 239, 18, \dodoi{10.3847/1538-4365/aae9f0}

\bibitem[{{Dey} {et~al.}(2019){Dey}, {Schlegel}, {Lang}, {Blum}, {Burleigh},
  {Fan}, {Findlay}, {Finkbeiner}, {Herrera}, {Juneau}, {Landriau}, {Levi},
  {McGreer}, {Meisner}, {Myers}, {Moustakas}, {Nugent}, {Patej}, {Schlafly},
  {Walker}, {Valdes}, {Weaver}, {Y{\`e}che}, {Zou}, {Zhou}, {Abareshi},
  {Abbott}, {Abolfathi}, {Aguilera}, {Alam}, {Allen}, {Alvarez}, {Annis},
  {Ansarinejad}, {Aubert}, {Beechert}, {Bell}, {BenZvi}, {Beutler}, {Bielby},
  {Bolton}, {Brice{\~n}o}, {Buckley-Geer}, {Butler}, {Calamida}, {Carlberg},
  {Carter}, {Casas}, {Castander}, {Choi}, {Comparat}, {Cukanovaite}, {Delubac},
  {DeVries}, {Dey}, {Dhungana}, {Dickinson}, {Ding}, {Donaldson}, {Duan},
  {Duckworth}, {Eftekharzadeh}, {Eisenstein}, {Etourneau}, {Fagrelius},
  {Farihi}, {Fitzpatrick}, {Font-Ribera}, {Fulmer}, {G{\"a}nsicke},
  {Gaztanaga}, {George}, {Gerdes}, {Gontcho}, {Gorgoni}, {Green}, {Guy},
  {Harmer}, {Hernandez}, {Honscheid}, {Huang}, {James}, {Jannuzi}, {Jiang},
  {Joyce}, {Karcher}, {Karkar}, {Kehoe}, {Kneib}, {Kueter-Young}, {Lan},
  {Lauer}, {Le Guillou}, {Le Van Suu}, {Lee}, {Lesser}, {Perreault Levasseur},
  {Li}, {Mann}, {Marshall}, {Mart{\'\i}nez-V{\'a}zquez}, {Martini}, {du Mas des
  Bourboux}, {McManus}, {Meier}, {M{\'e}nard}, {Metcalfe},
  {Mu{\~n}oz-Guti{\'e}rrez}, {Najita}, {Napier}, {Narayan}, {Newman}, {Nie},
  {Nord}, {Norman}, {Olsen}, {Paat}, {Palanque-Delabrouille}, {Peng},
  {Poppett}, {Poremba}, {Prakash}, {Rabinowitz}, {Raichoor}, {Rezaie},
  {Robertson}, {Roe}, {Ross}, {Ross}, {Rudnick}, {Safonova}, {Saha},
  {S{\'a}nchez}, {Savary}, {Schweiker}, {Scott}, {Seo}, {Shan}, {Silva},
  {Slepian}, {Soto}, {Sprayberry}, {Staten}, {Stillman}, {Stupak}, {Summers},
  {Sien Tie}, {Tirado}, {Vargas-Maga{\~n}a}, {Vivas}, {Wechsler}, {Williams},
  {Yang}, {Yang}, {Yapici}, {Zaritsky}, {Zenteno}, {Zhang}, {Zhang}, {Zhou}, \&
  {Zhou}}]{2019AJ....157..168D}
{Dey}, A., {Schlegel}, D.~J., {Lang}, D., {et~al.} 2019, \aj, 157, 168,
  \dodoi{10.3847/1538-3881/ab089d}

\bibitem[{{Dotter} {et~al.}(2008){Dotter}, {Chaboyer}, {Jevremovi{\'c}},
  {Kostov}, {Baron}, \& {Ferguson}}]{2008ApJS..178...89D}
{Dotter}, A., {Chaboyer}, B., {Jevremovi{\'c}}, D., {et~al.} 2008, \apjs, 178,
  89, \dodoi{10.1086/589654}

\bibitem[{Dressler {et~al.}(2006)Dressler, Hare, Bigelow, \&
  Osip}]{10.1117/12.670573}
Dressler, A., Hare, T., Bigelow, B.~C., \& Osip, D.~J. 2006, in Ground-based
  and Airborne Instrumentation for Astronomy, ed. I.~S. McLean \& M.~Iye, Vol.
  6269, International Society for Optics and Photonics (SPIE), 139 -- 151,
  \dodoi{10.1117/12.670573}

\bibitem[{Dressler {et~al.}(2011)Dressler, Bigelow, Hare, Sutin, Thompson,
  Burley, Epps, Oemler, Bagish, Birk, Clardy, Gunnels, Kelson, Shectman, \&
  Osip}]{Dressler_2011}
Dressler, A., Bigelow, B., Hare, T., {et~al.} 2011, Publications of the
  Astronomical Society of the Pacific, 123, 288, \dodoi{10.1086/658908}

\bibitem[{{Drlica-Wagner} {et~al.}(2015){Drlica-Wagner}, {Bechtol}, {Rykoff},
  {Luque}, {Queiroz}, {Mao}, {Wechsler}, {Simon}, {Santiago}, {Yanny},
  {Balbinot}, {Dodelson}, {Fausti Neto}, {James}, {Li}, {Maia}, {Marshall},
  {Pieres}, {Stringer}, {Walker}, {Abbott}, {Abdalla}, {Allam},
  {Benoit-L{\'e}vy}, {Bernstein}, {Bertin}, {Brooks}, {Buckley-Geer}, {Burke},
  {Carnero Rosell}, {Carrasco Kind}, {Carretero}, {Crocce}, {da Costa},
  {Desai}, {Diehl}, {Dietrich}, {Doel}, {Eifler}, {Evrard}, {Finley},
  {Flaugher}, {Fosalba}, {Frieman}, {Gaztanaga}, {Gerdes}, {Gruen}, {Gruendl},
  {Gutierrez}, {Honscheid}, {Kuehn}, {Kuropatkin}, {Lahav}, {Martini},
  {Miquel}, {Nord}, {Ogando}, {Plazas}, {Reil}, {Roodman}, {Sako}, {Sanchez},
  {Scarpine}, {Schubnell}, {Sevilla-Noarbe}, {Smith}, {Soares-Santos},
  {Sobreira}, {Suchyta}, {Swanson}, {Tarle}, {Tucker}, {Vikram}, {Wester},
  {Zhang}, {Zuntz}, \& {DES Collaboration}}]{dbr+15}
{Drlica-Wagner}, A., {Bechtol}, K., {Rykoff}, E.~S., {et~al.} 2015, \apj, 813,
  109, \dodoi{10.1088/0004-637X/813/2/109}

\bibitem[{{Drlica-Wagner} {et~al.}(2021){Drlica-Wagner}, {Carlin}, {Nidever},
  {Ferguson}, {Kuropatkin}, {Adam{\'o}w}, {Cerny}, {Choi}, {Esteves},
  {Mart{\'\i}nez-V{\'a}zquez}, {Mau}, {Miller}, {Mutlu-Pakdil}, {Neilsen},
  {Olsen}, {Pace}, {Riley}, {Sakowska}, {Sand}, {Santana-Silva}, {Tollerud},
  {Tucker}, {Vivas}, {Zaborowski}, {Zenteno}, {Abbott}, {Allam}, {Bechtol},
  {Bell}, {Bell}, {Bilaji}, {Bom}, {Carballo-Bello}, {Crnojevi{\'c}}, {Cioni},
  {Diaz-Ocampo}, {de Boer}, {Erkal}, {Gruendl}, {Hernandez-Lang}, {Hughes},
  {James}, {Johnson}, {Li}, {Mao}, {Mart{\'\i}nez-Delgado}, {Massana},
  {McNanna}, {Morgan}, {Nadler}, {No{\"e}l}, {Palmese}, {Peter}, {Rykoff},
  {S{\'a}nchez}, {Shipp}, {Simon}, {Smercina}, {Soares-Santos}, {Stringfellow},
  {Tavangar}, {van der Marel}, {Walker}, {Wechsler}, {Wu}, {Yanny},
  {Fitzpatrick}, {Huang}, {Jacques}, {Nikutta}, {Scott}, \& {Astro Data
  Lab}}]{drlica_2021}
{Drlica-Wagner}, A., {Carlin}, J.~L., {Nidever}, D.~L., {et~al.} 2021, \apjs,
  256, 2, \dodoi{10.3847/1538-4365/ac079d}

\bibitem[{{D'Souza} \& {Bell}(2022)}]{2022MNRAS.512..739D}
{D'Souza}, R., \& {Bell}, E.~F. 2022, \mnras, 512, 739,
  \dodoi{10.1093/mnras/stac404}

\bibitem[{{Erkal} {et~al.}(2019){Erkal}, {Belokurov}, {Laporte}, {Koposov},
  {Li}, {Grillmair}, {Kallivayalil}, {Price-Whelan}, {Evans}, {Hawkins},
  {Hendel}, {Mateu}, {Navarro}, {del Pino}, {Slater}, {Sohn}, \& {Orphan Aspen
  Treasury Collaboration}}]{2019MNRAS.487.2685E}
{Erkal}, D., {Belokurov}, V., {Laporte}, C.~F.~P., {et~al.} 2019, \mnras, 487,
  2685, \dodoi{10.1093/mnras/stz1371}

\bibitem[{{Fabrizio} {et~al.}(2012){Fabrizio}, {Merle}, {Th{\'e}venin},
  {Nonino}, {Bono}, {Stetson}, {Ferraro}, {Iannicola}, {Monelli}, {Walker},
  {Buonanno}, {Caputo}, {Corsi}, {Dall'Ora}, {Degl'Innocenti},
  {Fran{\c{c}}ois}, {Gilmozzi}, {Marconi}, {Pietrinferni}, {Prada Moroni},
  {Primas}, {Pulone}, {Ripepi}, \& {Romaniello}}]{fmt+12}
{Fabrizio}, M., {Merle}, T., {Th{\'e}venin}, F., {et~al.} 2012, \pasp, 124,
  519, \dodoi{10.1086/666480}

\bibitem[{{Foreman-Mackey} {et~al.}(2013){Foreman-Mackey}, {Hogg}, {Lang}, \&
  {Goodman}}]{2013PASP..125..306F}
{Foreman-Mackey}, D., {Hogg}, D.~W., {Lang}, D., \& {Goodman}, J. 2013, \pasp,
  125, 306, \dodoi{10.1086/670067}

\bibitem[{{Frebel} {et~al.}(2014){Frebel}, {Simon}, \&
  {Kirby}}]{2014ApJ...786...74F}
{Frebel}, A., {Simon}, J.~D., \& {Kirby}, E.~N. 2014, \apj, 786, 74,
  \dodoi{10.1088/0004-637X/786/1/74}

\bibitem[{{Gaia Collaboration} {et~al.}(2016){Gaia Collaboration}, {Prusti},
  {de Bruijne}, {Brown}, {Vallenari}, {Babusiaux}, {Bailer-Jones}, {Bastian},
  {Biermann}, {Evans}, {Eyer}, {Jansen}, {Jordi}, {Klioner}, {Lammers},
  {Lindegren}, {Luri}, {Mignard}, {Milligan}, {Panem}, {Poinsignon},
  {Pourbaix}, {Randich}, {Sarri}, {Sartoretti}, {Siddiqui}, {Soubiran},
  {Valette}, {van Leeuwen}, {Walton}, {Aerts}, {Arenou}, {Cropper}, {Drimmel},
  {H{\o}g}, {Katz}, {Lattanzi}, {O'Mullane}, {Grebel}, {Holland}, {Huc},
  {Passot}, {Bramante}, {Cacciari}, {Casta{\~n}eda}, {Chaoul}, {Cheek}, {De
  Angeli}, {Fabricius}, {Guerra}, {Hern{\'a}ndez}, {Jean-Antoine-Piccolo},
  {Masana}, {Messineo}, {Mowlavi}, {Nienartowicz}, {Ord{\'o}{\~n}ez-Blanco},
  {Panuzzo}, {Portell}, {Richards}, {Riello}, {Seabroke}, {Tanga},
  {Th{\'e}venin}, {Torra}, {Els}, {Gracia-Abril}, {Comoretto},
  {Garcia-Reinaldos}, {Lock}, {Mercier}, {Altmann}, {Andrae}, {Astraatmadja},
  {Bellas-Velidis}, {Benson}, {Berthier}, {Blomme}, {Busso}, {Carry},
  {Cellino}, {Clementini}, {Cowell}, {Creevey}, {Cuypers}, {Davidson}, {De
  Ridder}, {de Torres}, {Delchambre}, {Dell'Oro}, {Ducourant}, {Fr{\'e}mat},
  {Garc{\'\i}a-Torres}, {Gosset}, {Halbwachs}, {Hambly}, {Harrison}, {Hauser},
  {Hestroffer}, {Hodgkin}, {Huckle}, {Hutton}, {Jasniewicz}, {Jordan},
  {Kontizas}, {Korn}, {Lanzafame}, {Manteiga}, {Moitinho}, {Muinonen},
  {Osinde}, {Pancino}, {Pauwels}, {Petit}, {Recio-Blanco}, {Robin}, {Sarro},
  {Siopis}, {Smith}, {Smith}, {Sozzetti}, {Thuillot}, {van Reeven}, {Viala},
  {Abbas}, {Abreu Aramburu}, {Accart}, {Aguado}, {Allan}, {Allasia},
  {Altavilla}, {{\'A}lvarez}, {Alves}, {Anderson}, {Andrei}, {Anglada Varela},
  {Antiche}, {Antoja}, {Ant{\'o}n}, {Arcay}, {Atzei}, {Ayache}, {Bach},
  {Baker}, {Balaguer-N{\'u}{\~n}ez}, {Barache}, {Barata}, {Barbier}, {Barblan},
  {Baroni}, {Barrado y Navascu{\'e}s}, {Barros}, {Barstow}, {Becciani},
  {Bellazzini}, {Bellei}, {Bello Garc{\'\i}a}, {Belokurov}, {Bendjoya},
  {Berihuete}, {Bianchi}, {Bienaym{\'e}}, {Billebaud}, {Blagorodnova},
  {Blanco-Cuaresma}, {Boch}, {Bombrun}, {Borrachero}, {Bouquillon}, {Bourda},
  {Bouy}, {Bragaglia}, {Breddels}, {Brouillet}, {Br{\"u}semeister},
  {Bucciarelli}, {Budnik}, {Burgess}, {Burgon}, {Burlacu}, {Busonero}, {Buzzi},
  {Caffau}, {Cambras}, {Campbell}, {Cancelliere}, {Cantat-Gaudin}, {Carlucci},
  {Carrasco}, {Castellani}, {Charlot}, {Charnas}, {Charvet}, {Chassat},
  {Chiavassa}, {Clotet}, {Cocozza}, {Collins}, {Collins}, {Costigan}, {Crifo},
  {Cross}, {Crosta}, {Crowley}, {Dafonte}, {Damerdji}, {Dapergolas}, {David},
  {David}, {De Cat}, {de Felice}, {de Laverny}, {De Luise}, {De March}, {de
  Martino}, {de Souza}, {Debosscher}, {del Pozo}, {Delbo}, {Delgado},
  {Delgado}, {di Marco}, {Di Matteo}, {Diakite}, {Distefano}, {Dolding}, {Dos
  Anjos}, {Drazinos}, {Dur{\'a}n}, {Dzigan}, {Ecale}, {Edvardsson}, {Enke},
  {Erdmann}, {Escolar}, {Espina}, {Evans}, {Eynard Bontemps}, {Fabre},
  {Fabrizio}, {Faigler}, {Falc{\~a}o}, {Farr{\`a}s Casas}, {Faye}, {Federici},
  {Fedorets}, {Fern{\'a}ndez-Hern{\'a}ndez}, {Fernique}, {Fienga}, {Figueras},
  {Filippi}, {Findeisen}, {Fonti}, {Fouesneau}, {Fraile}, {Fraser}, {Fuchs},
  {Furnell}, {Gai}, {Galleti}, {Galluccio}, {Garabato}, {Garc{\'\i}a-Sedano},
  {Gar{\'e}}, {Garofalo}, {Garralda}, {Gavras}, {Gerssen}, {Geyer}, {Gilmore},
  {Girona}, {Giuffrida}, {Gomes}, {Gonz{\'a}lez-Marcos},
  {Gonz{\'a}lez-N{\'u}{\~n}ez}, {Gonz{\'a}lez-Vidal}, {Granvik}, {Guerrier},
  {Guillout}, {Guiraud}, {G{\'u}rpide}, {Guti{\'e}rrez-S{\'a}nchez}, {Guy},
  {Haigron}, {Hatzidimitriou}, {Haywood}, {Heiter}, {Helmi}, {Hobbs},
  {Hofmann}, {Holl}, {Holland}, {Hunt}, {Hypki}, {Icardi}, {Irwin}, {Jevardat
  de Fombelle}, {Jofr{\'e}}, {Jonker}, {Jorissen}, {Julbe}, {Karampelas},
  {Kochoska}, {Kohley}, {Kolenberg}, {Kontizas}, {Koposov}, {Kordopatis},
  {Koubsky}, {Kowalczyk}, {Krone-Martins}, {Kudryashova}, {Kull}, {Bachchan},
  {Lacoste-Seris}, {Lanza}, {Lavigne}, {Le Poncin-Lafitte}, {Lebreton},
  {Lebzelter}, {Leccia}, {Leclerc}, {Lecoeur-Taibi}, {Lemaitre}, {Lenhardt},
  {Leroux}, {Liao}, {Licata}, {Lindstr{\o}m}, {Lister}, {Livanou}, {Lobel},
  {L{\"o}ffler}, {L{\'o}pez}, {Lopez-Lozano}, {Lorenz}, {Loureiro},
  {MacDonald}, {Magalh{\~a}es Fernandes}, {Managau}, {Mann}, {Mantelet},
  {Marchal}, {Marchant}, {Marconi}, {Marie}, {Marinoni}, {Marrese},
  {Marschalk{\'o}}, {Marshall}, {Mart{\'\i}n-Fleitas}, {Martino}, {Mary},
  {Matijevi{\v{c}}}, {Mazeh}, {McMillan}, {Messina}, {Mestre}, {Michalik},
  {Millar}, {Miranda}, {Molina}, {Molinaro}, {Molinaro}, {Moln{\'a}r},
  {Moniez}, {Montegriffo}, {Monteiro}, {Mor}, {Mora}, {Morbidelli}, {Morel},
  {Morgenthaler}, {Morley}, {Morris}, {Mulone}, {Muraveva}, {Musella},
  {Narbonne}, {Nelemans}, {Nicastro}, {Noval}, {Ord{\'e}novic},
  {Ordieres-Mer{\'e}}, {Osborne}, {Pagani}, {Pagano}, {Pailler}, {Palacin},
  {Palaversa}, {Parsons}, {Paulsen}, {Pecoraro}, {Pedrosa}, {Pentik{\"a}inen},
  {Pereira}, {Pichon}, {Piersimoni}, {Pineau}, {Plachy}, {Plum}, {Poujoulet},
  {Pr{\v{s}}a}, {Pulone}, {Ragaini}, {Rago}, {Rambaux}, {Ramos-Lerate},
  {Ranalli}, {Rauw}, {Read}, {Regibo}, {Renk}, {Reyl{\'e}}, {Ribeiro},
  {Rimoldini}, {Ripepi}, {Riva}, {Rixon}, {Roelens}, {Romero-G{\'o}mez},
  {Rowell}, {Royer}, {Rudolph}, {Ruiz-Dern}, {Sadowski}, {Sagrist{\`a}
  Sell{\'e}s}, {Sahlmann}, {Salgado}, {Salguero}, {Sarasso}, {Savietto},
  {Schnorhk}, {Schultheis}, {Sciacca}, {Segol}, {Segovia}, {Segransan},
  {Serpell}, {Shih}, {Smareglia}, {Smart}, {Smith}, {Solano}, {Solitro},
  {Sordo}, {Soria Nieto}, {Souchay}, {Spagna}, {Spoto}, {Stampa}, {Steele},
  {Steidelm{\"u}ller}, {Stephenson}, {Stoev}, {Suess}, {S{\"u}veges}, {Surdej},
  {Szabados}, {Szegedi-Elek}, {Tapiador}, {Taris}, {Tauran}, {Taylor},
  {Teixeira}, {Terrett}, {Tingley}, {Trager}, {Turon}, {Ulla}, {Utrilla},
  {Valentini}, {van Elteren}, {Van Hemelryck}, {van Leeuwen}, {Varadi},
  {Vecchiato}, {Veljanoski}, {Via}, {Vicente}, {Vogt}, {Voss}, {Votruba},
  {Voutsinas}, {Walmsley}, {Weiler}, {Weingrill}, {Werner}, {Wevers},
  {Whitehead}, {Wyrzykowski}, {Yoldas}, {{\v{Z}}erjal}, {Zucker}, {Zurbach},
  {Zwitter}, {Alecu}, {Allen}, {Allende Prieto}, {Amorim},
  {Anglada-Escud{\'e}}, {Arsenijevic}, {Azaz}, {Balm}, {Beck}, {Bernstein},
  {Bigot}, {Bijaoui}, {Blasco}, {Bonfigli}, {Bono}, {Boudreault}, {Bressan},
  {Brown}, {Brunet}, {Bunclark}, {Buonanno}, {Butkevich}, {Carret}, {Carrion},
  {Chemin}, {Ch{\'e}reau}, {Corcione}, {Darmigny}, {de Boer}, {de Teodoro}, {de
  Zeeuw}, {Delle Luche}, {Domingues}, {Dubath}, {Fodor}, {Fr{\'e}zouls},
  {Fries}, {Fustes}, {Fyfe}, {Gallardo}, {Gallegos}, {Gardiol}, {Gebran},
  {Gomboc}, {G{\'o}mez}, {Grux}, {Gueguen}, {Heyrovsky}, {Hoar}, {Iannicola},
  {Isasi Parache}, {Janotto}, {Joliet}, {Jonckheere}, {Keil}, {Kim},
  {Klagyivik}, {Klar}, {Knude}, {Kochukhov}, {Kolka}, {Kos}, {Kutka}, {Lainey},
  {LeBouquin}, {Liu}, {Loreggia}, {Makarov}, {Marseille}, {Martayan},
  {Martinez-Rubi}, {Massart}, {Meynadier}, {Mignot}, {Munari}, {Nguyen},
  {Nordlander}, {Ocvirk}, {O'Flaherty}, {Olias Sanz}, {Ortiz}, {Osorio},
  {Oszkiewicz}, {Ouzounis}, {Palmer}, {Park}, {Pasquato}, {Peltzer}, {Peralta},
  {P{\'e}turaud}, {Pieniluoma}, {Pigozzi}, {Poels}, {Prat}, {Prod'homme},
  {Raison}, {Rebordao}, {Risquez}, {Rocca-Volmerange}, {Rosen}, {Ruiz-Fuertes},
  {Russo}, {Sembay}, {Serraller Vizcaino}, {Short}, {Siebert}, {Silva},
  {Sinachopoulos}, {Slezak}, {Soffel}, {Sosnowska}, {Strai{\v{z}}ys}, {ter
  Linden}, {Terrell}, {Theil}, {Tiede}, {Troisi}, {Tsalmantza}, {Tur},
  {Vaccari}, {Vachier}, {Valles}, {Van Hamme}, {Veltz}, {Virtanen}, {Wallut},
  {Wichmann}, {Wilkinson}, {Ziaeepour}, \& {Zschocke}}]{2016A&A...595A...1G}
{Gaia Collaboration}, {Prusti}, T., {de Bruijne}, J.~H.~J., {et~al.} 2016,
  \aap, 595, A1, \dodoi{10.1051/0004-6361/201629272}

\bibitem[{{Gaia Collaboration} {et~al.}(2021){Gaia Collaboration}, {Brown},
  {Vallenari}, {Prusti}, {de Bruijne}, {Babusiaux}, {Biermann}, {Creevey},
  {Evans}, {Eyer}, {Hutton}, {Jansen}, {Jordi}, {Klioner}, {Lammers},
  {Lindegren}, {Luri}, {Mignard}, {Panem}, {Pourbaix}, {Randich}, {Sartoretti},
  {Soubiran}, {Walton}, {Arenou}, {Bailer-Jones}, {Bastian}, {Cropper},
  {Drimmel}, {Katz}, {Lattanzi}, {van Leeuwen}, {Bakker}, {Cacciari},
  {Casta{\~n}eda}, {De Angeli}, {Ducourant}, {Fabricius}, {Fouesneau},
  {Fr{\'e}mat}, {Guerra}, {Guerrier}, {Guiraud}, {Jean-Antoine Piccolo},
  {Masana}, {Messineo}, {Mowlavi}, {Nicolas}, {Nienartowicz}, {Pailler},
  {Panuzzo}, {Riclet}, {Roux}, {Seabroke}, {Sordo}, {Tanga}, {Th{\'e}venin},
  {Gracia-Abril}, {Portell}, {Teyssier}, {Altmann}, {Andrae}, {Bellas-Velidis},
  {Benson}, {Berthier}, {Blomme}, {Brugaletta}, {Burgess}, {Busso}, {Carry},
  {Cellino}, {Cheek}, {Clementini}, {Damerdji}, {Davidson}, {Delchambre},
  {Dell'Oro}, {Fern{\'a}ndez-Hern{\'a}ndez}, {Galluccio}, {Garc{\'\i}a-Lario},
  {Garcia-Reinaldos}, {Gonz{\'a}lez-N{\'u}{\~n}ez}, {Gosset}, {Haigron},
  {Halbwachs}, {Hambly}, {Harrison}, {Hatzidimitriou}, {Heiter},
  {Hern{\'a}ndez}, {Hestroffer}, {Hodgkin}, {Holl}, {Jan{\ss}en}, {Jevardat de
  Fombelle}, {Jordan}, {Krone-Martins}, {Lanzafame}, {L{\"o}ffler}, {Lorca},
  {Manteiga}, {Marchal}, {Marrese}, {Moitinho}, {Mora}, {Muinonen}, {Osborne},
  {Pancino}, {Pauwels}, {Petit}, {Recio-Blanco}, {Richards}, {Riello},
  {Rimoldini}, {Robin}, {Roegiers}, {Rybizki}, {Sarro}, {Siopis}, {Smith},
  {Sozzetti}, {Ulla}, {Utrilla}, {van Leeuwen}, {van Reeven}, {Abbas}, {Abreu
  Aramburu}, {Accart}, {Aerts}, {Aguado}, {Ajaj}, {Altavilla}, {{\'A}lvarez},
  {{\'A}lvarez Cid-Fuentes}, {Alves}, {Anderson}, {Anglada Varela}, {Antoja},
  {Audard}, {Baines}, {Baker}, {Balaguer-N{\'u}{\~n}ez}, {Balbinot}, {Balog},
  {Barache}, {Barbato}, {Barros}, {Barstow}, {Bartolom{\'e}}, {Bassilana},
  {Bauchet}, {Baudesson-Stella}, {Becciani}, {Bellazzini}, {Bernet}, {Bertone},
  {Bianchi}, {Blanco-Cuaresma}, {Boch}, {Bombrun}, {Bossini}, {Bouquillon},
  {Bragaglia}, {Bramante}, {Breedt}, {Bressan}, {Brouillet}, {Bucciarelli},
  {Burlacu}, {Busonero}, {Butkevich}, {Buzzi}, {Caffau}, {Cancelliere},
  {C{\'a}novas}, {Cantat-Gaudin}, {Carballo}, {Carlucci}, {Carnerero},
  {Carrasco}, {Casamiquela}, {Castellani}, {Castro-Ginard}, {Castro Sampol},
  {Chaoul}, {Charlot}, {Chemin}, {Chiavassa}, {Cioni}, {Comoretto}, {Cooper},
  {Cornez}, {Cowell}, {Crifo}, {Crosta}, {Crowley}, {Dafonte}, {Dapergolas},
  {David}, {David}, {de Laverny}, {De Luise}, {De March}, {De Ridder}, {de
  Souza}, {de Teodoro}, {de Torres}, {del Peloso}, {del Pozo}, {Delbo},
  {Delgado}, {Delgado}, {Delisle}, {Di Matteo}, {Diakite}, {Diener},
  {Distefano}, {Dolding}, {Eappachen}, {Edvardsson}, {Enke}, {Esquej}, {Fabre},
  {Fabrizio}, {Faigler}, {Fedorets}, {Fernique}, {Fienga}, {Figueras},
  {Fouron}, {Fragkoudi}, {Fraile}, {Franke}, {Gai}, {Garabato},
  {Garcia-Gutierrez}, {Garc{\'\i}a-Torres}, {Garofalo}, {Gavras}, {Gerlach},
  {Geyer}, {Giacobbe}, {Gilmore}, {Girona}, {Giuffrida}, {Gomel}, {Gomez},
  {Gonzalez-Santamaria}, {Gonz{\'a}lez-Vidal}, {Granvik},
  {Guti{\'e}rrez-S{\'a}nchez}, {Guy}, {Hauser}, {Haywood}, {Helmi}, {Hidalgo},
  {Hilger}, {H{\l}adczuk}, {Hobbs}, {Holland}, {Huckle}, {Jasniewicz},
  {Jonker}, {Juaristi Campillo}, {Julbe}, {Karbevska}, {Kervella}, {Khanna},
  {Kochoska}, {Kontizas}, {Kordopatis}, {Korn}, {Kostrzewa-Rutkowska},
  {Kruszy{\'n}ska}, {Lambert}, {Lanza}, {Lasne}, {Le Campion}, {Le Fustec},
  {Lebreton}, {Lebzelter}, {Leccia}, {Leclerc}, {Lecoeur-Taibi}, {Liao},
  {Licata}, {Lindstr{\o}m}, {Lister}, {Livanou}, {Lobel}, {Madrero Pardo},
  {Managau}, {Mann}, {Marchant}, {Marconi}, {Marcos Santos}, {Marinoni},
  {Marocco}, {Marshall}, {Martin Polo}, {Mart{\'\i}n-Fleitas}, {Masip},
  {Massari}, {Mastrobuono-Battisti}, {Mazeh}, {McMillan}, {Messina},
  {Michalik}, {Millar}, {Mints}, {Molina}, {Molinaro}, {Moln{\'a}r},
  {Montegriffo}, {Mor}, {Morbidelli}, {Morel}, {Morris}, {Mulone}, {Munoz},
  {Muraveva}, {Murphy}, {Musella}, {Noval}, {Ord{\'e}novic}, {Orr{\`u}},
  {Osinde}, {Pagani}, {Pagano}, {Palaversa}, {Palicio}, {Panahi}, {Pawlak},
  {Pe{\~n}alosa Esteller}, {Penttil{\"a}}, {Piersimoni}, {Pineau}, {Plachy},
  {Plum}, {Poggio}, {Poretti}, {Poujoulet}, {Pr{\v{s}}a}, {Pulone}, {Racero},
  {Ragaini}, {Rainer}, {Raiteri}, {Rambaux}, {Ramos}, {Ramos-Lerate}, {Re
  Fiorentin}, {Regibo}, {Reyl{\'e}}, {Ripepi}, {Riva}, {Rixon}, {Robichon},
  {Robin}, {Roelens}, {Rohrbasser}, {Romero-G{\'o}mez}, {Rowell}, {Royer},
  {Rybicki}, {Sadowski}, {Sagrist{\`a} Sell{\'e}s}, {Sahlmann}, {Salgado},
  {Salguero}, {Samaras}, {Sanchez Gimenez}, {Sanna}, {Santove{\~n}a},
  {Sarasso}, {Schultheis}, {Sciacca}, {Segol}, {Segovia}, {S{\'e}gransan},
  {Semeux}, {Shahaf}, {Siddiqui}, {Siebert}, {Siltala}, {Slezak}, {Smart},
  {Solano}, {Solitro}, {Souami}, {Souchay}, {Spagna}, {Spoto}, {Steele},
  {Steidelm{\"u}ller}, {Stephenson}, {S{\"u}veges}, {Szabados}, {Szegedi-Elek},
  {Taris}, {Tauran}, {Taylor}, {Teixeira}, {Thuillot}, {Tonello}, {Torra},
  {Torra}, {Turon}, {Unger}, {Vaillant}, {van Dillen}, {Vanel}, {Vecchiato},
  {Viala}, {Vicente}, {Voutsinas}, {Weiler}, {Wevers}, {Wyrzykowski}, {Yoldas},
  {Yvard}, {Zhao}, {Zorec}, {Zucker}, {Zurbach}, \&
  {Zwitter}}]{2021A&A...649A...1G}
{Gaia Collaboration}, {Brown}, A.~G.~A., {Vallenari}, A., {et~al.} 2021, \aap,
  649, A1, \dodoi{10.1051/0004-6361/202039657}

\bibitem[{{Gaia Collaboration} {et~al.}(2022){Gaia Collaboration}, {Vallenari},
  {Brown}, {Prusti}, {de Bruijne}, {Arenou}, {Babusiaux}, {Biermann},
  {Creevey}, {Ducourant}, {Evans}, {Eyer}, {Guerra}, {Hutton}, {Jordi},
  {Klioner}, {Lammers}, {Lindegren}, {Luri}, {Mignard}, {Panem}, {Pourbaix},
  {Randich}, {Sartoretti}, {Soubiran}, {Tanga}, {Walton}, {Bailer-Jones},
  {Bastian}, {Drimmel}, {Jansen}, {Katz}, {Lattanzi}, {van Leeuwen}, {Bakker},
  {Cacciari}, {Casta{\~n}eda}, {De Angeli}, {Fabricius}, {Fouesneau},
  {Fr{\'e}mat}, {Galluccio}, {Guerrier}, {Heiter}, {Masana}, {Messineo},
  {Mowlavi}, {Nicolas}, {Nienartowicz}, {Pailler}, {Panuzzo}, {Riclet}, {Roux},
  {Seabroke}, {Sordo{\o}rcit}, {Th{\'e}venin}, {Gracia-Abril}, {Portell},
  {Teyssier}, {Altmann}, {Andrae}, {Audard}, {Bellas-Velidis}, {Benson},
  {Berthier}, {Blomme}, {Burgess}, {Busonero}, {Busso}, {C{\'a}novas}, {Carry},
  {Cellino}, {Cheek}, {Clementini}, {Damerdji}, {Davidson}, {de Teodoro},
  {Nu{\~n}ez Campos}, {Delchambre}, {Dell'Oro}, {Esquej},
  {Fern{\'a}ndez-Hern{\'a}ndez}, {Fraile}, {Garabato}, {Garc{\'\i}a-Lario},
  {Gosset}, {Haigron}, {Halbwachs}, {Hambly}, {Harrison}, {Hern{\'a}ndez},
  {Hestroffer}, {Hodgkin}, {Holl}, {Jan{\ss}en}, {Jevardat de Fombelle},
  {Jordan}, {Krone-Martins}, {Lanzafame}, {L{\"o}ffler}, {Marchal}, {Marrese},
  {Moitinho}, {Muinonen}, {Osborne}, {Pancino}, {Pauwels}, {Recio-Blanco},
  {Reyl{\'e}}, {Riello}, {Rimoldini}, {Roegiers}, {Rybizki}, {Sarro}, {Siopis},
  {Smith}, {Sozzetti}, {Utrilla}, {van Leeuwen}, {Abbas}, {{\'A}brah{\'a}m},
  {Abreu Aramburu}, {Aerts}, {Aguado}, {Ajaj}, {Aldea-Montero}, {Altavilla},
  {{\'A}lvarez}, {Alves}, {Anders}, {Anderson}, {Anglada Varela}, {Antoja},
  {Baines}, {Baker}, {Balaguer-N{\'u}{\~n}ez}, {Balbinot}, {Balog}, {Barache},
  {Barbato}, {Barros}, {Barstow}, {Bartolom{\'e}}, {Bassilana}, {Bauchet},
  {Becciani}, {Bellazzini}, {Berihuete}, {Bernet}, {Bertone}, {Bianchi},
  {Binnenfeld}, {Blanco-Cuaresma}, {Blazere}, {Boch}, {Bombrun}, {Bossini},
  {Bouquillon}, {Bragaglia}, {Bramante}, {Breedt}, {Bressan}, {Brouillet},
  {Brugaletta}, {Bucciarelli}, {Burlacu}, {Butkevich}, {Buzzi}, {Caffau},
  {Cancelliere}, {Cantat-Gaudin}, {Carballo}, {Carlucci}, {Carnerero},
  {Carrasco}, {Casamiquela}, {Castellani}, {Castro-Ginard}, {Chaoul},
  {Charlot}, {Chemin}, {Chiaramida}, {Chiavassa}, {Chornay}, {Comoretto},
  {Contursi}, {Cooper}, {Cornez}, {Cowell}, {Crifo}, {Cropper}, {Crosta},
  {Crowley}, {Dafonte}, {Dapergolas}, {David}, {David}, {de Laverny}, {De
  Luise}, {De March}, {De Ridder}, {de Souza}, {de Torres}, {del Peloso}, {del
  Pozo}, {Delbo}, {Delgado}, {Delisle}, {Demouchy}, {Dharmawardena}, {Di
  Matteo}, {Diakite}, {Diener}, {Distefano}, {Dolding}, {Edvardsson}, {Enke},
  {Fabre}, {Fabrizio}, {Faigler}, {Fedorets}, {Fernique}, {Fienga}, {Figueras},
  {Fournier}, {Fouron}, {Fragkoudi}, {Gai}, {Garcia-Gutierrez},
  {Garcia-Reinaldos}, {Garc{\'\i}a-Torres}, {Garofalo}, {Gavel}, {Gavras},
  {Gerlach}, {Geyer}, {Giacobbe}, {Gilmore}, {Girona}, {Giuffrida}, {Gomel},
  {Gomez}, {Gonz{\'a}lez-N{\'u}{\~n}ez}, {Gonz{\'a}lez-Santamar{\'\i}a},
  {Gonz{\'a}lez-Vidal}, {Granvik}, {Guillout}, {Guiraud},
  {Guti{\'e}rrez-S{\'a}nchez}, {Guy}, {Hatzidimitriou}, {Hauser}, {Haywood},
  {Helmer}, {Helmi}, {Sarmiento}, {Hidalgo}, {Hilger}, {H{\l}adczuk}, {Hobbs},
  {Holland}, {Huckle}, {Jardine}, {Jasniewicz}, {Jean-Antoine Piccolo},
  {Jim{\'e}nez-Arranz}, {Jorissen}, {Juaristi Campillo}, {Julbe}, {Karbevska},
  {Kervella}, {Khanna}, {Kontizas}, {Kordopatis}, {Korn}, {K{\'o}sp{\'a}l},
  {Kostrzewa-Rutkowska}, {Kruszy{\'n}ska}, {Kun}, {Laizeau}, {Lambert},
  {Lanza}, {Lasne}, {Le Campion}, {Lebreton}, {Lebzelter}, {Leccia}, {Leclerc},
  {Lecoeur-Taibi}, {Liao}, {Licata}, {Lindstr{\o}m}, {Lister}, {Livanou},
  {Lobel}, {Lorca}, {Loup}, {Madrero Pardo}, {Magdaleno Romeo}, {Managau},
  {Mann}, {Manteiga}, {Marchant}, {Marconi}, {Marcos}, {Marcos Santos},
  {Mar{\'\i}n Pina}, {Marinoni}, {Marocco}, {Marshall}, {Polo},
  {Mart{\'\i}n-Fleitas}, {Marton}, {Mary}, {Masip}, {Massari},
  {Mastrobuono-Battisti}, {Mazeh}, {McMillan}, {Messina}, {Michalik}, {Millar},
  {Mints}, {Molina}, {Molinaro}, {Moln{\'a}r}, {Monari}, {Mongui{\'o}},
  {Montegriffo}, {Montero}, {Mor}, {Mora}, {Morbidelli}, {Morel}, {Morris},
  {Muraveva}, {Murphy}, {Musella}, {Nagy}, {Noval}, {Oca{\~n}a}, {Ogden},
  {Ordenovic}, {Osinde}, {Pagani}, {Pagano}, {Palaversa}, {Palicio},
  {Pallas-Quintela}, {Panahi}, {Payne-Wardenaar}, {Pe{\~n}alosa Esteller},
  {Penttil{\"a}}, {Pichon}, {Piersimoni}, {Pineau}, {Plachy}, {Plum}, {Poggio},
  {Pr{\v{s}}a}, {Pulone}, {Racero}, {Ragaini}, {Rainer}, {Raiteri}, {Rambaux},
  {Ramos}, {Ramos-Lerate}, {Re Fiorentin}, {Regibo}, {Richards}, {Rios Diaz},
  {Ripepi}, {Riva}, {Rix}, {Rixon}, {Robichon}, {Robin}, {Robin}, {Roelens},
  {Rogues}, {Rohrbasser}, {Romero-G{\'o}mez}, {Rowell}, {Royer}, {Ruz Mieres},
  {Rybicki}, {Sadowski}, {S{\'a}ez N{\'u}{\~n}ez}, {Sagrist{\`a} Sell{\'e}s},
  {Sahlmann}, {Salguero}, {Samaras}, {Sanchez Gimenez}, {Sanna},
  {Santove{\~n}a}, {Sarasso}, {Schultheis}, {Sciacca}, {Segol}, {Segovia},
  {S{\'e}gransan}, {Semeux}, {Shahaf}, {Siddiqui}, {Siebert}, {Siltala},
  {Silvelo}, {Slezak}, {Slezak}, {Smart}, {Snaith}, {Solano}, {Solitro},
  {Souami}, {Souchay}, {Spagna}, {Spina}, {Spoto}, {Steele},
  {Steidelm{\"u}ller}, {Stephenson}, {S{\"u}veges}, {Surdej}, {Szabados},
  {Szegedi-Elek}, {Taris}, {Taylo}, {Teixeira}, {Tolomei}, {Tonello}, {Torra},
  {Torra}, {Torralba Elipe}, {Trabucchi}, {Tsounis}, {Turon}, {Ulla}, {Unger},
  {Vaillant}, {van Dillen}, {van Reeven}, {Vanel}, {Vecchiato}, {Viala},
  {Vicente}, {Voutsinas}, {Weiler}, {Wevers}, {Wyrzykowski}, {Yoldas}, {Yvard},
  {Zhao}, {Zorec}, {Zucker}, \& {Zwitter}}]{2022arXiv220800211G}
{Gaia Collaboration}, {Vallenari}, A., {Brown}, A.~G.~A., {et~al.} 2022, arXiv
  e-prints, arXiv:2208.00211.
\newblock \doarXiv{2208.00211}

\bibitem[{{Gallart} {et~al.}(2021){Gallart}, {Monelli}, {Ruiz-Lara},
  {Calamida}, {Cassisi}, {Cignoni}, {Anderson}, {Battaglia}, {Bermejo-Climent},
  {Bernard}, {Mart{\'\i}nez-V{\'a}zquez}, {Mayer}, {Salvadori}, {Monachesi},
  {Navarro}, {Shen}, {Surot}, {Tosi}, {Bajaj}, \& {Strinfellow}}]{Gallart_2021}
{Gallart}, C., {Monelli}, M., {Ruiz-Lara}, T., {et~al.} 2021, \apj, 909, 192,
  \dodoi{10.3847/1538-4357/abddbe}

\bibitem[{{Geringer-Sameth} {et~al.}(2015){Geringer-Sameth}, {Koushiappas}, \&
  {Walker}}]{GeringerSameth2015ApJ...801...74G}
{Geringer-Sameth}, A., {Koushiappas}, S.~M., \& {Walker}, M. 2015, \apj, 801,
  74, \dodoi{10.1088/0004-637X/801/2/74}

\bibitem[{{Harris}(2010)}]{harris_2010}
{Harris}, W.~E. 2010, arXiv e-prints, arXiv:1012.3224,
  \dodoi{10.48550/arXiv.1012.3224}

\bibitem[{{Hayashi} {et~al.}(2020){Hayashi}, {Chiba}, \&
  {Ishiyama}}]{2020ApJ...904...45H}
{Hayashi}, K., {Chiba}, M., \& {Ishiyama}, T. 2020, \apj, 904, 45,
  \dodoi{10.3847/1538-4357/abbe0a}

\bibitem[{{Jenkins} {et~al.}(2021){Jenkins}, {Li}, {Pace}, {Ji}, {Koposov}, \&
  {Mutlu-Pakdil}}]{jlp+21}
{Jenkins}, S.~A., {Li}, T.~S., {Pace}, A.~B., {et~al.} 2021, \apj, 920, 92,
  \dodoi{10.3847/1538-4357/ac1353}

\bibitem[{{Ji} {et~al.}(2016{\natexlab{a}}){Ji}, {Frebel}, {Simon}, \&
  {Chiti}}]{jfs+16_ret}
{Ji}, A.~P., {Frebel}, A., {Simon}, J.~D., \& {Chiti}, A. 2016{\natexlab{a}},
  \apj, 830, 93, \dodoi{10.3847/0004-637X/830/2/93}

\bibitem[{{Ji} {et~al.}(2016{\natexlab{b}}){Ji}, {Frebel}, {Simon}, \&
  {Geha}}]{2016ApJ...817...41J}
{Ji}, A.~P., {Frebel}, A., {Simon}, J.~D., \& {Geha}, M. 2016{\natexlab{b}},
  \apj, 817, 41, \dodoi{10.3847/0004-637X/817/1/41}

\bibitem[{{Kim} {et~al.}(2015){Kim}, {Jerjen}, {Mackey}, {Da Costa}, \&
  {Milone}}]{kjm+15}
{Kim}, D., {Jerjen}, H., {Mackey}, D., {Da Costa}, G.~S., \& {Milone}, A.~P.
  2015, \apjl, 804, L44, \dodoi{10.1088/2041-8205/804/2/L44}

\bibitem[{{Kim} {et~al.}(2016){Kim}, {Jerjen}, {Geha}, {Chiti}, {Milone}, {Da
  Costa}, {Mackey}, {Frebel}, \& {Conn}}]{kjg+16}
{Kim}, D., {Jerjen}, H., {Geha}, M., {et~al.} 2016, \apj, 833, 16,
  \dodoi{10.3847/0004-637X/833/1/16}

\bibitem[{{Kirby} {et~al.}(2013{\natexlab{a}}){Kirby}, {Boylan-Kolchin},
  {Cohen}, {Geha}, {Bullock}, \& {Kaplinghat}}]{kbc+13}
{Kirby}, E.~N., {Boylan-Kolchin}, M., {Cohen}, J.~G., {et~al.}
  2013{\natexlab{a}}, \apj, 770, 16, \dodoi{10.1088/0004-637X/770/1/16}

\bibitem[{{Kirby} {et~al.}(2013{\natexlab{b}}){Kirby}, {Cohen}, {Guhathakurta},
  {Cheng}, {Bullock}, \& {Gallazzi}}]{kcg+13}
{Kirby}, E.~N., {Cohen}, J.~G., {Guhathakurta}, P., {et~al.}
  2013{\natexlab{b}}, \apj, 779, 102, \dodoi{10.1088/0004-637X/779/2/102}

\bibitem[{{Kirby} {et~al.}(2015{\natexlab{a}}){Kirby}, {Cohen}, {Simon}, \&
  {Guhathakurta}}]{Kirby2015ApJ...814L...7K}
{Kirby}, E.~N., {Cohen}, J.~G., {Simon}, J.~D., \& {Guhathakurta}, P.
  2015{\natexlab{a}}, \apjl, 814, L7, \dodoi{10.1088/2041-8205/814/1/L7}

\bibitem[{{Kirby} {et~al.}(2017){Kirby}, {Cohen}, {Simon}, {Guhathakurta},
  {Thygesen}, \& {Duggan}}]{Kirby2017ApJ...838...83K}
{Kirby}, E.~N., {Cohen}, J.~G., {Simon}, J.~D., {et~al.} 2017, \apj, 838, 83,
  \dodoi{10.3847/1538-4357/aa6570}

\bibitem[{{Kirby} {et~al.}(2015{\natexlab{b}}){Kirby}, {Simon}, \&
  {Cohen}}]{ksc+15}
{Kirby}, E.~N., {Simon}, J.~D., \& {Cohen}, J.~G. 2015{\natexlab{b}}, \apj,
  810, 56, \dodoi{10.1088/0004-637X/810/1/56}

\bibitem[{{Koch} \& {Rich}(2014)}]{koch_2014}
{Koch}, A., \& {Rich}, R.~M. 2014, \apj, 794, 89,
  \dodoi{10.1088/0004-637X/794/1/89}

\bibitem[{{Koch} {et~al.}(2009){Koch}, {Wilkinson}, {Kleyna}, {Irwin},
  {Zucker}, {Belokurov}, {Gilmore}, {Fellhauer}, \& {Evans}}]{Koch_2008}
{Koch}, A., {Wilkinson}, M.~I., {Kleyna}, J.~T., {et~al.} 2009, \apj, 690, 453,
  \dodoi{10.1088/0004-637X/690/1/453}

\bibitem[{{Koposov} {et~al.}(2015){Koposov}, {Casey}, {Belokurov}, {Lewis},
  {Gilmore}, {Worley}, {Hourihane}, {Randich}, {Bensby}, {Bragaglia},
  {Bergemann}, {Carraro}, {Costado}, {Flaccomio}, {Francois}, {Heiter}, {Hill},
  {Jofre}, {Lando}, {Lanzafame}, {de Laverny}, {Monaco}, {Morbidelli},
  {Sbordone}, {Mikolaitis}, \& {Ryde}}]{kcb+15}
{Koposov}, S.~E., {Casey}, A.~R., {Belokurov}, V., {et~al.} 2015, \apj, 811,
  62, \dodoi{10.1088/0004-637X/811/1/62}

\bibitem[{{Koposov} {et~al.}(2018){Koposov}, {Walker}, {Belokurov}, {Casey},
  {Geringer-Sameth}, {Mackey}, {Da Costa}, {Erkal}, {Jethwa}, {Mateo},
  {Olszewski}, \& {Bailey}}]{kwb+18}
{Koposov}, S.~E., {Walker}, M.~G., {Belokurov}, V., {et~al.} 2018, \mnras, 479,
  5343, \dodoi{10.1093/mnras/sty1772}

\bibitem[{{Li} {et~al.}(2021){Li}, {Hammer}, {Babusiaux}, {Pawlowski}, {Yang},
  {Arenou}, {Du}, \& {Wang}}]{2021ApJ...916....8L}
{Li}, H., {Hammer}, F., {Babusiaux}, C., {et~al.} 2021, \apj, 916, 8,
  \dodoi{10.3847/1538-4357/ac0436}

\bibitem[{Li {et~al.}(2017)Li, Simon, Drlica-Wagner, Bechtol, Wang,
  Garc{\'{\i}}a-Bellido, Frieman, Marshall, James, Strigari, Pace, Balbinot,
  Zhang, Abbott, Allam, Benoit-L{\'{e}}vy, Bernstein, Bertin, Brooks, Burke,
  Rosell, Kind, Carretero, Cunha, D'Andrea, da~Costa, DePoy, Desai, Diehl,
  Eifler, Flaugher, Goldstein, Gruen, Gruendl, Gschwend, Gutierrez, Krause,
  Kuehn, Lin, Maia, March, Menanteau, Miquel, Plazas, Romer, Sanchez, Santiago,
  Schubnell, Sevilla-Noarbe, Smith, Sobreira, Suchyta, Tarle, Thomas, Tucker,
  Walker, Wechsler, Wester, \& and}]{Li_2017}
Li, T.~S., Simon, J.~D., Drlica-Wagner, A., {et~al.} 2017, The Astrophysical
  Journal, 838, 8, \dodoi{10.3847/1538-4357/aa6113}

\bibitem[{{Li} {et~al.}(2018{\natexlab{a}}){Li}, {Simon}, {Kuehn}, {Pace},
  {Erkal}, {Bechtol}, {Yanny}, {Drlica-Wagner}, {Marshall}, {Lidman},
  {Balbinot}, {Carollo}, {Jenkins}, {Mart{\'\i}nez-V{\'a}zquez}, {Shipp},
  {Stringer}, {Vivas}, {Walker}, {Wechsler}, {Abdalla}, {Allam}, {Annis},
  {Avila}, {Bertin}, {Brooks}, {Buckley-Geer}, {Burke}, {Carnero Rosell},
  {Carrasco Kind}, {Carretero}, {Cunha}, {D'Andrea}, {da Costa}, {Davis}, {De
  Vicente}, {Doel}, {Eifler}, {Evrard}, {Flaugher}, {Frieman},
  {Garc{\'\i}a-Bellido}, {Gaztanaga}, {Gerdes}, {Gruen}, {Gruendl}, {Gschwend},
  {Gutierrez}, {Hartley}, {Hollowood}, {Honscheid}, {James}, {Krause}, {Maia},
  {March}, {Menanteau}, {Miquel}, {Plazas}, {Sanchez}, {Santiago}, {Scarpine},
  {Schindler}, {Schubnell}, {Sevilla-Noarbe}, {Smith}, {Smith},
  {Soares-Santos}, {Sobreira}, {Suchyta}, {Swanson}, {Tarle}, {Tucker}, \& {DES
  Collaboration}}]{2018ApJ...866...22L}
{Li}, T.~S., {Simon}, J.~D., {Kuehn}, K., {et~al.} 2018{\natexlab{a}}, \apj,
  866, 22, \dodoi{10.3847/1538-4357/aadf91}

\bibitem[{{Li} {et~al.}(2018{\natexlab{b}}){Li}, {Simon}, {Pace}, {Torrealba},
  {Kuehn}, {Drlica-Wagner}, {Bechtol}, {Vivas}, {van der Marel}, {Wood},
  {Yanny}, {Belokurov}, {Jethwa}, {Zucker}, {Lewis}, {Kron}, {Nidever},
  {S{\'a}nchez-Conde}, {Ji}, {Conn}, {James}, {Martin}, {Martinez-Delgado},
  {No{\"e}l}, \& {MagLiteS Collaboration}}]{Li2018ApJ...857..145L}
{Li}, T.~S., {Simon}, J.~D., {Pace}, A.~B., {et~al.} 2018{\natexlab{b}}, \apj,
  857, 145, \dodoi{10.3847/1538-4357/aab666}

\bibitem[{{Lindegren} {et~al.}(2021){Lindegren}, {Klioner}, {Hern{\'a}ndez},
  {Bombrun}, {Ramos-Lerate}, {Steidelm{\"u}ller}, {Bastian}, {Biermann}, {de
  Torres}, {Gerlach}, {Geyer}, {Hilger}, {Hobbs}, {Lammers}, {McMillan},
  {Stephenson}, {Casta{\~n}eda}, {Davidson}, {Fabricius}, {Gracia-Abril},
  {Portell}, {Rowell}, {Teyssier}, {Torra}, {Bartolom{\'e}}, {Clotet},
  {Garralda}, {Gonz{\'a}lez-Vidal}, {Torra}, {Abbas}, {Altmann}, {Anglada
  Varela}, {Balaguer-N{\'u}{\~n}ez}, {Balog}, {Barache}, {Becciani}, {Bernet},
  {Bertone}, {Bianchi}, {Bouquillon}, {Brown}, {Bucciarelli}, {Busonero},
  {Butkevich}, {Buzzi}, {Cancelliere}, {Carlucci}, {Charlot}, {Cioni},
  {Crosta}, {Crowley}, {del Peloso}, {del Pozo}, {Drimmel}, {Esquej}, {Fienga},
  {Fraile}, {Gai}, {Garcia-Reinaldos}, {Guerra}, {Hambly}, {Hauser},
  {Jan{\ss}en}, {Jordan}, {Kostrzewa-Rutkowska}, {Lattanzi}, {Liao}, {Licata},
  {Lister}, {L{\"o}ffler}, {Marchant}, {Masip}, {Mignard}, {Mints}, {Molina},
  {Mora}, {Morbidelli}, {Murphy}, {Pagani}, {Panuzzo}, {Pe{\~n}alosa Esteller},
  {Poggio}, {Re Fiorentin}, {Riva}, {Sagrist{\`a} Sell{\'e}s}, {Sanchez
  Gimenez}, {Sarasso}, {Sciacca}, {Siddiqui}, {Smart}, {Souami}, {Spagna},
  {Steele}, {Taris}, {Utrilla}, {van Reeven}, \&
  {Vecchiato}}]{2021A&A...649A...2L}
{Lindegren}, L., {Klioner}, S.~A., {Hern{\'a}ndez}, J., {et~al.} 2021, \aap,
  649, A2, \dodoi{10.1051/0004-6361/202039709}

\bibitem[{{Longeard} {et~al.}(2018){Longeard}, {Martin}, {Starkenburg},
  {Ibata}, {Collins}, {Geha}, {Laevens}, {Rich}, {Aguado}, {Arentsen},
  {Carlberg}, {C{\^o}t{\'e}}, {Hill}, {Jablonka}, {Gonz{\'a}lez Hern{\'a}ndez},
  {Navarro}, {S{\'a}nchez-Janssen}, {Tolstoy}, {Venn}, \& {Youakim}}]{lms+18}
{Longeard}, N., {Martin}, N., {Starkenburg}, E., {et~al.} 2018, \mnras, 480,
  2609, \dodoi{10.1093/mnras/sty1986}

\bibitem[{{Longeard} {et~al.}(2021){Longeard}, {Martin}, {Ibata},
  {Starkenburg}, {Jablonka}, {Aguado}, {Carlberg}, {C{\^o}t{\'e}},
  {Gonz{\'a}lez Hern{\'a}ndez}, {Lucchesi}, {Malhan}, {Navarro},
  {S{\'a}nchez-Janssen}, {Thomas}, {Venn}, \& {McConnachie}}]{lmi+21}
{Longeard}, N., {Martin}, N., {Ibata}, R.~A., {et~al.} 2021, \mnras, 503, 2754,
  \dodoi{10.1093/mnras/stab604}

\bibitem[{{Majewski} {et~al.}(2003){Majewski}, {Skrutskie}, {Weinberg}, \&
  {Ostheimer}}]{msw+03}
{Majewski}, S.~R., {Skrutskie}, M.~F., {Weinberg}, M.~D., \& {Ostheimer}, J.~C.
  2003, \apj, 599, 1082, \dodoi{10.1086/379504}

\bibitem[{{Mateo} {et~al.}(2008){Mateo}, {Olszewski}, \& {Walker}}]{mow+08}
{Mateo}, M., {Olszewski}, E.~W., \& {Walker}, M.~G. 2008, \apj, 675, 201,
  \dodoi{10.1086/522326}

\bibitem[{{McConnachie} \& {C{\^o}t{\'e}}(2010)}]{2010ApJ...722L.209M}
{McConnachie}, A.~W., \& {C{\^o}t{\'e}}, P. 2010, \apjl, 722, L209,
  \dodoi{10.1088/2041-8205/722/2/L209}

\bibitem[{{McConnachie} \& {Venn}(2020)}]{McConnachie2020RNAAS...4..229M}
{McConnachie}, A.~W., \& {Venn}, K.~A. 2020, Research Notes of the American
  Astronomical Society, 4, 229, \dodoi{10.3847/2515-5172/abd18b}

\bibitem[{{McMillan}(2017)}]{2017MNRAS.465...76M}
{McMillan}, P.~J. 2017, \mnras, 465, 76, \dodoi{10.1093/mnras/stw2759}

\bibitem[{{Minor} {et~al.}(2010){Minor}, {Martinez}, {Bullock}, {Kaplinghat},
  \& {Trainor}}]{2010ApJ...721.1142M}
{Minor}, Q.~E., {Martinez}, G., {Bullock}, J., {Kaplinghat}, M., \& {Trainor},
  R. 2010, \apj, 721, 1142, \dodoi{10.1088/0004-637X/721/2/1142}

\bibitem[{{Mu{\~n}oz} {et~al.}(2018){Mu{\~n}oz}, {C{\^o}t{\'e}}, {Santana},
  {Geha}, {Simon}, {Oyarz{\'u}n}, {Stetson}, \&
  {Djorgovski}}]{2018ApJ...860...66M}
{Mu{\~n}oz}, R.~R., {C{\^o}t{\'e}}, P., {Santana}, F.~A., {et~al.} 2018, \apj,
  860, 66, \dodoi{10.3847/1538-4357/aac16b}

\bibitem[{{Mucciarelli} {et~al.}(2017){Mucciarelli}, {Bellazzini}, {Ibata},
  {Romano}, {Chapman}, \& {Monaco}}]{mbi+17}
{Mucciarelli}, A., {Bellazzini}, M., {Ibata}, R., {et~al.} 2017, \aap, 605,
  A46, \dodoi{10.1051/0004-6361/201730707}

\bibitem[{{Mutlu-Pakdil} {et~al.}(2018){Mutlu-Pakdil}, {Sand}, {Carlin},
  {Spekkens}, {Caldwell}, {Crnojevi{\'c}}, {Hughes}, {Willman}, \&
  {Zaritsky}}]{msc+18}
{Mutlu-Pakdil}, B., {Sand}, D.~J., {Carlin}, J.~L., {et~al.} 2018, \apj, 863,
  25, \dodoi{10.3847/1538-4357/aacd0e}

\bibitem[{{Mutlu-Pakdil} {et~al.}(2021){Mutlu-Pakdil}, {Sand}, {Crnojevi{\'c}},
  {Drlica-Wagner}, {Caldwell}, {Guhathakurta}, {Seth}, {Simon}, {Strader}, \&
  {Toloba}}]{msc+21}
{Mutlu-Pakdil}, B., {Sand}, D.~J., {Crnojevi{\'c}}, D., {et~al.} 2021, \apj,
  918, 88, \dodoi{10.3847/1538-4357/ac0db8}

\bibitem[{{Oemler} {et~al.}(2017){Oemler}, {Clardy}, {Kelson}, {Walth}, \&
  {Villanueva}}]{oemler_2017}
{Oemler}, A., {Clardy}, K., {Kelson}, D., {Walth}, G., \& {Villanueva}, E.
  2017, {COSMOS: Carnegie Observatories System for MultiObject Spectroscopy},
  Astrophysics Source Code Library, record ascl:1705.001.
\newblock \doeprint{1705.001}

\bibitem[{{Okamoto} {et~al.}(2008){Okamoto}, {Arimoto}, {Yamada}, \&
  {Onodera}}]{oay+08}
{Okamoto}, S., {Arimoto}, N., {Yamada}, Y., \& {Onodera}, M. 2008, \aap, 487,
  103, \dodoi{10.1051/0004-6361:20078232}

\bibitem[{{Pace} {et~al.}(2022){Pace}, {Erkal}, \& {Li}}]{2022arXiv220505699P}
{Pace}, A.~B., {Erkal}, D., \& {Li}, T.~S. 2022, \apj, 940, 136,
  \dodoi{10.3847/1538-4357/ac997b}

\bibitem[{{Pace} \& {Strigari}(2019)}]{Pace2019MNRAS.482.3480P}
{Pace}, A.~B., \& {Strigari}, L.~E. 2019, \mnras, 482, 3480,
  \dodoi{10.1093/mnras/sty2839}

\bibitem[{{Pianta} {et~al.}(2022){Pianta}, {Capuzzo-Dolcetta}, \&
  {Carraro}}]{Pianta2022ApJ...939....3P}
{Pianta}, C., {Capuzzo-Dolcetta}, R., \& {Carraro}, G. 2022, \apj, 939, 3,
  \dodoi{10.3847/1538-4357/ac9303}

\bibitem[{{Safarzadeh} \& {Spergel}(2020)}]{2020ApJ...893...21S}
{Safarzadeh}, M., \& {Spergel}, D.~N. 2020, \apj, 893, 21,
  \dodoi{10.3847/1538-4357/ab7db2}

\bibitem[{{Schlafly} \& {Finkbeiner}(2011)}]{schlafly_2011}
{Schlafly}, E.~F., \& {Finkbeiner}, D.~P. 2011, \apj, 737, 103,
  \dodoi{10.1088/0004-637X/737/2/103}

\bibitem[{{Shanks} {et~al.}(2015){Shanks}, {Metcalfe}, {Chehade}, {Findlay},
  {Irwin}, {Gonzalez-Solares}, {Lewis}, {Yoldas}, {Mann}, {Read}, {Sutorius},
  \& {Voutsinas}}]{2015MNRAS.451.4238S}
{Shanks}, T., {Metcalfe}, N., {Chehade}, B., {et~al.} 2015, \mnras, 451, 4238,
  \dodoi{10.1093/mnras/stv1130}

\bibitem[{Simon(2019)}]{doi:10.1146/annurev-astro-091918-104453}
Simon, J.~D. 2019, Annual Review of Astronomy and Astrophysics, 57, 375,
  \dodoi{10.1146/annurev-astro-091918-104453}

\bibitem[{{Simon} \& {Geha}(2007)}]{sg+07}
{Simon}, J.~D., \& {Geha}, M. 2007, \apj, 670, 313, \dodoi{10.1086/521816}

\bibitem[{{Simon} {et~al.}(2011){Simon}, {Geha}, {Minor}, {Martinez}, {Kirby},
  {Bullock}, {Kaplinghat}, {Strigari}, {Willman}, {Choi}, {Tollerud}, \&
  {Wolf}}]{sgm+11}
{Simon}, J.~D., {Geha}, M., {Minor}, Q.~E., {et~al.} 2011, \apj, 733, 46,
  \dodoi{10.1088/0004-637X/733/1/46}

\bibitem[{{Simon} {et~al.}(2015){Simon}, {Drlica-Wagner}, {Li}, {Nord}, {Geha},
  {Bechtol}, {Balbinot}, {Buckley-Geer}, {Lin}, {Marshall}, {Santiago},
  {Strigari}, {Wang}, {Wechsler}, {Yanny}, {Abbott}, {Bauer}, {Bernstein},
  {Bertin}, {Brooks}, {Burke}, {Capozzi}, {Carnero Rosell}, {Carrasco Kind},
  {D'Andrea}, {da Costa}, {DePoy}, {Desai}, {Diehl}, {Dodelson}, {Cunha},
  {Estrada}, {Evrard}, {Fausti Neto}, {Fernandez}, {Finley}, {Flaugher},
  {Frieman}, {Gaztanaga}, {Gerdes}, {Gruen}, {Gruendl}, {Honscheid}, {James},
  {Kent}, {Kuehn}, {Kuropatkin}, {Lahav}, {Maia}, {March}, {Martini}, {Miller},
  {Miquel}, {Ogando}, {Romer}, {Roodman}, {Rykoff}, {Sako}, {Sanchez},
  {Schubnell}, {Sevilla}, {Smith}, {Soares-Santos}, {Sobreira}, {Suchyta},
  {Swanson}, {Tarle}, {Thaler}, {Tucker}, {Vikram}, {Walker}, {Wester}, \& {DES
  Collaboration}}]{sdl+15}
{Simon}, J.~D., {Drlica-Wagner}, A., {Li}, T.~S., {et~al.} 2015, \apj, 808, 95,
  \dodoi{10.1088/0004-637X/808/1/95}

\bibitem[{Simon {et~al.}(2017)Simon, Li, Drlica-Wagner, Bechtol, Marshall,
  James, Wang, Strigari, Balbinot, Kuehn, Walker, Abbott, Allam, Annis,
  Benoit-L{\'{e}}vy, Brooks, Buckley-Geer, Burke, Rosell, Kind, Carretero,
  Cunha, D'Andrea, da~Costa, DePoy, Desai, Doel, Fernandez, Flaugher, Frieman,
  Garc{\'{\i}}a-Bellido, Gaztanaga, Goldstein, Gruen, Gutierrez, Kuropatkin,
  Maia, Martini, Menanteau, Miller, Miquel, Neilsen, Nord, Ogando, Plazas,
  Romer, Rykoff, Sanchez, Santiago, Scarpine, Schubnell, Sevilla-Noarbe, Smith,
  Sobreira, Suchyta, Swanson, Tarle, Whiteway, \& and}]{Simon_2017}
Simon, J.~D., Li, T.~S., Drlica-Wagner, A., {et~al.} 2017, The Astrophysical
  Journal, 838, 11, \dodoi{10.3847/1538-4357/aa5be7}

\bibitem[{{Simon} {et~al.}(2020){Simon}, {Li}, {Erkal}, {Pace},
  {Drlica-Wagner}, {James}, {Marshall}, {Bechtol}, {Hansen}, {Kuehn}, {Lidman},
  {Allam}, {Annis}, {Avila}, {Bertin}, {Brooks}, {Burke}, {Rosell}, {Carrasco
  Kind}, {Carretero}, {da Costa}, {De Vicente}, {Desai}, {Doel}, {Eifler},
  {Everett}, {Fosalba}, {Frieman}, {Garc{\'\i}a-Bellido}, {Gaztanaga},
  {Gerdes}, {Gruen}, {Gruendl}, {Gschwend}, {Gutierrez}, {Hollowood},
  {Honscheid}, {Krause}, {Kuropatkin}, {MacCrann}, {Maia}, {March}, {Miquel},
  {Palmese}, {Paz-Chinch{\'o}n}, {Plazas}, {Reil}, {Roodman}, {Sanchez},
  {Santiago}, {Scarpine}, {Schubnell}, {Serrano}, {Smith}, {Suchyta}, {Tarle},
  {Walker}, \& {DES Collaboration}}]{Simon2020ApJ...892..137S}
{Simon}, J.~D., {Li}, T.~S., {Erkal}, D., {et~al.} 2020, \apj, 892, 137,
  \dodoi{10.3847/1538-4357/ab7ccb}

\bibitem[{{Spencer} {et~al.}(2017){Spencer}, {Mateo}, {Walker}, {Olszewski},
  {McConnachie}, {Kirby}, \& {Koch}}]{smw+17}
{Spencer}, M.~E., {Mateo}, M., {Walker}, M.~G., {et~al.} 2017, \aj, 153, 254,
  \dodoi{10.3847/1538-3881/aa6d51}

\bibitem[{{Strigari}(2018)}]{2018RPPh...81e6901S}
{Strigari}, L.~E. 2018, Reports on Progress in Physics, 81, 056901,
  \dodoi{10.1088/1361-6633/aaae16}

\bibitem[{{Tarumi} {et~al.}(2021){Tarumi}, {Yoshida}, \&
  {Frebel}}]{2021ApJ...914L..10T}
{Tarumi}, Y., {Yoshida}, N., \& {Frebel}, A. 2021, \apjl, 914, L10,
  \dodoi{10.3847/2041-8213/ac024e}

\bibitem[{{Torrealba} {et~al.}(2016){Torrealba}, {Koposov}, {Belokurov}, \&
  {Irwin}}]{tkb+16b}
{Torrealba}, G., {Koposov}, S.~E., {Belokurov}, V., \& {Irwin}, M. 2016,
  \mnras, 459, 2370, \dodoi{10.1093/mnras/stw733}

\bibitem[{Torrealba {et~al.}(2016)Torrealba, Koposov, Belokurov, Irwin,
  Collins, Spencer, Ibata, Mateo, Bonaca, \& Jethwa}]{Torrealba_2016}
Torrealba, G., Koposov, S.~E., Belokurov, V., {et~al.} 2016, Monthly Notices of
  the Royal Astronomical Society, 463, 712, \dodoi{10.1093/mnras/stw2051}

\bibitem[{{Torrealba} {et~al.}(2018){Torrealba}, {Belokurov}, {Koposov},
  {Bechtol}, {Drlica-Wagner}, {Olsen}, {Vivas}, {Yanny}, {Jethwa}, {Walker},
  {Li}, {Allam}, {Conn}, {Gallart}, {Gruendl}, {James}, {Johnson}, {Kuehn},
  {Kuropatkin}, {Martin}, {Martinez-Delgado}, {Nidever}, {No{\"e}l}, {Simon},
  {Stringfellow}, \& {Tucker}}]{tbk+18}
{Torrealba}, G., {Belokurov}, V., {Koposov}, S.~E., {et~al.} 2018, \mnras, 475,
  5085, \dodoi{10.1093/mnras/sty170}

\bibitem[{{Walker} {et~al.}(2009{\natexlab{a}}){Walker}, {Mateo}, \&
  {Olszewski}}]{wmo+09b}
{Walker}, M.~G., {Mateo}, M., \& {Olszewski}, E.~W. 2009{\natexlab{a}}, \aj,
  137, 3100, \dodoi{10.1088/0004-6256/137/2/3100}

\bibitem[{Walker {et~al.}(2006)Walker, Mateo, Olszewski, Bernstein, Wang, \&
  Woodroofe}]{Walker_2006}
Walker, M.~G., Mateo, M., Olszewski, E.~W., {et~al.} 2006, The Astronomical
  Journal, 131, 2114, \dodoi{10.1086/500193}

\bibitem[{{Walker} {et~al.}(2009{\natexlab{b}}){Walker}, {Mateo}, {Olszewski},
  {Pe{\~n}arrubia}, {Evans}, \& {Gilmore}}]{wmo+09}
{Walker}, M.~G., {Mateo}, M., {Olszewski}, E.~W., {et~al.} 2009{\natexlab{b}},
  \apj, 704, 1274, \dodoi{10.1088/0004-637X/704/2/1274}

\bibitem[{{Walsh} {et~al.}(2007){Walsh}, {Jerjen}, \&
  {Willman}}]{2007ApJ...662L..83W}
{Walsh}, S.~M., {Jerjen}, H., \& {Willman}, B. 2007, \apjl, 662, L83,
  \dodoi{10.1086/519684}

\bibitem[{Walsh {et~al.}(2008)Walsh, Willman, Sand, Harris, Seth, Zaritsky, \&
  Jerjen}]{Walsh_2008}
Walsh, S.~M., Willman, B., Sand, D., {et~al.} 2008, The Astrophysical Journal,
  688, 245, \dodoi{10.1086/592076}

\bibitem[{{Weisz} {et~al.}(2014){Weisz}, {Dolphin}, {Skillman}, {Holtzman},
  {Gilbert}, {Dalcanton}, \& {Williams}}]{weisz_2014}
{Weisz}, D.~R., {Dolphin}, A.~E., {Skillman}, E.~D., {et~al.} 2014, \apj, 789,
  148, \dodoi{10.1088/0004-637X/789/2/148}

\bibitem[{{Willman} {et~al.}(2011){Willman}, {Geha}, {Strader}, {Strigari},
  {Simon}, {Kirby}, {Ho}, \& {Warres}}]{wgs+11}
{Willman}, B., {Geha}, M., {Strader}, J., {et~al.} 2011, \aj, 142, 128,
  \dodoi{10.1088/0004-6256/142/4/128}

\bibitem[{{Wolf} {et~al.}(2010){Wolf}, {Martinez}, {Bullock}, {Kaplinghat},
  {Geha}, {Mu{\~n}oz}, {Simon}, \& {Avedo}}]{2010MNRAS.406.1220W}
{Wolf}, J., {Martinez}, G.~D., {Bullock}, J.~S., {et~al.} 2010, \mnras, 406,
  1220, \dodoi{10.1111/j.1365-2966.2010.16753.x}

\bibitem[{{Zaritsky} \& {Behroozi}(2022)}]{2022arXiv221202948Z}
{Zaritsky}, D., \& {Behroozi}, P. 2022, arXiv e-prints, arXiv:2212.02948.
\newblock \doarXiv{2212.02948}

\end{thebibliography}



\end{document}